\documentclass[pdflatex,sn-nature]{sn-jnl}
\usepackage{graphicx}%
\usepackage{multirow}%
\usepackage{fix-cm}  
\usepackage{amsmath,amssymb,amsfonts}%
\usepackage{amsthm}%
\usepackage{mathrsfs}%
\usepackage[title]{appendix}%
\usepackage{xcolor}%
\usepackage{textcomp}%
\usepackage{manyfoot}%
\usepackage{booktabs}%
\usepackage{tabularx}%
\usepackage{array} 
\usepackage{algorithm}%
\usepackage{algorithmicx}%
\usepackage{algpseudocode}%
\usepackage{listings}%
\usepackage{comment}
\usepackage{subcaption}
\usepackage{placeins}
\usepackage[colorinlistoftodos]{todonotes}
\theoremstyle{thmstyleone}%
\theoremstyle{thmstyletwo}%
\theoremstyle{thmstylethree}%
\renewcommand{\thesection}{\Roman{section}} 
\renewcommand{\thesubsection}{\thesection.\Alph{subsection}} 

\begin{document}

\title[Article Title]{Solid-state transcapacitor : a new gain element for logic, memory and interconnects}

\author*{\fnm{Amrita} \sur{Mathuriya}$^{*}$}
\email{amrita.mathuriya@keplercompute.com}
\author{\fnm{Roza} \sur{Kotlyar}}
\author{\fnm{Neal} \sur{Reynolds}}
\author{\fnm{Rafael} \sur{Rios}}
\author{\fnm{Alan} \sur{Kalitsov}}

\author{\fnm{Peter B.} \sur{Meisenheimer}}
\author{\fnm{James} \sur{Clarkson}}
\author{\fnm{Noriyuki} \sur{Sato}}
\author{\fnm{Tanay} \sur{Gosavi}}
\author{\fnm{Ramamoorthy} \sur{Ramesh}}
\author{\fnm{Dmitri E.} \sur{Nikonov}}
\author*{\fnm{Sasikanth} \sur{Manipatruni}*}
\email{sm448@cornell.edu}

\affil{
\orgname{
Kepler Computing}
}

\abstract{Today's transistors dictate the voltage and charge scales for both logic and memory systems\cite{horowitz20141, dallyhbm}. While AI computation is widely understood to be limited by memory energy\cite{shalf2020future}, the dominant share of the energy is expended in the dense interconnects inside and to the memory\cite{adhinarayanan2025folded}, whose energy needs are set by the transistors\cite{kim2010equalized, li2022interconnect}.   The energy scaling challenges of modern transistors can be attributed to simultaneously meeting high current density, high current/impedance modulation, and the inability to lower voltages in CMOS\cite{cao2023future, horowitz2005scaling}. Hence, a new logic element that lowers the voltage and charge needs is a priority, not only for lowering logic power but also memory access power. Here, we propose a novel 3-terminal logic element for low energy computing, a solid-state transcapacitor (TCAP). A TCAP is a solid state displacement current modulator realized by a gate which controls the charge-voltage relationship of the channel.  Unlike transistors, TCAPs eliminate the dissipative transport current, are not bound by the Boltzmann current modulation limit, and operate with displacement currents limited only by the polarization response and contact resistance. Since there is no partially conducting path for complementary circuits, they can operate with a smaller ON/OFF impedance modulation. Hence, TCAP circuits may simultaneously overcome the voltage, current density, and current modulation limits of CMOS. 
We describe a solid state TCAP using a piezoelectric transcapacitor in which a gate-controlled stressor modulates the capacitance of a polar channel via electromechanical coupling. This device achieves inversion and gain—essential for logic—and is functionally equivalent to a 1T-1C memory cell, enabling dense memory. 
Using voltage scaling, reactive-capacitive energy recovery circuits\cite{singh2009static}, and the high polarization densities of polar materials, the logic based on TCAP offers a pathway to 100× lower energy consumption with a delay comparable to ultimately scaled CMOS devices\cite{yang2025multi}.
We describe the material scaling path for TCAP and other alternative modes of TCAP via metal-insulator transition of electrodes and transverse ferroelectric modulation\cite{guptatrans}. 
This approach provides a new potential pathway for low-energy computing beyond the limits of transistors using electro-mechanics\cite{he2020tunable} and multiferroics\cite{fiebig2016evolution}\cite{Gupta2026transswitching}.
}

\maketitle

\section{Introduction}
\label{sec:intro}

Even though the thermodynamic limits to computing efficiency\cite{landauer1976} are much higher\cite{bennett1985fundamental}, today's logic circuits and memory interconnects operate at a significantly lower level due to the physical constraints of the CMOS transistor.  The CMOS logic energy and memory (interconnect) scaling is fundamentally bound by the simultaneous constraints of lowering the voltage, improving the gate modulation per unit charge, and meeting the drive current per unit length. A high $I_{\text{on}}/I_{\text{off}}$ ratio is an additional pre-requisite for complementary logic design due to the resistive division in CMOS circuits, balancing the energy of active switching versus leakage in the non-active devices.  As we look for options for the next 20-30 years of scaling (requiring a 100x energy reduction), the impressive array of complex device options in play today \cite{yang2025multi,cao2023future,UygarAvci} are all bound by either the Boltzmann limit of 60 mV/decade or by a low drive current.\cite{appenzeller2004band}. The best case is therefore continued density improvement with limited voltage and charge scaling. 

We especially note that the memory access energy per unit length, which limits AI, is bound by\cite{shalf2020future} $\frac{\pi \epsilon_0}{\ln(d/a)} $ $V^2$ and is set by fundamental constants (capacitance per unit length, set by $\epsilon_0$ and line fill factor (d/a)) and the operating voltage of CMOS. The ability to lower the interconnect energy per unit length with lower swing and optimally repeated interconnects is ultimately bound by the threshold voltage,
$V_t$, of the transistors used in the transmitters and receivers\cite{kim2010equalized, li2022interconnect}.  Energy breakdown of the AI memory systems\cite{shalf2020future, dallyhbm} show that the AI memory access energy per bit is actually dominated by interconnects in the package and routing inside DRAM. The length scale matching of memory arrays and linear bandwidth density targets prohibit the use of optics due to the mismatch between wavelength ($\approx 1550 nm$) and memory ($<30-100 nm$).  Hence, it is essential to look for new modes of gain elements for both compute and memory that can lower the voltage and the net charge exchanged between the supply and the ground.      

In this paper, we propose the solid-state transcapacitor, a  3-terminal gain element for logic. A transcapacitor is a three terminal circuit element relating large signal voltage at a capacitive gate with a capacitive channel (See Table~\ref{tab:bjt_mosfet_tcap}). Depending on the details of the device operation, 
a transcapacitor can modify the effective channel capacitance from a low value, $C_{L}$, 
to a high value $C_{H}$ 
or modify the polar response function $P(V_{CH})$ (polarization vs potential difference across the channel). 
A variety of potential pathways for TCAPs are feasible, 
notably with air gaps and micro-electro-mechanical 
devices\cite{galisultanov2017capacitive}. 
But airgap systems must meet the gain requirement as well as overcome device adhesive force scaling~\cite{pawashe2013scaling}. 
Solid state transcapacitors meeting device scalability are possible, comprising the channel or electrode modulation (phase change, polar to charge, magnetic or strain order parameters). 

\begin{figure}[htbp]
\centering
\begin{subfigure}{0.25\textwidth}
\includegraphics[width=\textwidth]{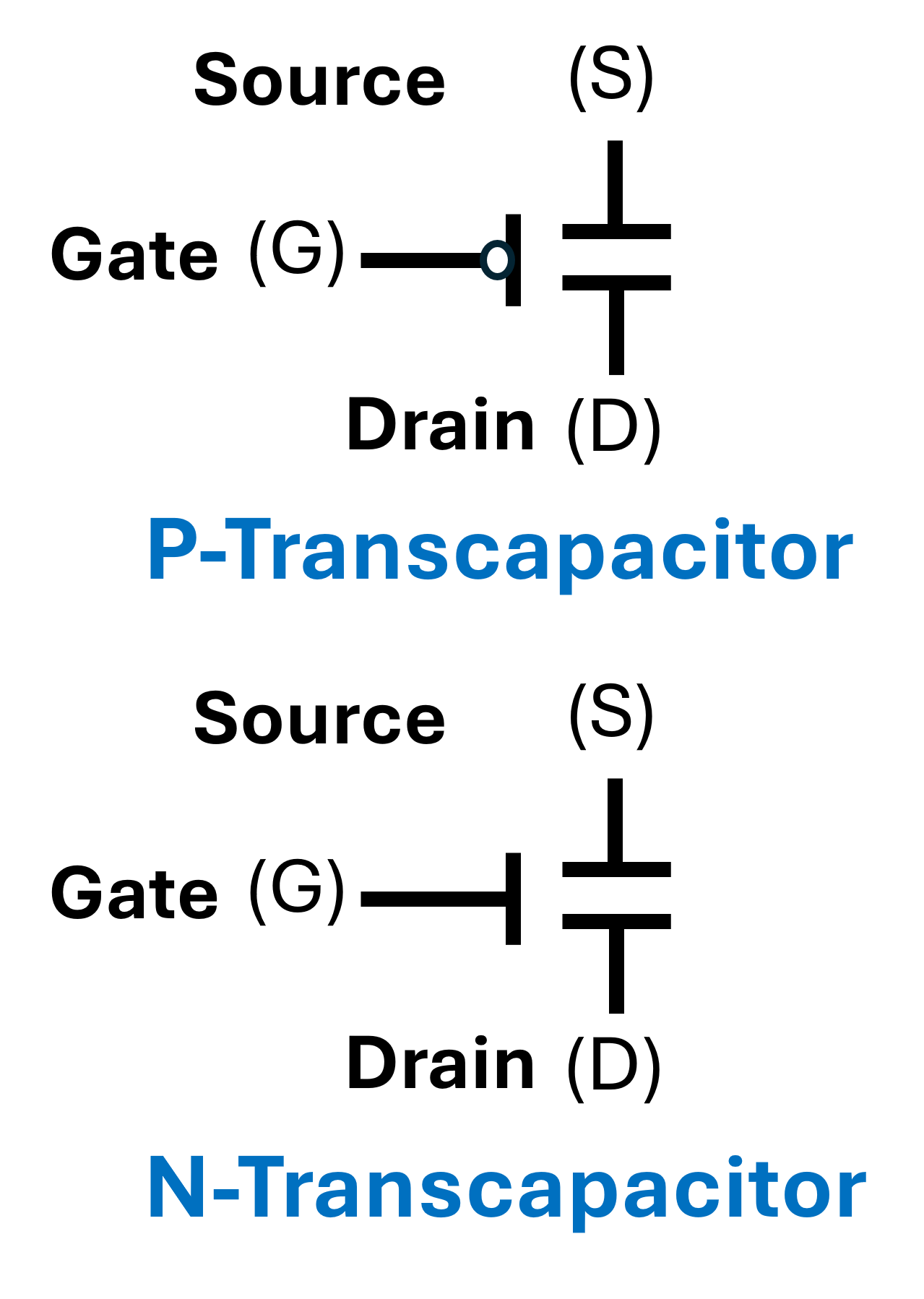}
\caption{}
\label{fig:Figure1_symbols}
\end{subfigure}%
\begin{subfigure}{0.37\textwidth}
\includegraphics[width=\textwidth]{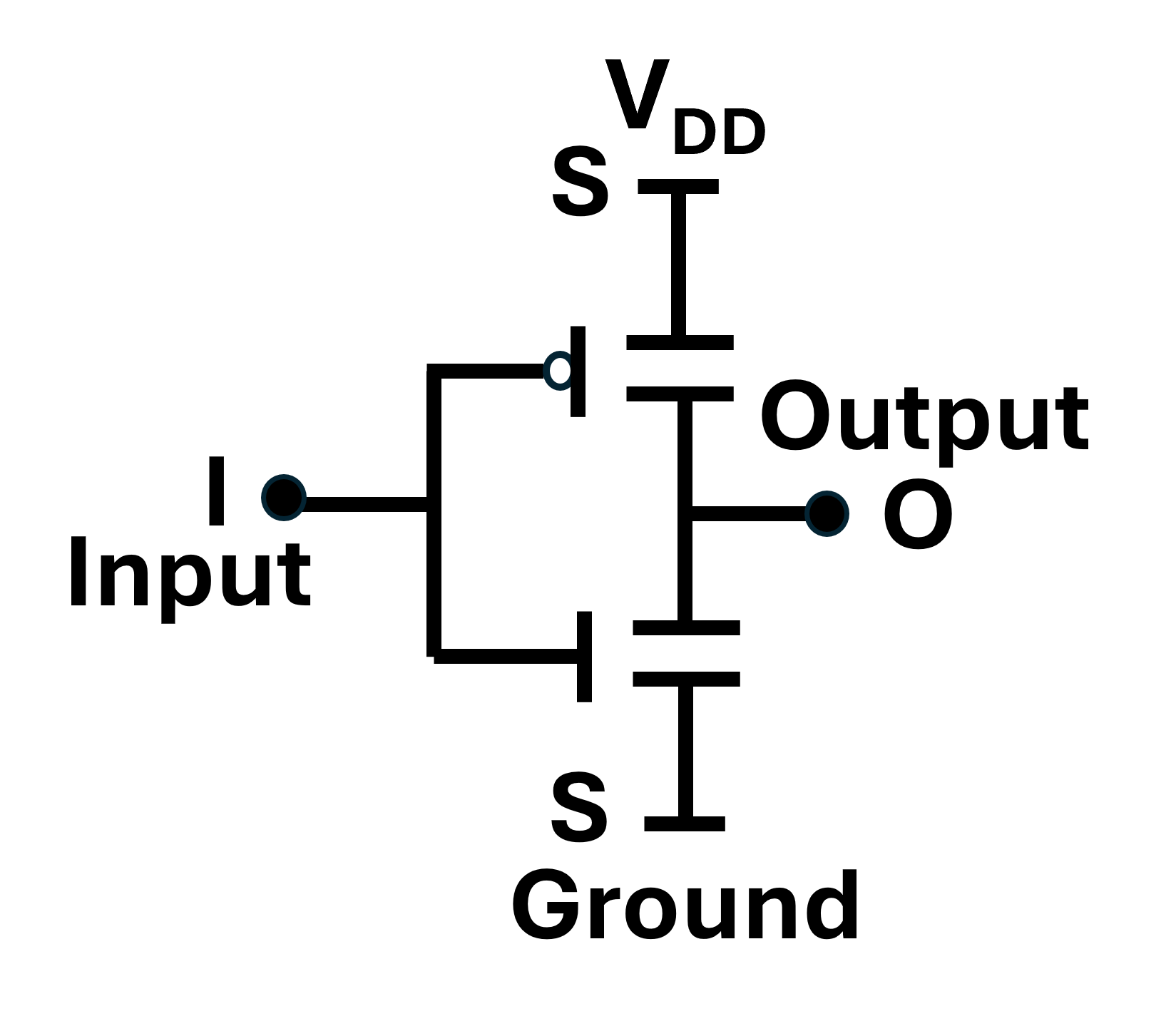}
\caption{}
\label{fig:Figure1_unloadedInv}
\end{subfigure}%
\begin{subfigure}{0.32\textwidth}
\includegraphics[width=\textwidth]{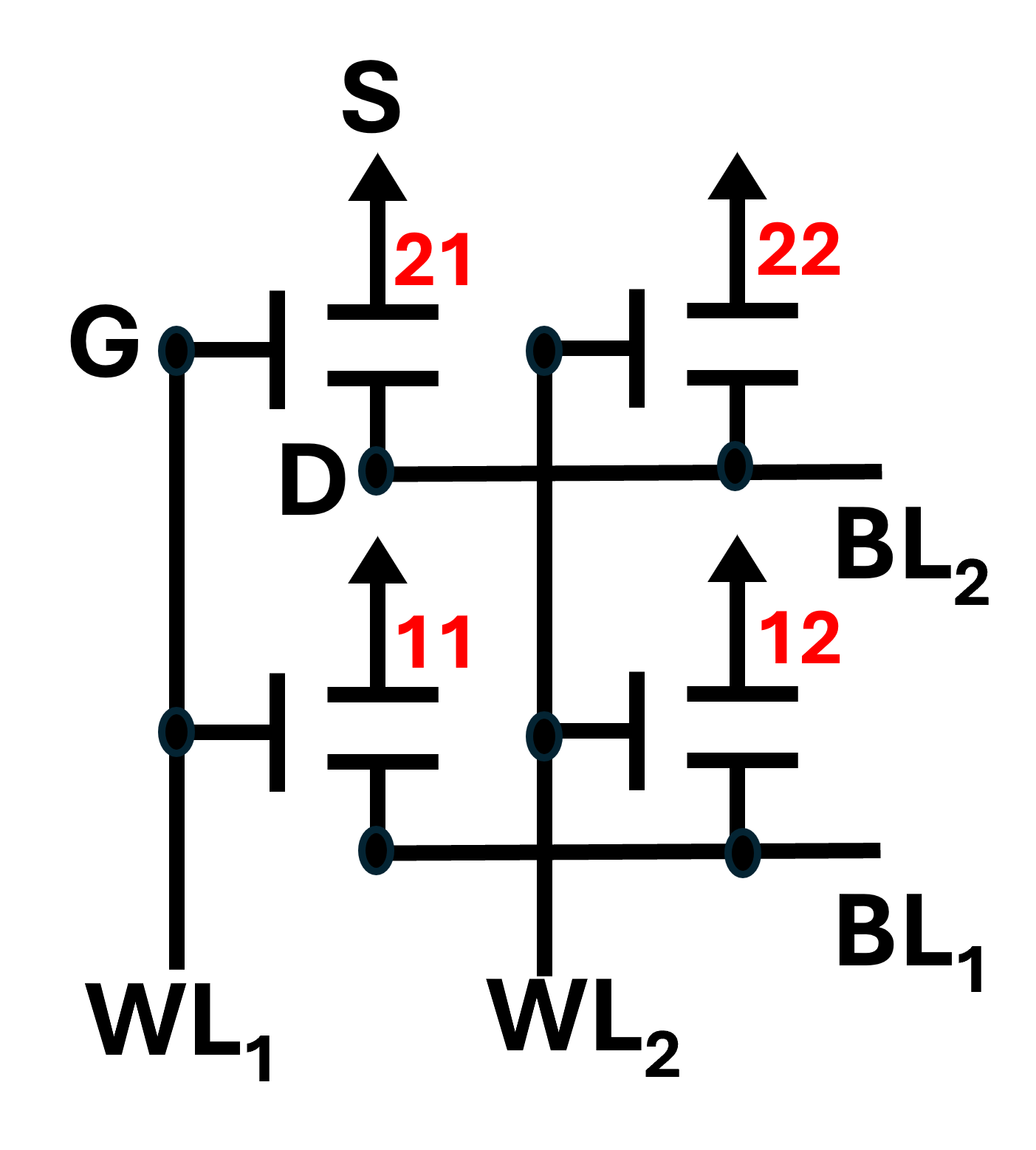}
\caption{}
\label{fig:Figure1_2x2_memory}
\end{subfigure}
\vspace{0.3em}
\begin{subfigure}{0.49\textwidth}
\includegraphics[width=\textwidth]{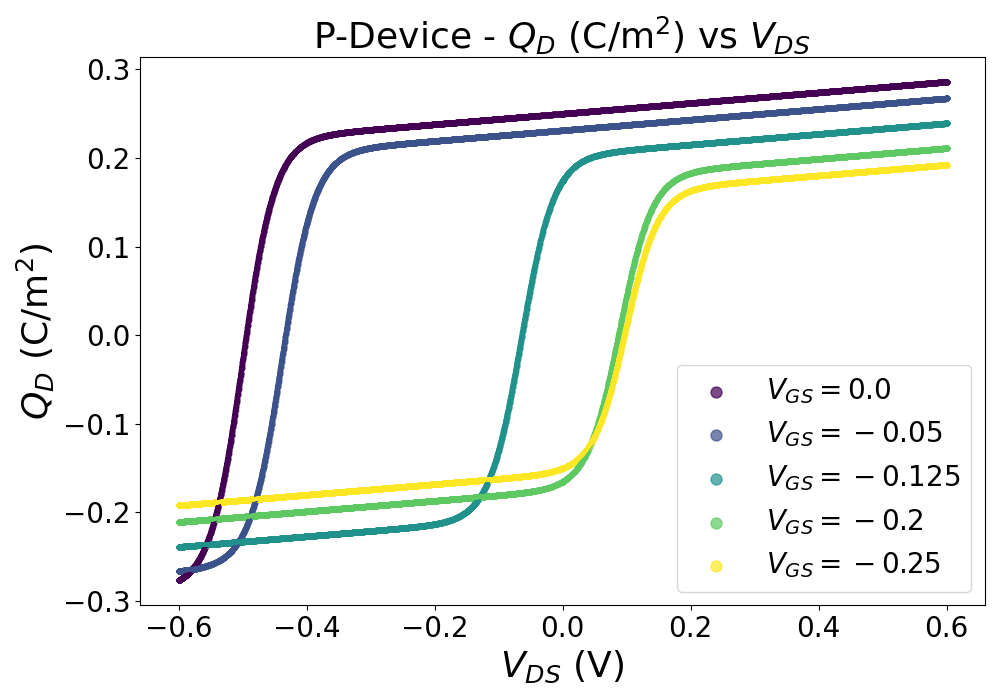}
\caption{}
\label{fig1:QDvsVDS_N}
\end{subfigure}%
\begin{subfigure}{0.49\textwidth}
\includegraphics[width=\textwidth]{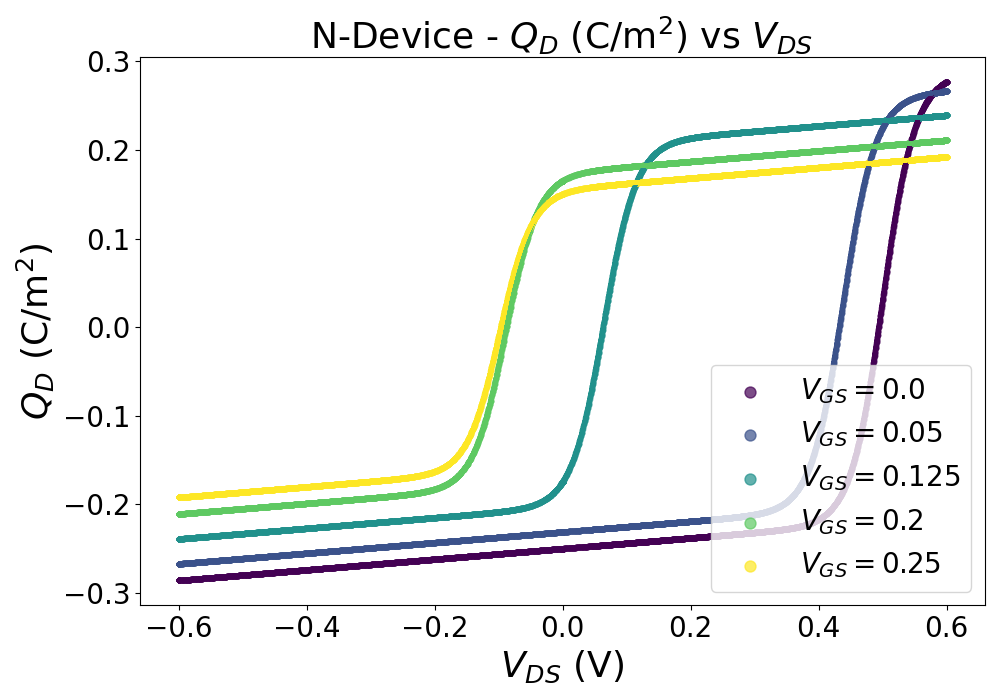}
\caption{}
\label{fig1:QDvsVDS_P}
\end{subfigure}
\vspace{0.3em}
\begin{subfigure}{0.49\textwidth}
\includegraphics[width=\textwidth]{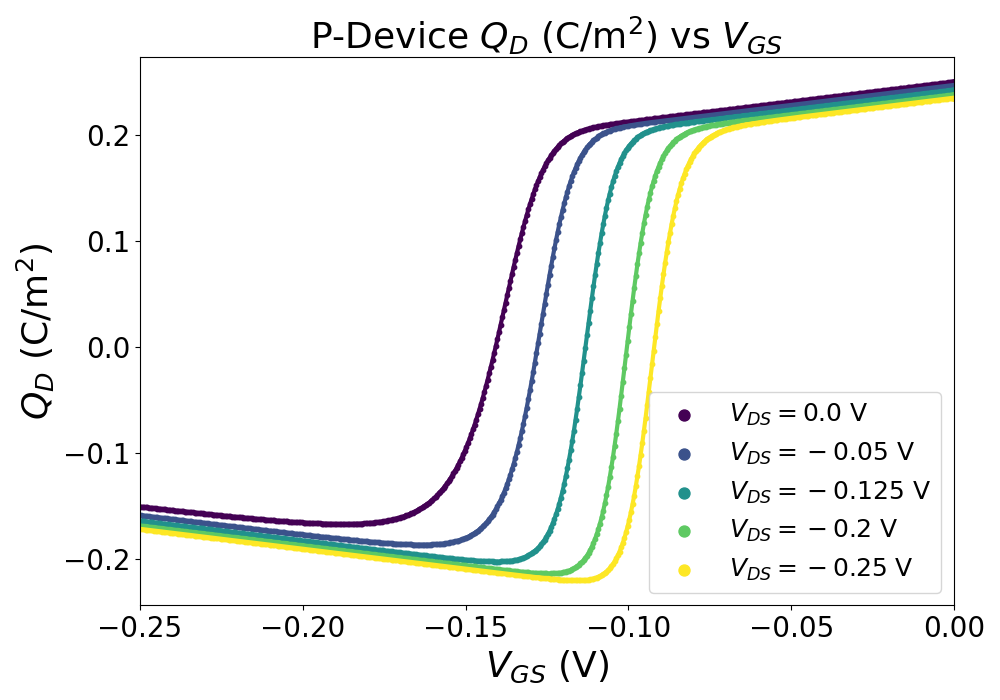}
\caption{}
\label{fig:Figure1_sub6}
\end{subfigure}%
\begin{subfigure}{0.49\textwidth}
\includegraphics[width=\textwidth]{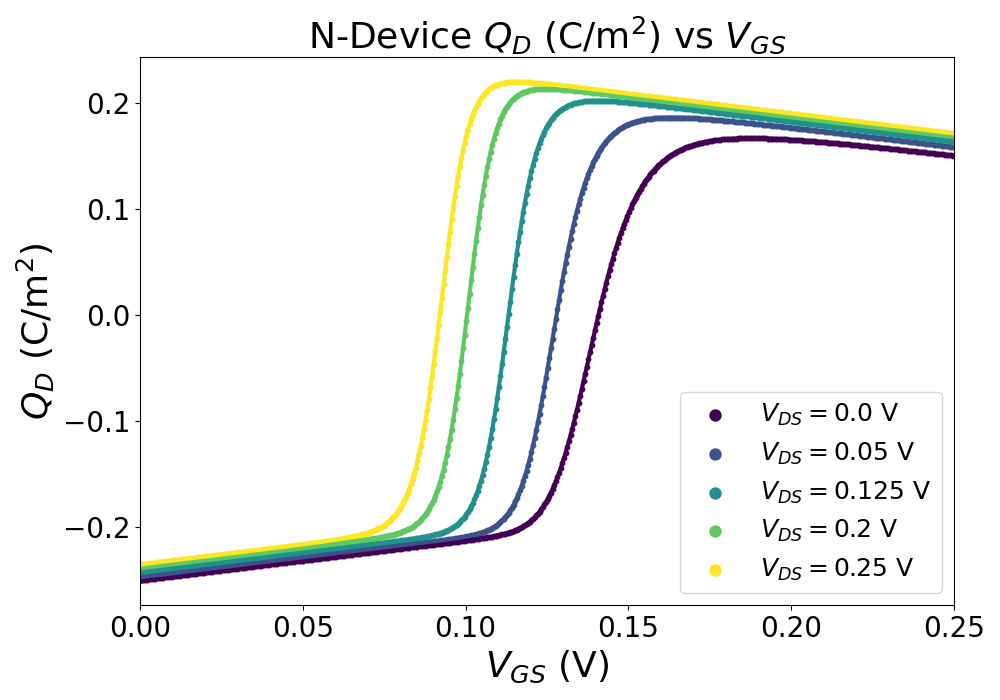}
\caption{}
\label{fig:Figure1_sub7}
\end{subfigure}
\caption{(a) Circuit symbols for transcapacitors, (b) An Unloaded inverter with TCAPs, (c) 2x2 memory array circuit with TCAPs,
(d-e) $Q_D$ vs. $V_{DS}$ for P and N devices respectively, 
(f-g) $Q_D$ vs $V_{GS}$ transfer plots for P and N devices respectively.}
    \label{fig:Figure1}
\end{figure}

As a circuit element, a solid state TCAP operates by modulating the effective capacitance (impedance) or P(V) of the channel and drives the output via charge and current balance of the nodes. This is in exact analog to transistor as a current driver. A complementary TCAP (CTCAP) inverter comprises a P-TCAP and an N-TCAP connected in series with a shared gate node.  
A CTCAP inverter (See Fig.~\ref{fig:Figure1}) operates similarly to a CMOS  inverter\cite{sakurai2002alpha}, with the output voltage given by the impedance division or charge transfer from  the supply voltage.  The instantaneous transient current through the devices is given by
\begin{equation}
\frac{dQ}{dt} = V \frac{dC}{dt} + C \frac{dV}{dt} = V \epsilon_0 \frac{A}{d} \frac{d\epsilon}{dt} + C \frac{dV}{dt},
\end{equation}
where $Q$ is the charge, $V$ is voltage, $C$ is capacitance,
$A$ is the area, $d$ is the thickness of the capacitor,
$\epsilon$ is the dielectric constant.

In Section~\ref{sec:EdCompar} we present the method of benchmarking the area, energy and delay of CMOS and beyond-CMOS devices and circuits, see the explanations and notation there.
Here we summarize the results of it.
The ratio of delays of CMOS and transcapacitors is
\begin{equation}
\frac{\tau_{tr}}{\tau_{tc}}
= \frac{C_L V_{DD} (0.5+F_1 F_2) }
{I_{d0}
(R_{ic} \Delta Q / \Delta V
+\tau_{ferroic}) }
=\frac{\tau_{\rm cmos} (0.5 + F_1 F_2)}
{ \left(
\frac{\Delta Q}{I_{\rm disp}}
+\tau_{ferroic}
\right)}
\label{eq:delay_ratio}
\end{equation}
The ratio of energies is 
\begin{equation}
\frac{E_{tr}}{E_{tc}}  = 
\frac{C_L V_{DD}^2
(0.5 + F_2 F_3)}
{(2 V_c \Delta Q_{fe} + 
4 R_{ic} \Delta Q^2 / \tau_{tc})}
\label{eq:energy_ratio}
\end{equation}
$\tau_{tr}$ is limited by the drive current $I_{d0}$ of the transistor. 
$\tau_{ferroic}$ is the nucleation time of the channel polar material if a ferroic order is used. For non-ferroic channels, 
the charging time of the transcapacitor 
is limited only by the displacement 
current ${I_{\text{disp}}}$
set by the contact resistance of the channel capacitor.  
Substituting typical values of quantities, we obtain that the ratio of the delays varies from 15 to 0.23, 
while $\tau_{ferroic}$ varies from 0 to 20ps.
The ratio of the energies varies from 
329 to 19, while $V_c$ varies from 0 to 50mV.
See the definitions and the values of parameters used in 
Table~\ref{tab:compare_cmos_tcap}.
This is in agreement with the benchmarking estimates and detailed CMOS advanced node circuit simulations in Section~\ref{sec:EdCompar}. 
TCAP could provide superior current density over CMOS by utilizing a different mode of current not constrained by the channel density of states.  The characteristic currents for TCAP are given by the peak displacement currents ($\mathbf{J}_d = \frac{\partial}{\partial t}(\varepsilon_0 \mathbf{E} + \mathbf{P})$), while the transistor peak current is limited by the density of states. Here $J$ is the current per unit of width $w$ of the device
\begin{equation}
\frac{J_{tr}}{J_{tc}} 
\approx
\frac{e^2 \cdot DOS \cdot v_e V_{DD}w R_{ic}}
{\Delta V},
\label{eq:current_ratio}
\end{equation}
where $DOS$ is the 2D density of states in the channel,
and $v_e$ is the average electron velocity in the channel. 

\begin{figure}[htbp]
\centering
\begin{subfigure}[t]{0.39\linewidth}
\includegraphics[width=\linewidth]{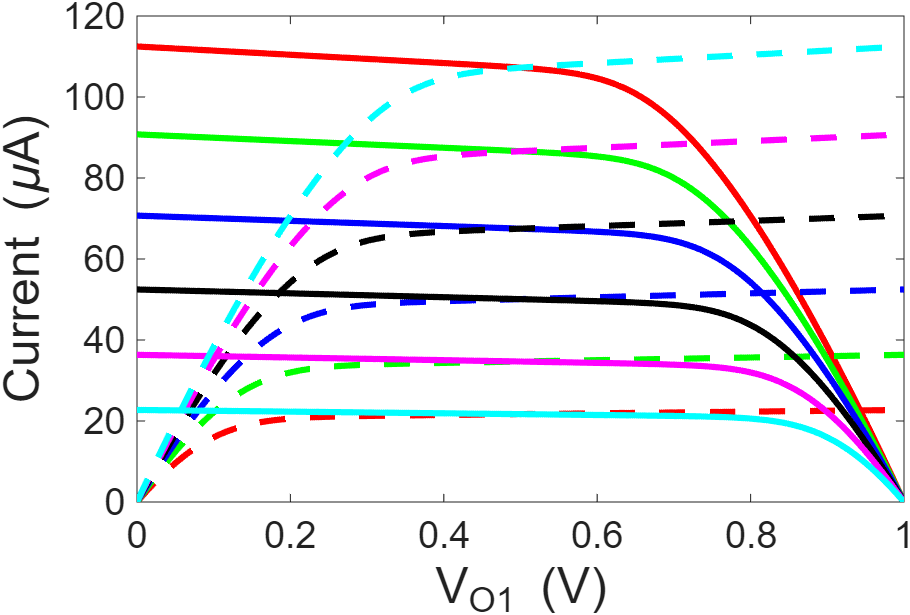}
\caption{}
\label{fig:fig2_coms_loadline}
\end{subfigure}
\hfill
\begin{subfigure}[t]{0.4\linewidth}
\includegraphics[width=\linewidth]{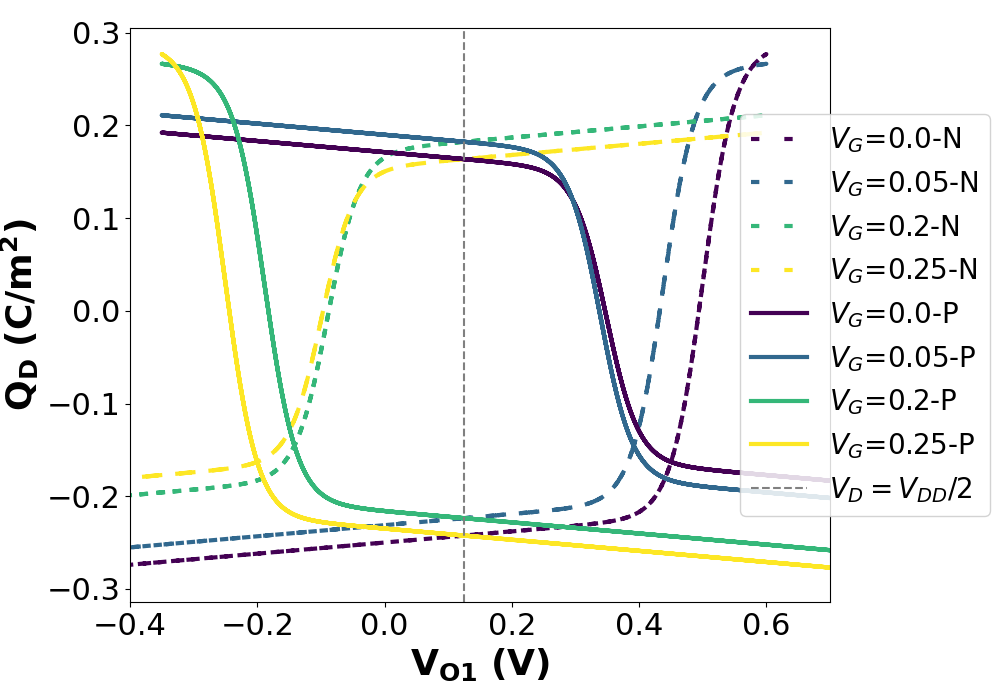}
\caption{}    
\label{fig:fig2_tcap_loadline}
\end{subfigure}
\centering
\begin{subfigure}[t]{0.39\linewidth}
\includegraphics[width=\linewidth]{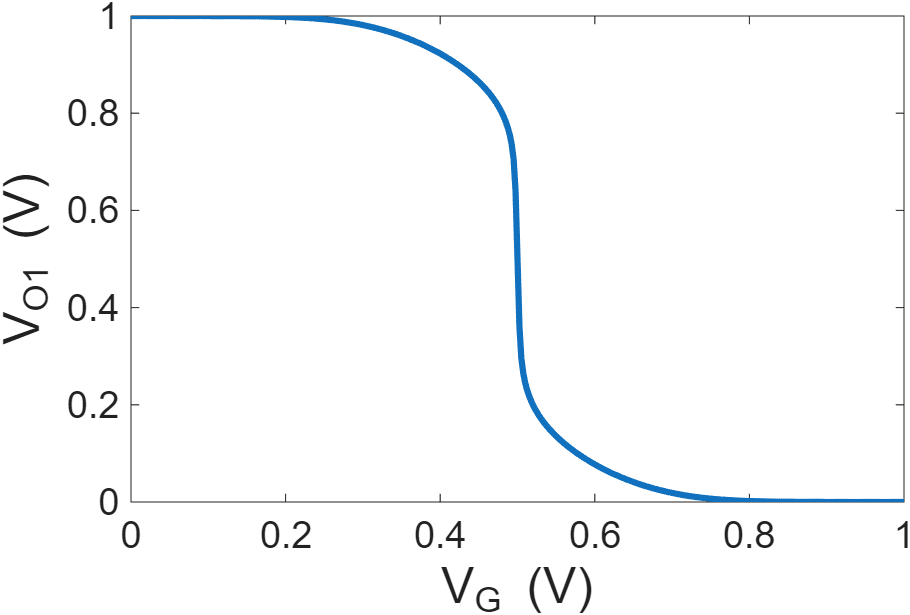}
\caption{}
\label{fig:fig2_coms_transferFunction}
\end{subfigure}
\hfill
\begin{subfigure}[t]{0.4\linewidth}
\includegraphics[width=\linewidth]{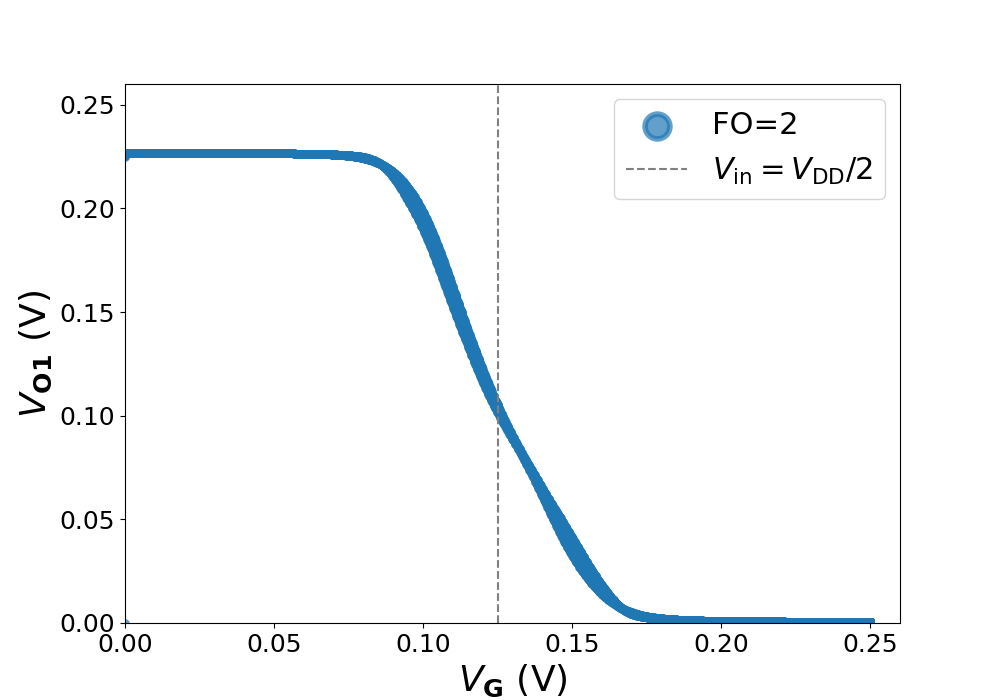}
\caption{}    
\label{fig:fig2_tcap_transferFunction}
\end{subfigure}
\centering
\begin{subfigure}[t]{0.39\linewidth}
\includegraphics[width=\linewidth]{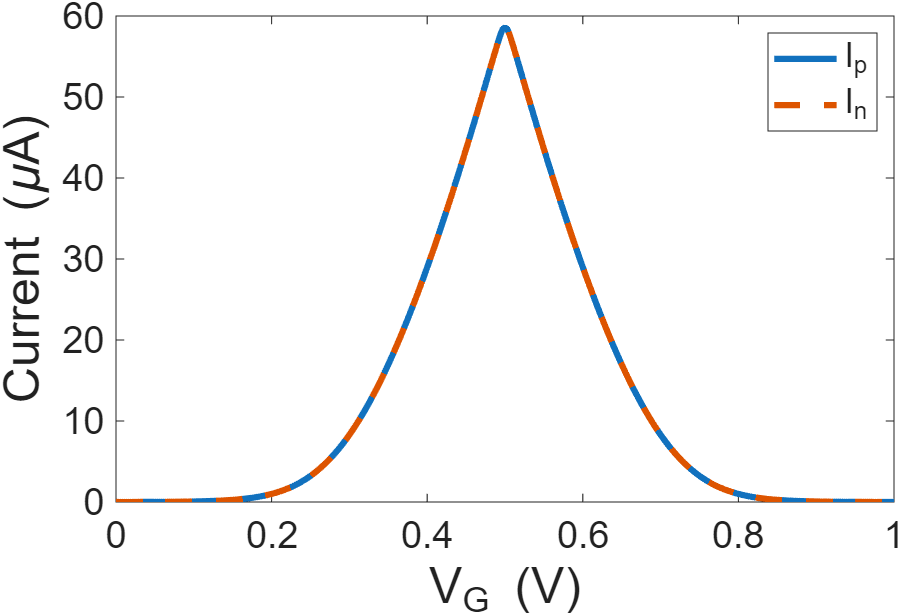}
\caption{}
\label{fig:fig2_cmos_current}
\end{subfigure}
\hfill
\begin{subfigure}[t]{0.4\linewidth}
\includegraphics[width=\linewidth]{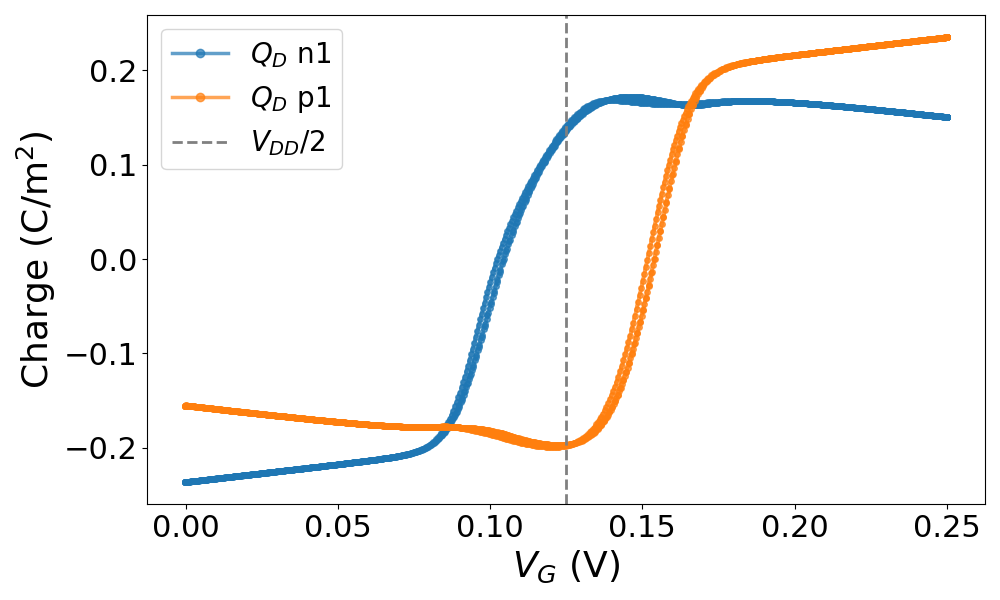}
\caption{}
\label{fig:fig2_tcap_current}
\end{subfigure}
\centering
\begin{subfigure}[t]{0.45\linewidth}
\includegraphics[width=\linewidth]{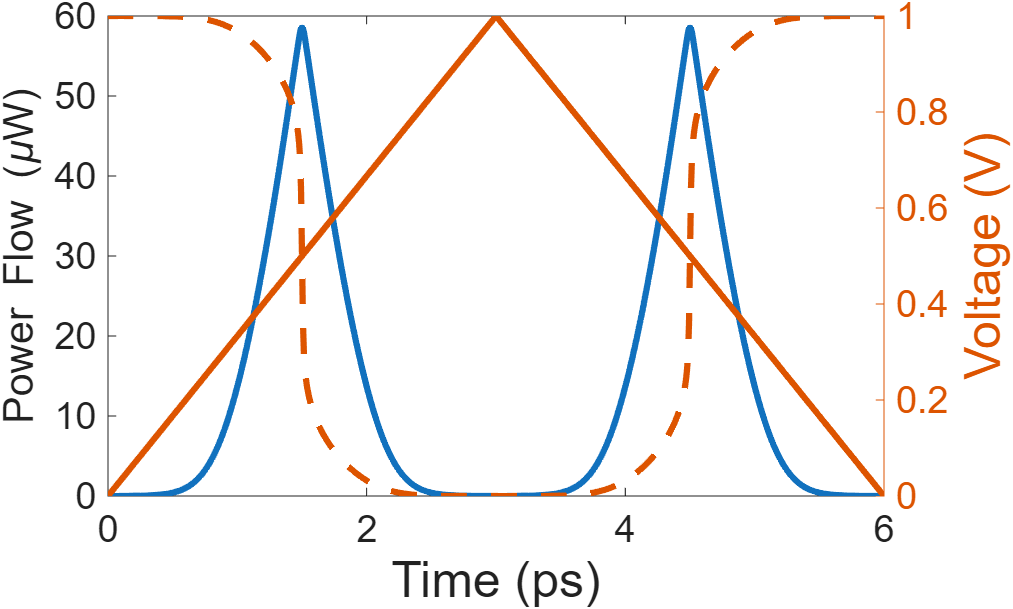}
\caption{}
\label{fig:fig2_cmos_powerFlow}
\end{subfigure}
\hfill
\begin{subfigure}[t]{0.4\linewidth}
\includegraphics[width=\linewidth]{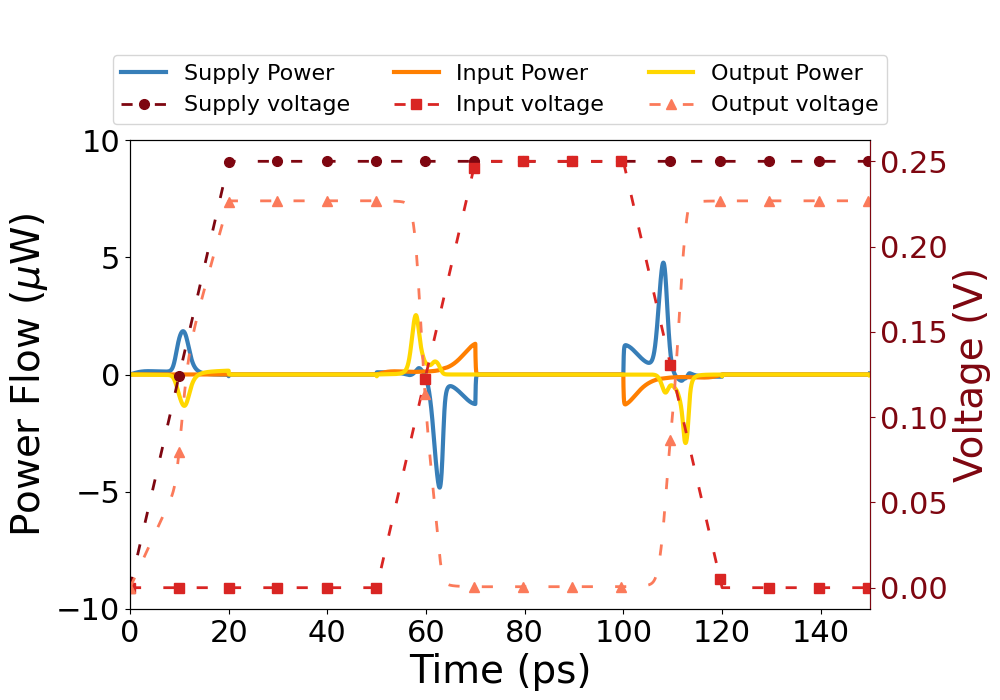}
\caption{} 
\label{fig:fig2_tcap_powerFlow}
\end{subfigure}
\centering
\begin{subfigure}[t]{0.39\linewidth}
\includegraphics[width=\linewidth]{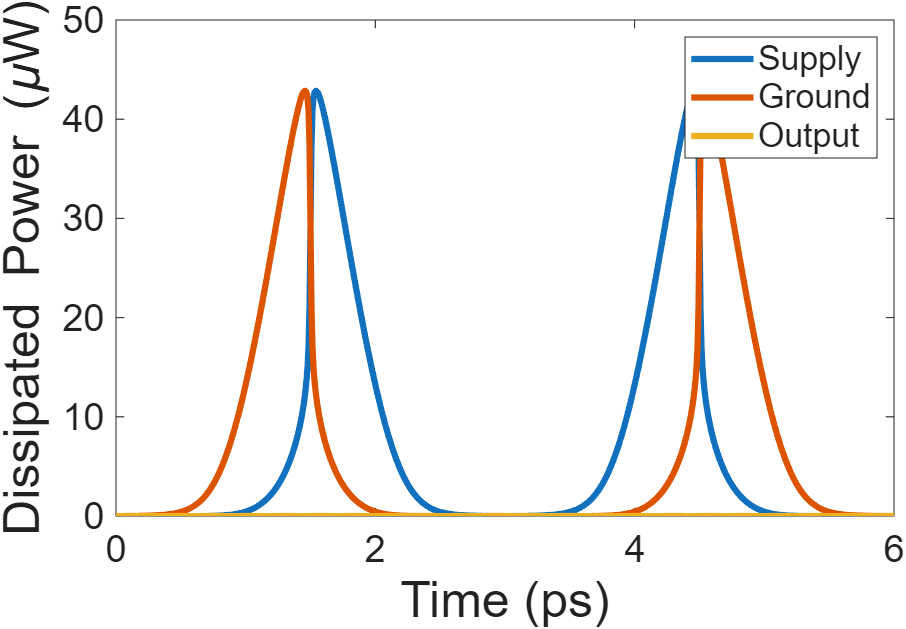}
\caption{}
\label{fig:fig2_cmos_dissipatedPower}
\end{subfigure}
\hfill
\begin{subfigure}[t]{0.4\linewidth}
\includegraphics[width=\linewidth]{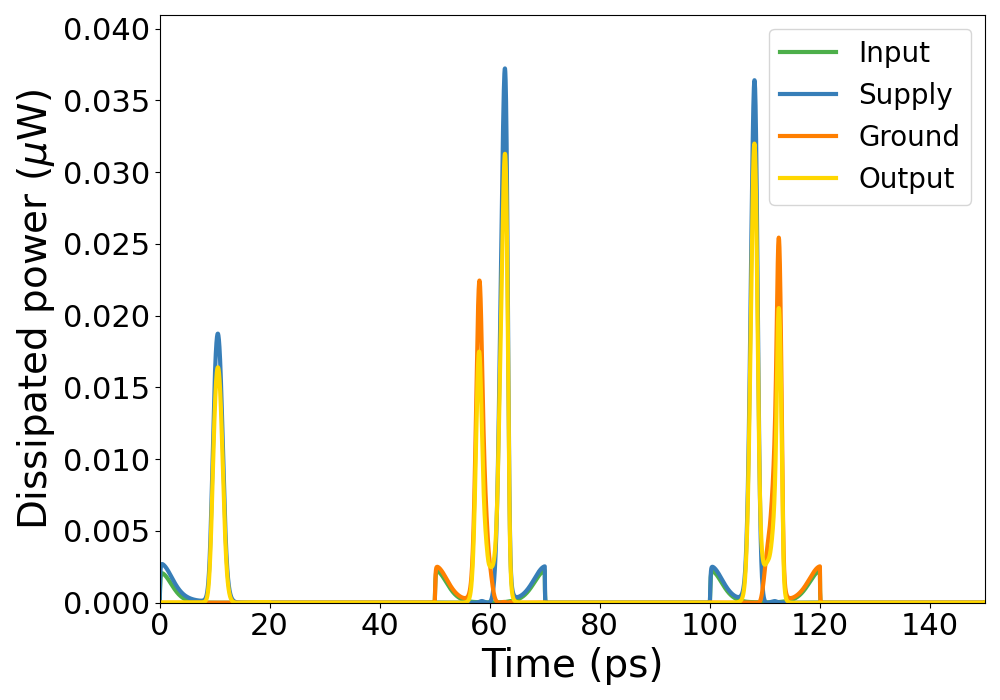}
\caption{} 
\label{fig:fig2_tcap_dissipatedPower}
\end{subfigure}
\caption{
\textbf{
Comparison of the operation characteristics of a transistor-based (left column) and transcapacitor-based (right column) inverters.}
a, b, Load line analysis of the current and charge, respectively, in the pull-up and pull-down devices.
c, d, Transfer characteristic, i.e., output vs. input voltage.
e, f, Resulting currents or charges/ vs. input voltage.  
g, h, Power flows vs. time as well as voltages vs. time. 
i, j, Dissipated power via Joule heating in wires vs. time.}
\label{fig:inverter_compar}
\end{figure}

Using the charge neutrality of the three terminal device and a simplified model for a transcapacitor (e.g., operating via a shift in the polarization-voltage characteristics), we can derive the basic analytical model for TCAP inverter (described in section~\ref{tcapAnalyticalMain} and ~\ref{sec:analyticalSpiceShrinkModel}), similar to the Sakurai-Newton (SN) model
\cite{sakurai2002alpha}, 
which we used for the CMOS inverter.
Please see Sections~\ref{sec:EdCMOS}
and~\ref{sec:EdTpol} for details.
A side-by-side comparison of the characteristics of inverters based on transistors (CMOS) and transcapacitor (TCAP) is shown in 
Fig.~\ref{fig:inverter_compar}. 

Panels (a) and (b) compare the loadline analysis of the inverters with possible output voltages as the horizontal axis. 
The intersection of the 
current or charge
curves give a static solution for the output voltage vs. the input voltage (aka the transfer characteristic).
It is obtained for the unloaded case of CMOS and fanout of 3 for the TCAP.
The difference is that currents are balanced for CMOS as opposed to charges being balanced for TCAP.
In both cases, the solid lines represent the pull-up ('p') devices and the dashed lines 
represent pull-down ('n') devices of the complementary gate. 
They are plotted in CMOS
for a set $0:0.2:1V$ of input (gate) voltages
(0V red, 0.2V green, 0.4 blue, 0.6 black, 0.8 magenta, 1V cyan); 
and in TCAP for  a set $0, 0.05, 0.2;0.25V$ of input (gate) voltages.

The CMOS inverter transfer characteristic,
panel (c), 
is well known from textbooks such as 
\cite{hu2010modern}: with a sharp slope in the middle and approaching the supply and ground voltage and the edges of the range. 
One can notice that the TCAP inverter characteristic, panel (d),
does not completely level off at the edges of the range. 
However it still has a slope $>1$ in the middle which makes it suitable for regenerating voltage and suppressing the noise.

The resulting currents in the inverters vs. the input voltage are shown in panels (e) and (f).
Panels (g) and (h) show the flow of power from the terminals of the inverter stage vs. time as the input voltage is varied. 
Panels (i) and (j) show the power dissipated via the Joule heating
in the resistances of the wires in the circuit. 
One can observe that the dissipated power in CMOS is comparable with the power flows. However in TCap, the peak power flow is 
around 140x greater than the peak dissipated power. This is the consequence of 
capacitive power recovery
in the TCAP circuit.

We ported a compact models for a transcapacitor with tunable capacitance to 
a SPICE environment for an
advanced node  (N5) and tested the theoretical improvements if such a device were to be integrated, keeping the parasitics and metal interconnects the same, described in section~\ref{sec:analyticalCLCH}, showing that the losses in parasitics are minimal. 

We show that implementing a gain element with a capacitive device can recover energy back to the supply, details given in section~\ref{sec:reactive}. 
This is analogous to reactive power recycling in power systems to shape and control the currents in 
power grids, as well as to provide reactive power\cite{singh2009static}. 
In the context of gain elements for logic and interconnects, a TCAP circuit will reshuffle the energy between the complementary N and P side circuits (energy is stored in the N or P side where the charge state is high), and part of the energy is fed back to the power supply capacitance. 
The savings in energy from capacitive (reactive) power shuffling is given by 
Eq.~\ref{eq:energy_ratio}.

We further describe a transcapacitor, which is a solid state transcapacitor with signal gain that operates via electromechanical tuning of the channel with a piezo-electric gate capacitor. 
Previously, piezoelectric effects have been used for logic 
devices~\cite{newns2012low,newns2012piezoelectronic,solomon2015pathway,weinstein2010resonant}, 
however they are used towards building a transistor type of device,
i.e., the one based of varying resistance.
Strain remains the strongest tuning parameter of the Landau energy landscape
owing to the strong piezoelectric/electrostrictive coupling in typical polar materials. Principle of this stress coupling mechanism to build a transcapacitor device is described in fig.~\ref{fig:piezoStressCoupling}.

We synthesize a few figure of merits (FOMs) for a TCAP device to follow at the circuit level in Section~\ref{sec:FOMs}. 
These enable it to perform as a switch and satisfy the 5 essential requirements to build a digital logic switch as described in~\cite{waser2012nanoelectronics}, such as inversion and gain. 
These are subsequently listed in Table~\ref{tab:logic_req}. In case of TCAP, gain which is one of the most critical metric to achieve is represented by the ratio of drain to gate charges. 
The charge gain for a piezoelectric transcapacitor with an analytical model using linear piezoelectric materials (see Section~\ref{sec:PiezoTpolModel}) 
is given by 
\begin{equation}
CG = 
\frac{\Delta D_c}{\Delta D_g}
= \frac
{d_c  d_g}
{ d_g^2 - \epsilon _g \epsilon_0
\left(1/C_g+h_c / (h_g C_c) \right)}
\end{equation}
It can reach the value of 4.9 for typical values of the material parameters 
(see Section~\ref{sec:PiezoTpolModel})
in a transcapacitor consisting of linear piezoelectrics as channel and stressor materials. 

If one of layers has  an energy barrier to switch polarization, such as in a ferroelectric channel, larger charge gains can be obtained as discussed in Section~\ref{sec:switch_mechanism}. Detailed self consistent finite element simulations demonstrate an operation of  such a MEMS device.   Signal gain ensures non-reciprocity for logic functionality. We report the first experimental demonstration of a key  principle needed for the piezo-transcapacitor as discussed in Section~\ref{sec:experim_demo}. 
We show that when stress is applied to $BaTiO_3$ (a typical ferroelectric), its charge voltage response change significantly as needed by the piezo-transcapacitor device. 
Moreover, we demonstrate inversion based on the principle of capacitance based voltage division as needed in a TCAP based inverter circuit using external stress application on one of the two $BaTiO_3$ capacitors. 
Finally, we show detailed benchmarking of TCAP logic with CMOS.

\begin{table}[ht]
\caption{BJT, MOSFET and Transcapacitor comparison}
\label{tab:bjt_mosfet_tcap}
\centering
\scriptsize
\begin{tabular}{lccc}
\toprule
Feature & BJT & MOSFET & Transcapacitor \\ 
\midrule
Gate    &  Resistor & Capacitor      & Capacitor \\
Channel     &  Variable resistor     & Variable resistor  & Variable capacitor \\
Channel material &  Semiconductor & Semiconductor      & Ferroelectric/Paraelectric \\
Voltage division  & Resistive & Resistive          & Capacitive \\
\bottomrule
\end{tabular}
\end{table}

\begin{table}[ht]
\caption{Energy comparison MOSFET vs. Transcapacitor }
\label{tab:energy_comparison}
\centering
\begin{tabular}{lcc}
\toprule
Energy consumption & MOSFET & Transcapacitor \\ 
\midrule
Short circuit current & Yes & No \\
Joule heating & Yes & Very small (partially reactive) \\
Voltage & 0.25 to 1V range & 50 mV to higher \\
Discharging to ground & Yes & No \\
Intrinsic damping & No & Yes \\
\bottomrule
\end{tabular}
\end{table}

\section{Transcapacitor as a concept}
\label{tcap_as_a_concept}

We introduce a fully capacitive solid-state switch based on the concept of modulating displacement currents. A TCAP has three terminals named gate, source and drain similar to MOSFET, where the gate terminal serves as an input, source as a reference and drain as the output, as shown in the Fig.~\ref{fig:Figure1_symbols}. 

\subsection{Transcapacitor with ferroelectric materials}
It is possible to build various incarnations of a TCAP device using polar class materials with varied physical mechanisms as described in Table~\ref{tab:tpol_types}. 
 Ferroelectrics are the most suitable candidates for enabling low voltage TCAP devices because of the following characteristics. a) Ferroelectrics are inherently insulating, polarization change resulting in displacement currents b) their crystals have well defined energy landscapes c) they can be switched at low voltages d) capacitance of ferroelectrics can be controlled by the voltage applied across it e) they have rich physics with various coupling mechanisms such as with stress, magnetism, temperature etc. 

\begin{figure}[ht!]
  \centering
  \begin{subfigure}[t]{0.20\linewidth}
    \includegraphics[width=\linewidth]{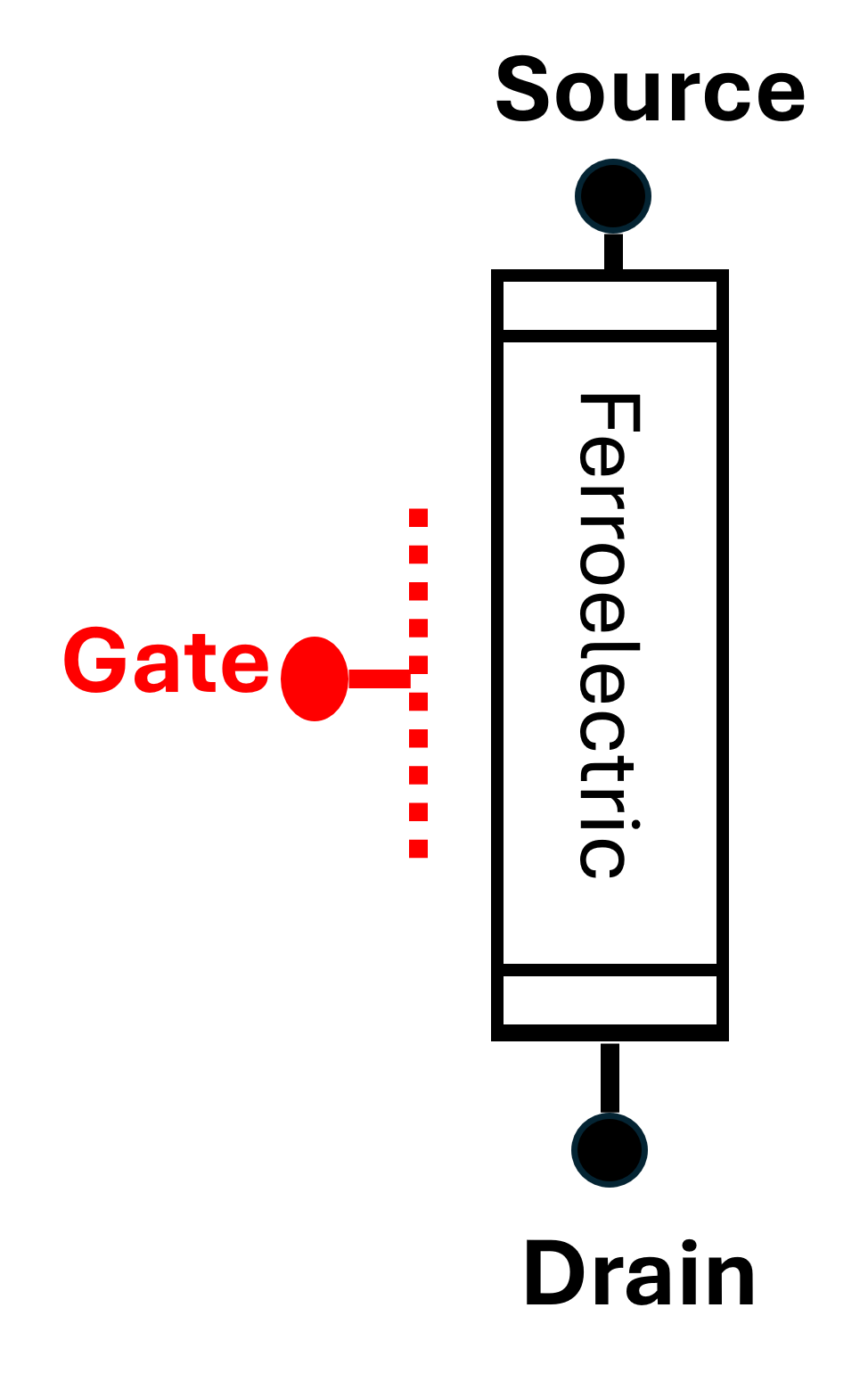}
\caption{}\label{fig:symbolicFETranspol}
  \end{subfigure} \hfill
  \begin{subfigure}[t]{0.45\linewidth}
    \includegraphics[width=\linewidth]{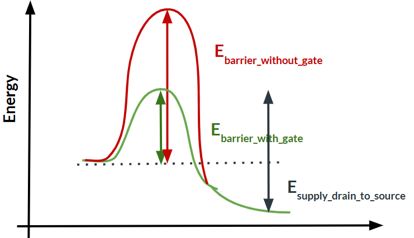}
    \caption{}\label{fig3:SymbolicEnergyBarrier}
  \end{subfigure}\hfill
\hfill
        \begin{subfigure}[t]{0.30\linewidth}
    \includegraphics[width=\linewidth]{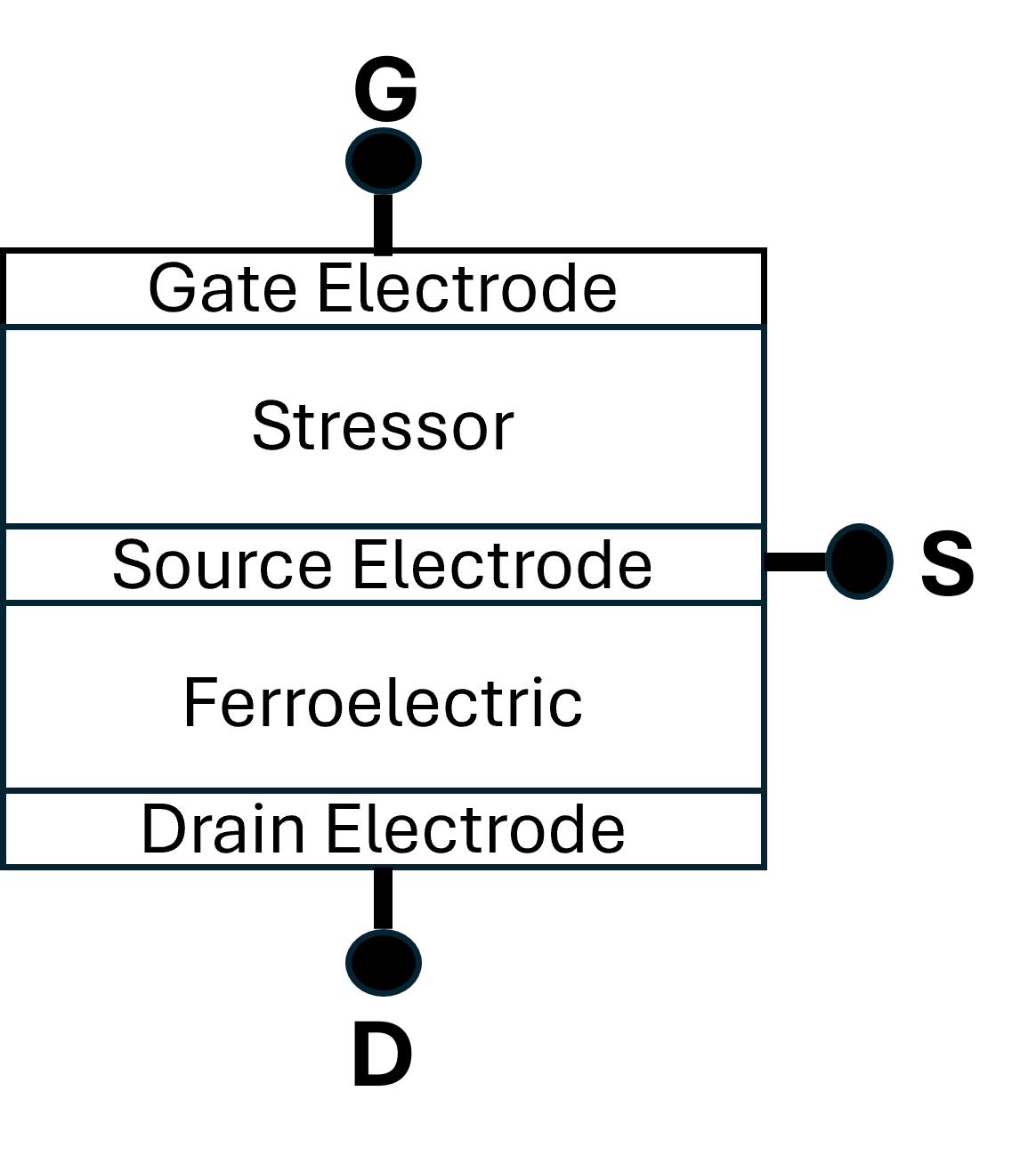}
    \caption{}\label{fig:piezoConcept}
  \end{subfigure}
  \vspace{1pt} 
    \begin{subfigure}[t]{0.33\linewidth}
    \includegraphics[width=\linewidth]{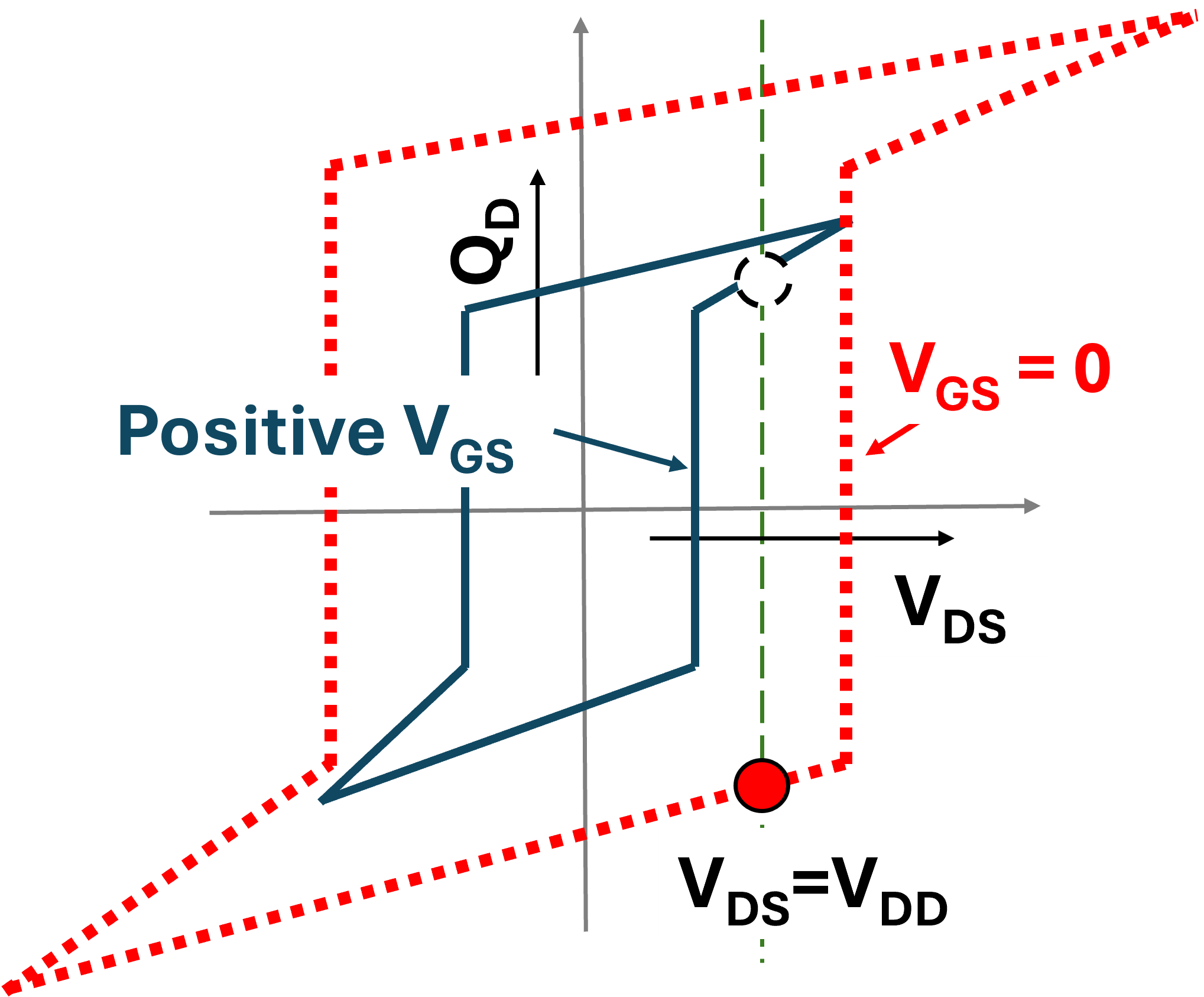}
    \caption{}\label{fig:HystShrinkFEtoFENDevice}
  \end{subfigure}\hfill
      \begin{subfigure}[t]{0.33\linewidth}
    \includegraphics[width=\linewidth]{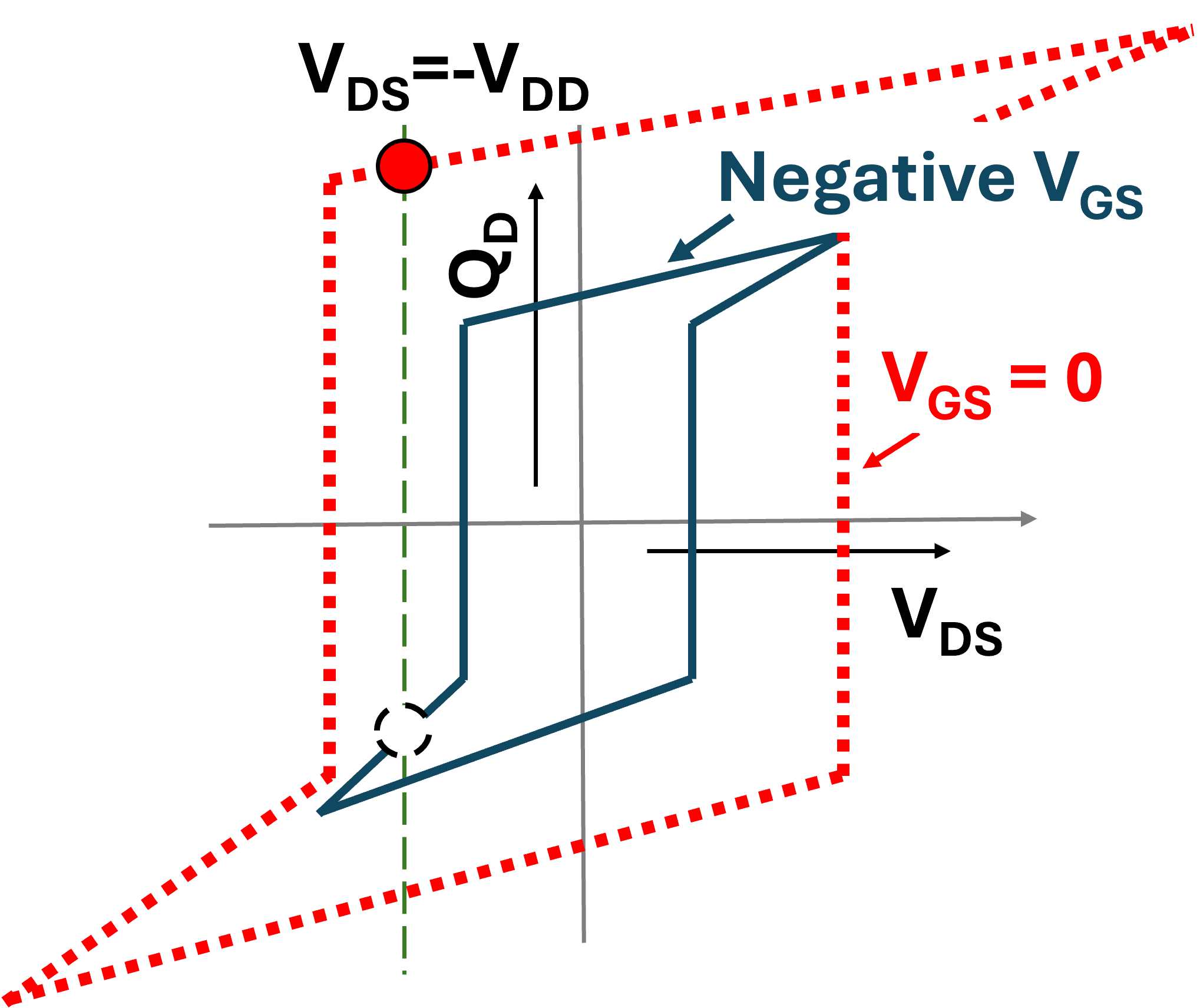}
\caption{}\label{fig:HystShrinkFEtoFEPDevice}
  \end{subfigure}\hfill
    \begin{subfigure}[t]{0.33\linewidth}
    \includegraphics[width=\linewidth]{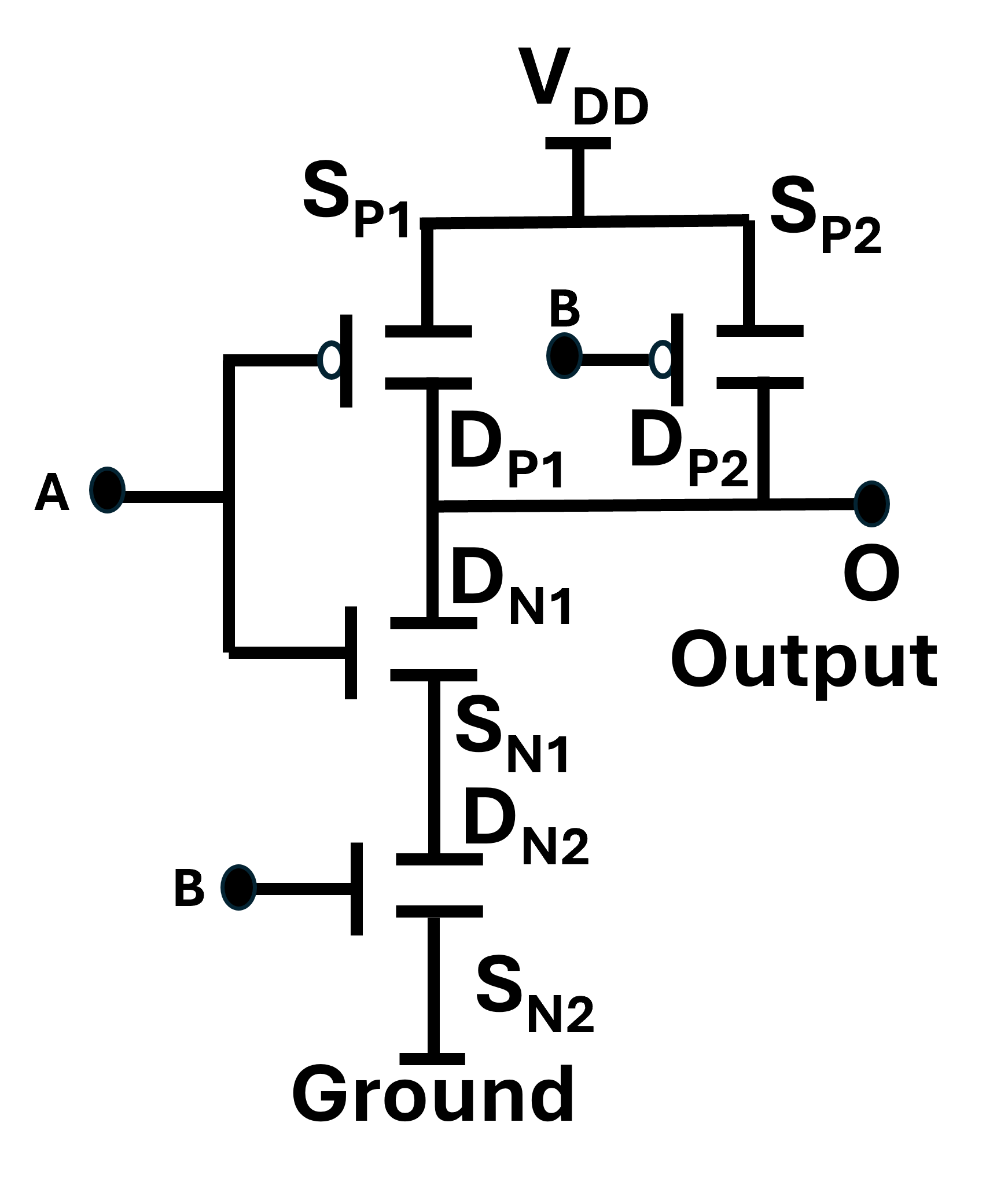}
    \caption{}\label{fig:NandGateCircuit}
  \end{subfigure}
\caption{\textbf{Transcapacitor basics.}
\textbf{a)} symbolic TCAP device with FE material, \textbf{b)} Concept of lowering the energy barrier of an FE with the gate, \textbf{c)} piezo-transcapacitor concept. \textbf{d-e)} hysteresis shrink model as an effect of the gate for N and P device respectively, \textbf{f)} NAND2 gate circuit with TCAP.} 
\label{fig:1}
\end{figure}

To build gate control in a ferroelectric channel device, two main mechanisms can be used which are, a) modulate the properties/energy landscape of the FE material b) modulate the properties such as carrier concentration in the electrodes attached to it. This philosophy leads to various potential incarnations of a transcapacitor with varied types of transfer functions satisfying the requirement of achieving inversion, as shown in Table~\ref{tab:tpol_types}. In this article, we focus on the mechanism of modulating the energy barrier critically using both the gate and drain fields, similar to a MOSFET device as symbolically represented with 
Figs.~\ref{fig:symbolicFETranspol} 
and~\ref{fig3:SymbolicEnergyBarrier}. 




A transcapacitor is a capacitor controlled capacitive channel device. It is of utmost importance to separate out drain and gate electrodes electrostatically, to avoid a follow up (buffer) behavior. Hence in the device implementation, we make use of coupling mechanism to flow additional input energy to the channel capacitor, enabling critical gate control of the device. 
For our leading candidate called piezo-transcapacitor as described in Fig.~\ref{fig:piezoConcept}, we use stress coupling as a way to reduce the energy barrier of switching a ferroelectric and provide gate control~\ref{fig:piezoStressCoupling}. The device as shown in
Fig.~\ref{fig:piezoConcept}, keeps gate and drain electrodes separated through a reference source electrode which is tied to a fixed supply voltage, and keeps the electrostatic coupling between the two terminals to minimal. 


Our simulations with Comsol indicate shrinkage of hysteresis of $BaTiO_3$ with applied stress as shown in
Fig.~\ref{fig:deviceFig1_energybarriers_h}. 
This is in agreement with the work done by~\cite{choi2004enhancement}, where by engineering the stress in thin BTO films, both its critical voltage ($V_c$) and polarization reduce. 
With reduced $V_c$ obtained by applying a gate field, we enable switching of the FE channel at a lower voltage than otherwise.  This essentially builds a gate controlled FE channel with appropriate operating voltages. 
Our experiments described in section~\ref{sec:experim_demo} show that, for a para-electric channel the stress application results into the non-linear charge-voltage response, reducing its non-linearity or converting into a linear dielectric response. This establishes gate control over a para-electric channel enabling building of logic circuits. 

The concept of building a switch with this gate controlled shrunk hysteresis scheme is depicted in
Fig.~\ref{fig:HystShrinkFEtoFENDevice}
and~\ref{fig:HystShrinkFEtoFEPDevice}. The device is considered off with the original hysteresis and produces a shrunk hysteresis in the on state with the gate-source voltage. 
Depending upon the strain-voltage relation of the gate stressor material and the charge-voltage response  with applied stress of the channel, both P- and N-TCAPs can be implemented with the same device as is done in this article. As shown in figure~\ref{fig:HystShrinkFEtoFENDevice}, a positive gate-source voltage ($V_{GS}$) leads to shrunk hysteresis for the N-device. Further, FE channel of the N-device is initialized with negative polarization at the drain terminal and goes through the right branch of the hysteresis as  a positive drain-source voltage ($V_{DS}$) is applied, releasing its full capacity charge once switched. This charge at the drain is used to drive the input of the next stage, for cascading multiple devices. Similarly, P-device is initialized with a positive polarization and works in the negative $V_{GS}$ and $V{DS}$ voltage regimes. For extremely low voltage ferroelectrics (para-electrics), the same initialization effect for the channel is obtained by shifting the $QV$ response using work-function offsets of various electrodes of the device and the stress application resulting in a charge release at the drain terminal.

To exploit the hysteresis–shrink phenomenon as a gate-enabled switch, we restrict the \emph{operating voltage}  \(V_{DS}\) applied across the ferroelectric as:
\begin{equation}
    -V_{C_\mathrm{Original}} \;<\; V_{DS} \;<\; V_{C_\mathrm{Original}}
    \label{eq:on_off_window}
\end{equation}
where \(V_{C_\mathrm{Original}}\) denotes the FE’s critical voltage measured at \(V_{GS}=0\). If \(|V_{DS}|\) exceeds \(V_{C_\mathrm{Original}}\) the gate ceases to have control: the drain–source voltage alone can fully switch the polarization state. In the operating $V_{DS}$ range, the TCAP device works as a switch.  
 
\subsection{FE Transcapacitor as a memory element}
With the gate controlled FE channel as depicted symbolically in Figures~\ref{fig:HystShrinkFEtoFENDevice} and~\ref{fig:HystShrinkFEtoFEPDevice}, it is easy to see that a TCAP with a non-volatile ferroelectric capacitor is also a 1-bit memory element with a built-in selector. The gate voltage (selector using $V_{GS}$) enables writing of the memory capacitor with either positive or negative polarization at a reduced voltage. 
In the operating $V_{DS}$ range, the TCAP device works as a 1-bit memory element similar to 1T-1C FRAM\cite{scott1989ferroelectric}, where 1C is a FE capacitor. 
With a single device able to represent both a selector a capacitor, in comparison to a typical FERAM bit with one transistor and one FE capacitor, TCAP based memories have potential to be much more scalable and amenable to 3D integration similar to 3-terminal NAND memory~\cite{fukuzumi2007optimal}. 

For the charges at drain, gate and source terminals we use $Q_D$, $Q_G$ and $Q_S$ as notations respectively. Charge at a certain $V_{GS}$ and $V_{DS}$ is denoted by $Q(V_{GS}, V_{DS})$.
A figure of merit (FOM) for TCAP based memory operations can be defined as follows. 

\textbf{ON/OFF Ratio.}
This metric defines a TCAP's ability to 
produce a distinguishable charge output in response to the applied voltages. For a sensible switch, this ratio should always be $> 1$. 
A higher on/off ratio of the released charges at the drain terminal defines the effectiveness of the gate voltage in controlling the switch.
The charges are referenced to the value at the extreme levels of voltage, $Q_D(0,0)$.
The device is considered ``on'' when $V_{GS}=\pm V_{DD}$ and ``off'' when $V_{GS} = 0$.
At the device level, this ratio is
\begin{equation}
\mathrm{Q}_{ON/OFF} (V_{DS}) 
= \frac{(\Delta Q_D)_{{ON}}}{(\Delta Q_D)_{{OFF}}}
= \frac{ Q_D\!\bigl(V_{DD},\,V_{DS}\bigr) - Q_D(0,0)}
{ Q_D\!\bigl(0,\,V_{DS}\bigr) - Q_D(0,0)}.
\end{equation}
For a FE channel with lower dielectric cap vs. higher, enables larger ON/OFF Ratio.  This device level ratio is also useful for a TCAP logic element, where a higher ratio means lower self loading, resulting in higher voltage margins at the output nodes. Please note that the range of voltages $V_{GS}$ and $V_{DS}$ are independent of each other for a memory circuit, in contrast to logic where they are required to be the same to enable concatenation of multiple stages. This means that the maximum value $V_{DD}$, would be different for both $V_{GS}$ and $V_{DS}$, which will be addressed by $WL$ and $BL$ voltages in section~\ref{memory_circuits_details} and equation~\ref{eq:QonQoff_deviceSim}. 

\subsection{Transcapacitor as a digital logic switch}
\label{sec:transcap_digital_switch}
Synthesizing digital logic requires that the underlying switch must  satisfy five essential requirements, see 
\cite{waser2012nanoelectronics}. These requirements are highlighted in the Table~\ref{tab:logic_req} with the details of Piezo-transcapacitor's (Piezo-TCAP) way of fulfilling those. 
We define an analytical model of TCAP that shows a fully working logic circuit.
Further, we prove inversion and building a universal gate with detailed COMSOL simulations.

\begin{table}[htbp]
    \centering
    \renewcommand{\arraystretch}{1.5} 
    \begin{tabularx}{\textwidth}{|X|X|}
        \hline
        1. Building a universal logic gate with inversion & Shown using NAND2 gate and inverter. \\ \hline
        2. Concatenation of successive logic gates. & Input and output are both voltage signals. \\ \hline
        3. Minimal feedback from the output to the input. & Presence of a hysteresis in either the gate or channel enables irreversibility. \\ \hline
        4. Nonlinearity of input vs. output (signal gain). & Achieved with the non-linear charge-voltage response of both the gate and channel. \\ \hline
        5. Energy gain or fanout & Enabled by the gate (stressor) change of charge being lower than the channel charge change as a requisite. Represented by \textbf{CG} for TCAP. \\ \hline
    \end{tabularx}
    \caption{5 Requirements for a digital logic element as satisfied with Piezo-TCAP }
    \label{tab:logic_req}
\end{table}

\subsubsection{TCAP based logic circuits}
Similar to CMOS, two complementary circuit elements called N-TCAP and P-TCAP, are used to enable push-pull mechanism between them to resolve the logic levels of 0 and 1. 
In TCAP based logic circuits, source of N-TCAP and P-TCAP devices are tied to ground (0) and $V_{DD}$ respectively as shown in
Figs.~\ref{fig:Figure1_unloadedInv} with an inverter and a NAND2 gate~\ref{fig:NandGateCircuit}. This voltage biasing makes N-TCAP and P-TCAP operate in mainly positive $V_{GS}$ and $V_{DS}$ voltage ranges and negative voltages ranges respectively similar to MOSFET. With this voltage biasing and proper initialization of the FE channel polarization,  the same TCAP device implementation can serve both as a P-TCAP and an N-TCAP. Cascading multiple logic stages requires the range of input and output voltages to be the same.
In a piezo-TCAP circuit, since both the input and output signals are electrical voltages and have the same range, it is natural to make multi-stage circuits, satisfying the requirement of concatenation in Table~\ref{tab:logic_req}.

\textbf{Inversion: }Dominant electrical elements in a TCAP circuit are capacitors. And hence, voltage at any node, follows two fundamental principles, 1) charge conservation and, 2) voltage division based on the capacitances connected to the node. A circuit figure of an unloaded inverter and an inverter driving another inverter is shown in
Fig.~\ref{fig:deviceFig_circuitinv}, showing the dominant capacitances affecting the voltage at the output floating node $V_{out1}$. We can see that the capacitance based voltage division at $V_{out1}$ for an unloaded inverter accounts for not only $C_{DS}$, but also $C_{DG}$. If $C_{DG}$ is dominated by electrostatics (which would be the case if both the gate and drain electrodes were directly attached to the FE channel), the output voltage $V_{out1}$ may essentially just follow $V_{inp}$ and wouldn't produce an inversion. To avoid this voltage follower behavior, we use stress coupling to provide extra energy from input to the FE channel and reduce its energy barrier, leading to an inverter circuit. 

Similar to CMOS, when the input signal $V_{inp}$ is zero, it leads to switch-on the P-TCAP with $V_{GS}=-V_{DD}$, and turn-off the N-TCAP with $V_{GS}=0$. In this configuration,  the P-TCAP gets the ability to release a substantial charge from its drain terminal at a lower $V_{DS}$ due to the inward shift of its hysteresis branch from left to right. With a substantial shift of the branch getting close to zero,  voltage division at $V_{out1}$ forces $V_{DS}$ on the P-device being close to zero. This leads to $V_{out1}$ near $V_{DD}$, essentially inverting the input signal $V_{inp}$. Charge from the P-device become available to charge up the input gate load of the next stage inverter. Maximum charge released here could be approximately equal to twice the polarization of the FE channel. 

\textbf{Circuits with TCAPs:} Building a universal gate such as a NAND2, NOR2, is one of the essential requirements to make full digital logic as 
listed in the Table~\ref{tab:logic_req}.
Now that we see, how an inverter resolves the logic levels, it is easy to extend the idea to other types of multi-input gates such as a NAND2 or NOR2 and even more complex gates such as XOR2 and AOI. The circuits of these gates such as NAND2 (shown in
Fig.~\ref{fig:NandGateCircuit}) can be built following the same principles.

\textbf{Gain in TCAP circuits: } Gain in a circuit element is essentially its ability to drive higher output load compared to its own input load. In circuits, the ratio of output load capacitance to input load capacitance is called fanout (FO).
This is listed as one of the essential requirements to build a digital logic in Table~\ref{tab:logic_req}.  
The circuit and device operating mechanisms should basically enable pulling of energy from the supply (in addition to the input energy) to drive the output higher load. This additional energy from the supply would enable $FO>1$. 

Unlike CMOS, where this pull of energy from the supply voltage is enabled by modulation of the semiconductor channel's resistance, followed by a resistive division at the output node; in a TCAP circuit, modulation of the charges/capacitances of the FE channels creates the pull for the supply energy. This modulation is basically driven by the input voltage signals and is followed by capacitance based voltage division at the output node. In contrast to CMOS, where the gates are capacitors and channels are resistors, a TCAP is a fully capacitive device with both the gate and channel being capacitors. This essentially means that all capacitances connected at a output node of both the driving circuit and the driven circuit participate in this voltage division. Hence, it is important to keep the self-loading of the driving devices to minimal values, and reduce the input (gate) capacitances of the driven devices to maximize the voltage ranges and get the highest signal resolution at the output nodes. For our piezo-TCAP, the main source of self-loading is the dielectric capacitance/charge of the FE channel. And, output load in TCAP circuits, is a function of charges on the gates of the driven devices. These are to be minimized compared to the saturation polarization of the FE channels. 

\textbf{Non-linear transfer characteristics: } Small signal voltage gain is another essential requirement as mentioned in Table~\ref{tab:logic_req}. This is to recover any potential voltage signal loss in the successive logic stages with higher fanouts. For our piezo-TCAP devices, gate's strain-voltage response plays a huge role to achieve this, as explained in the subsequent section. Non-linear response of both the gate and the channel enable this small signal gain in the piezo-TCAP circuits. 

\textbf{Minimal feedback from output to input: } It is one of the essential logic requirements as listed in table~\ref{tab:logic_req} enabling irreversibility.  TCAP devices built with low voltage FE class materials contain hysteresis, possibly causing small energy losses during the circuit operation. We hypothesize that the presence of a hysteresis in the device (either in the gate or the channel or both) stops the reversal of computations and results in the minimal feedback from output to input. In our preliminary COMSOL simulations, to be discussed in future articles, we notice that the presence of a hysteresis in the FE channel creates this directionality once the polarization of the FE channel switches. Since an opposite polarity voltage is required to switch the device back to its original polarization state, the circuit provides a stable output voltage, achieving directionality. 

To meet the above listed requirements, a TCAP circuit element should follow certain characteristics which enable an inverting element and gain and maximize its performance. We crystallize these in a few figure of merits (FOMs) described in section~\ref{sec:FOMs}.

\subsubsection{Analytical model for a FE based logic TCAP}\label{tcapAnalyticalMain}

We created an analytical model of a logic TCAP, imitating the P-V response of a ferroelectric channel behavior as given by the shrink of its hysteresis. 
For simplicity, we represent only the single appropriate branch of the P-V hysteresis loop of the channel using a tanh function (right branch for N and left branch for P). $V_{GS}$ application causes a shift of the hysteresis branch inwards towards the center for both the polarization and voltage axis as shown in figure~\ref{fig1:QDvsVDS_N} and~\ref{fig1:QDvsVDS_P}.   This essentially represents shrinkage of the hysteresis with applied stress for the piezo-TCAP. Equations for the model are described in the supplementary section~\ref{sec:analyticalSpiceShrinkModel}. The shift amount with $VGS=\pm V_{DD}$ is kept minimal while essentially aligning the branch to a full polarization switch while having $V_{DS}=0$. This is done to enable maximum voltage margins at the output node, while releasing the maximum charge out of the drain of a driving device. 

The amount of shift voltage applied to $V_{DS}$ is  a non-linear function of $V_{GS}$. It is an imitation of  what an anti-ferroelectric (AFE) stressor such as $PbZrO_3$~\cite{pan2024clamping} could produce on a ferroelectric channel through the stress coupling mechanism. 
 In an AFE material, the strain is low near zero with a small field till the material starts becoming polar, and finally saturates  quickly to a maximum value following the completion of AFE to FE phase transition.  
In our model,  the shift caused by $V_{GS}$ is really small when $V_{GS} <= V_{TH,G}$, and rises up sharply beyond that, finally saturating to a fixed value for N-TCAP. This non-linear dependence of the channel response on $V_{GS}$ results in a non-linear transfer characteristics of the TCAP inverter circuit with  small signal voltage gain, an essential requirement to build a digital logic circuit as listed in Table~\ref{tab:logic_req}. Ideally, the charge-voltage dependence of the gate should also be derived similarly accounting for $V_{TH,G}$, however due to numerical convergence issues in the Spice simulation, we make this a simple non-linear shape with tanh function to still keep non-linearity. We do not believe this causes a substantial change in the circuit operations. To demonstrate a chain of TCAP circuits in action, we implement this model in Spice and study the output voltages, charges and power dissipation as shown in 
Fig.~\ref{fig:InvChainCircuitSuppl}.

\subsubsection{Figure of merits for a transcapacitor}\label{sec:FOMs}
Similar to the transistors where a few figure of merits (FOMs) such as on/off ratio of the source-drain currents allow us to compare different transistor implementations, we can define FOMs for TCAP devices. These are essentially guidelines defined at the circuit element level and hence they allow us to compare different devices implemented with the same or different physics mechanisms and also guide us to discover new TCAP implementations such as the ones described in 
Table~\ref{tab:tpol_types}.  

\textbf{Transcapacitance (TC)} 
The first figure of merit targets the ability of a TCAP element to be used for inversion. 
It quantifies how effectively the gate voltage modulates/controls the charge at the drain terminal in the correct direction, reflecting $C_{DG}$ which is an  off-diagonal term of the matrix of differential capacitances. Since these capacitances are possibly non-linear, we use effective capacitances defined in the operating voltage regime. Transcapacitance (\textbf{TC}) is negative of effective $C_{DG}$. A positive TC reflects that a positive change in $V_{GS}$ results in the increase of drain charge value (N-TCAP), opposite of what electrostatic coupling provides.

\begin{equation}
\mathrm{TC = }- \textrm{effective}
~C_{DG} =  
\frac{\Delta Q_D}{\Delta V_{GS}},  \textrm{at fixed} V_{DS}
\end{equation}

For an \emph{inverting} element, we require that \emph{ transcapacitance} $\mathrm{TC}>0$ 
(which implies that predominantly $C_{DG} < 0$). For a P-device, decrease in $V_{GS}$ results into a decrease in the drain charge, exhibiting a positive \textbf{TC} as well.  When the electrostatic coupling between the gate and the drain dominates,
$\mathrm{TC}$ becomes negative and the device may not exhibit inversion.
Therefore, to implement a logic NOT operation, we envision transcapacitors based on coupling mechanisms other than electrostatic ones.
Fig.~\ref{fig:symbolicFETranspol} symbolically depicts such a device: the dotted gate terminal indicates that the gate influences the ferroelectric only \emph{indirectly} through a coupling mechanism. 

\textbf{Charge gain (CG):} 
It quantifies the driving ability of a device, i.e., the ability to supply charge to the output load in response to the charge taken at the input node. 
\textbf{CG} for the device level at a certain $V_{DS}$ is defined as following, as $V_{GS}$ goes from $0$ to $V_{DD}$: 
\begin{equation}
\mathrm{CG} (V_{DS})
= \frac{\Delta Q_D}{\Delta Q_G}, \textit{at fixed $V_{DS}$}
\label{eq:CG_Device}
\end{equation}

The above equation can be used to compare devices implementations of TCAP, testing their inherent ability to provide gain or fanout. At the circuit level, the equation needs to be modified to account for the appropriate voltage changes and the present fanout. As an example, for an inverter driving another single inverter, a representative \textbf{CG} would be given by the following equation, where $Q_G$ corresponds to the input charge of the driven devices and $Q_D$ corresponds to the output charge of the driving device. 
\begin{equation}
\mathrm{CG}
= \frac{ Q_D\!\bigl(V_{DD},0\bigr) - Q_D\!\bigl(0,V_{DD}\bigr) }
 { Q_G\!\bigl(V_{DD},0\bigr) - Q_G\!\bigl(0,V_{DD}\bigr) }
\end{equation}
One can expand the above definition to include fanout and self-loading of the driving devices. A greater charge gain also means that a perturbation of charge at the drain node would lead to a smaller perturbation of charge at the gate node.
This contributes to the requirement {\#3}, 
minimum feedback from the input to the output as specified in
Table~\ref{tab:logic_req}.
In a transcapacitor,
the input capacitance depends on the specific coupling mechanism and its
strength—both are highly implementation-dependent and will be examined in detail
later for the piezo-transcapacitor variant. 

\textbf{Voltage margin gain (VMG)} 
is the ratio of the output voltage maximum swing
to the input voltage maximum swing at a given fanout (FO).
\begin{equation}
\mathrm{VMG}
  = \frac{\Delta V_{DS}}{\Delta V_{GS}}, \textit{at fixed FO}.
\end{equation}

\textbf{VMG} characterizes the ability of the logic gate
to maintain the voltage swing with multiple cascading stages. For the logic gates to be cascade-able (the requirement {\#4}, concatenation in
Table~\ref{tab:logic_req}, and maintain the signal gain over multiple stages. VMG should be such that it is recoverable to its full value, with the non-linear transfer characteristics of a lower fanout loaded inverter. \textbf{VMG} depends upon the fanout and the maximum fanout value at which this is still fulfilled, determines the maximum load that the logic gate can drive.

\section{Stress-assist transcapacitor engineering and operation}
\label{sec:device}
 
In our proposed novel 3-terminal solid state transcapacitor device gate modulates the polarization/capacitance across 2-terminals to achieve desired logic or selector memory functions. Electrostatic crosstalk between gate  and source / drain electrodes which is causing a detrimental gate follow effect on source/drain potential needs to be excluded from the device operation. Therefore, the gate action on source/drain charge is implemented through an intermediate electro-mechanical coupling effect of stress modulation of polarization in ferroelectric or paraelectric channel by transferred stress created in gate stressor under application of gate voltage across stressor. The source electrode both transfers stress created in gate capacitor and screens gate electric field from the device channel as illustrated in Fig. \ref{fig:piezoConcept}.

\subsection{Switching mechanism and models}
\label{sec:switch_mechanism}

\begin{figure}[htbp]
  \centering
  \begin{subfigure}[t]{0.23\linewidth}
    \includegraphics[width=\linewidth]{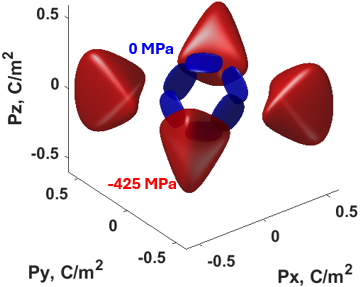}
\caption{}\label{fig:deviceFig1_energybarriers_a}
  \end{subfigure}\hfill
  \begin{subfigure}[t]{0.23\linewidth}
    \includegraphics[width=\linewidth]{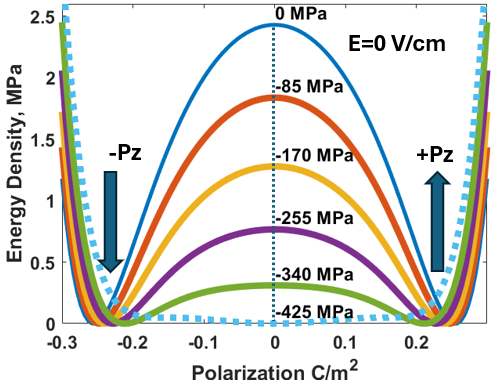}
\caption{}\label{fig:deviceFig1_energybarriers_b}
  \end{subfigure}\hfill
  \begin{subfigure}[t]{0.23\linewidth}
    \includegraphics[width=\linewidth]{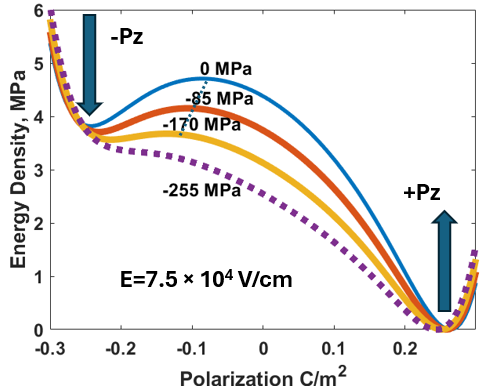}
\caption{}\label{fig:deviceFig1_energybarriers_c}
  \end{subfigure}\hfill
  \begin{subfigure}[t]{0.23\linewidth}
    \includegraphics[width=\linewidth]{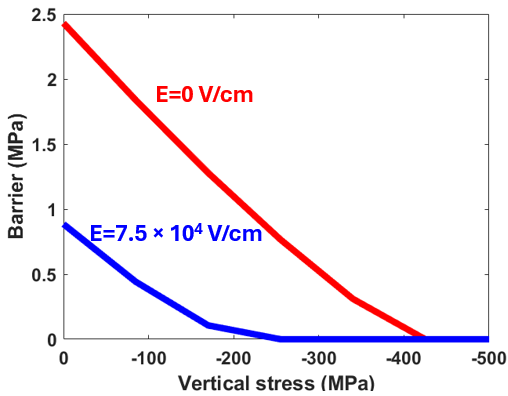}
\caption{}\label{fig:deviceFig1_energybarriers_d}
  \end{subfigure}
  \\
  \begin{subfigure}[t]{0.23\linewidth}
    \includegraphics[width=\linewidth]{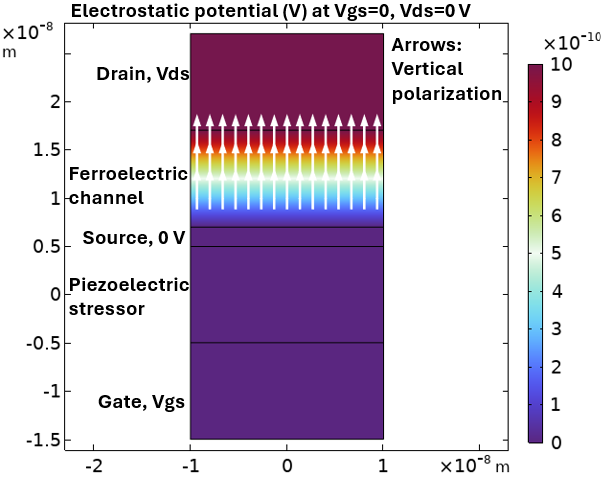}
\caption{}\label{fig:deviceFig1_energybarriers_e}
  \end{subfigure}\hfill
   \begin{subfigure}[t]{0.23\linewidth}
    \includegraphics[width=\linewidth]{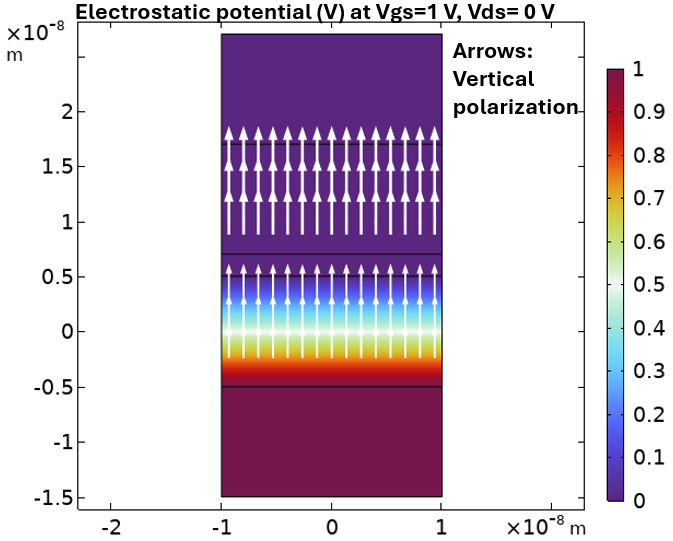}
\caption{}\label{fig:deviceFig1_energybarriers_f}
  \end{subfigure}\hfill
  \begin{subfigure}[t]{0.23\linewidth}
    \includegraphics[width=\linewidth]{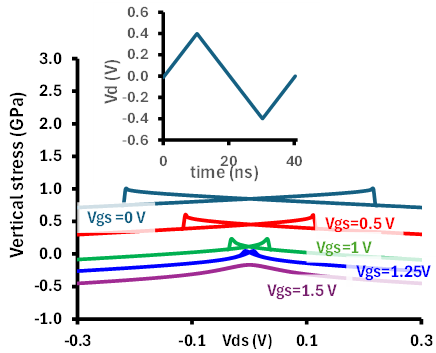}
\caption{}\label{fig:deviceFig1_energybarriers_g}
  \end{subfigure}\hfill
  \begin{subfigure}[t]{0.23\linewidth}
    \includegraphics[width=\linewidth]{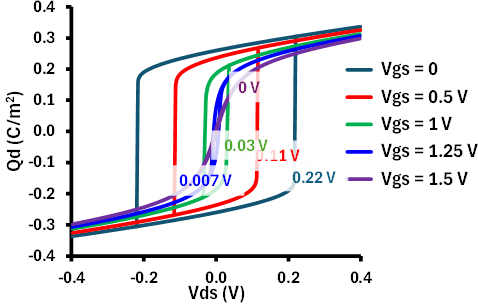}
\caption{}\label{fig:deviceFig1_energybarriers_h}
  \end{subfigure}
  \\
  \begin{subfigure}[t]{0.35\linewidth}
    \includegraphics[width=\linewidth]{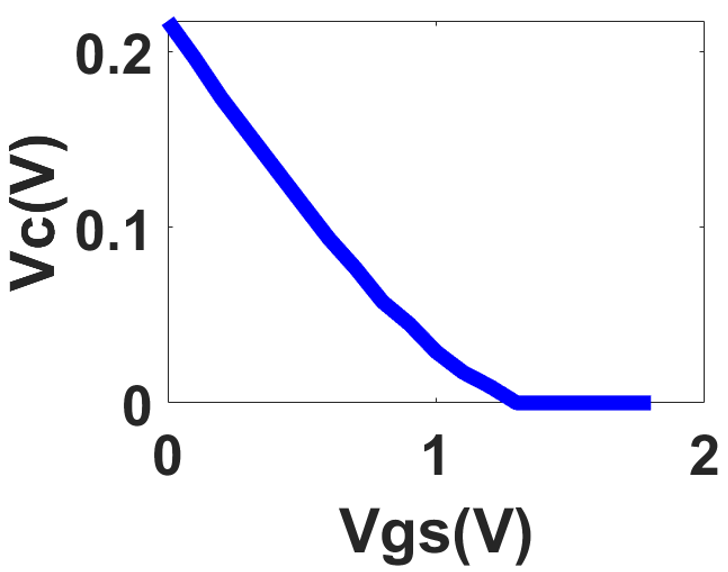}
\caption{}\label{fig:deviceFig1_energybarriers_i}
  \end{subfigure}\hfill
  \begin{subfigure}[t]{0.55\linewidth}
    \includegraphics[width=\linewidth]{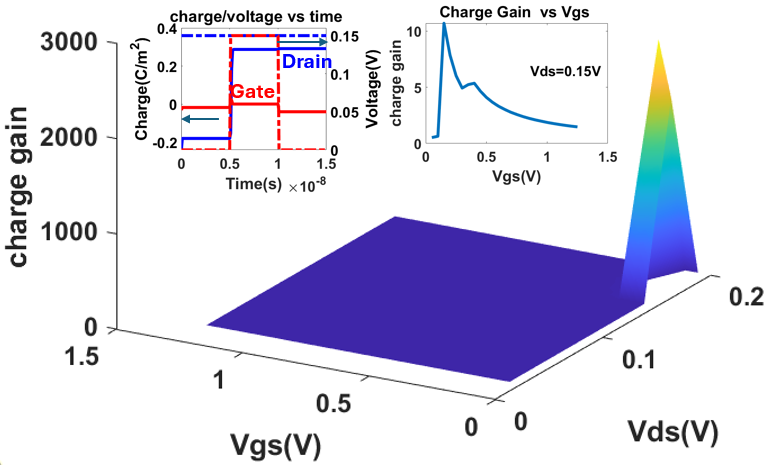}
\caption{}\label{fig:deviceFig1_energybarriers_j}
  \end{subfigure}
  \\
  \begin{subfigure}[t]{0.23\linewidth}
    \includegraphics[width=\linewidth]{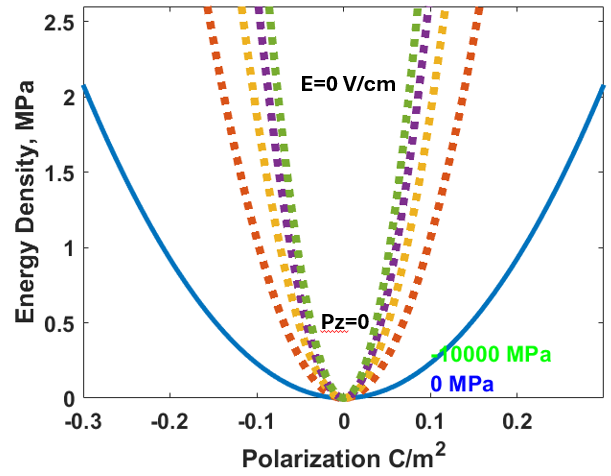}
\caption{}\label{fig:deviceFig1_energybarriers_k}
  \end{subfigure}\hfill
   \begin{subfigure}[t]{0.23\linewidth}
    \includegraphics[width=\linewidth]{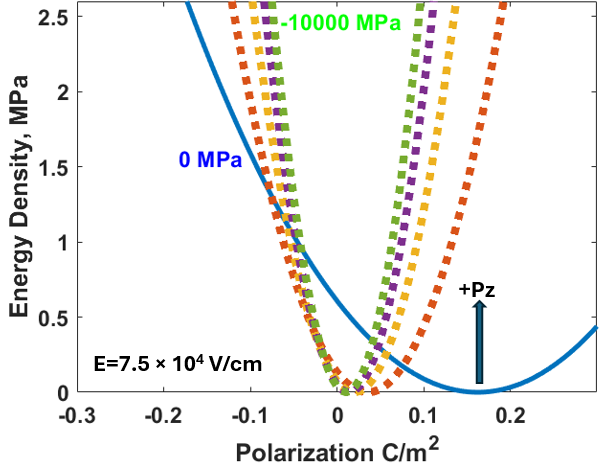}
\caption{}\label{fig:deviceFig1_energybarriers_l}
  \end{subfigure}\hfill
  \begin{subfigure}[t]{0.23\linewidth}
    \includegraphics[width=\linewidth]{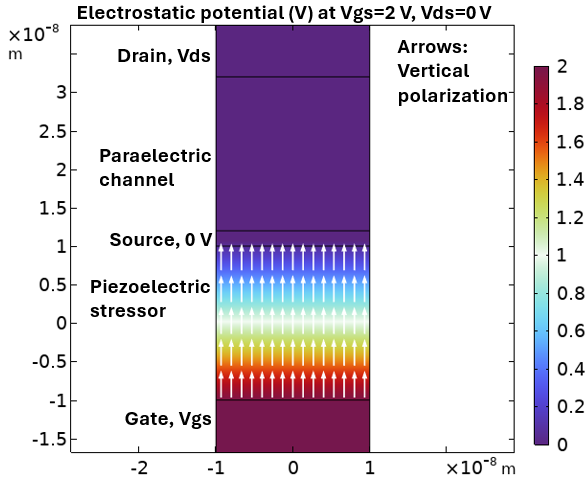}
\caption{}\label{fig:deviceFig1_energybarriers_m}
  \end{subfigure}\hfill
  \begin{subfigure}[t]{0.23\linewidth}
    \includegraphics[width=\linewidth]{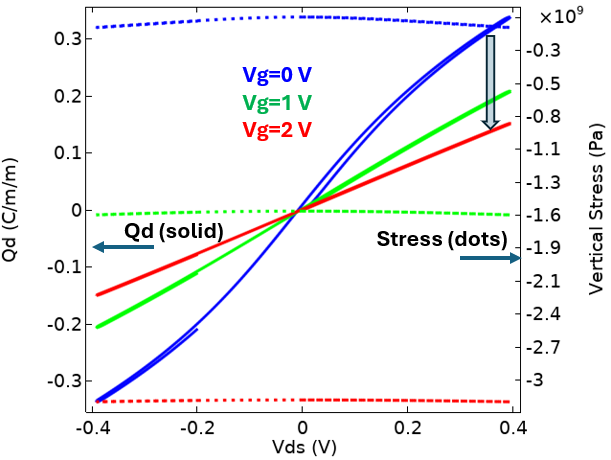}
\caption{}\label{fig:deviceFig1_energybarriers_n}
  \end{subfigure}
  \caption{\textbf{Switching mechanism and device operation.}
\textbf{a} - \textbf{d}: lowering of energy barrier for polarization switching due to electric field and stress two-assist mechanism in ferroelectric channel. \textbf{a},  energy landscapes in unstressed BTO and under compressive vertical (along z-direction) elastic stress of -425 MPa. \textbf{b} – \textbf{c}, Energy profiles  vs  stress under 0 and 7.5$\times 10^4$ V/cm field along z-direction. \textbf{d}, Corresponding energy barrier lowering for switching. \textbf{e} - \textbf{j}: device operation in ferroelectric channel. \textbf{e} – \textbf{f}, Polarization and electrostatic potential  vs V$_{GS}$. \textbf{g} – \textbf{i}, Channel vertical stress, drain charge hysteresis, coercive voltage are modulated by V$_{GS}$ (V$_{DS}$ signal is plotted in inset in \textbf{g}). Charge gain is obtained in \textbf{j}. \textbf{k} - \textbf{l}: reduction of phase space for vertical polarization under electric field with stress in two-assist mechanism in paraelectric BTO channel. \textbf{m} - \textbf{n}: drain charge reduces due to transferred stress in paraelectric channel by applying V$_{GS}$ across PZT stressor.}
\label{fig:deviceFig1_energybarriers}
\end{figure}

A logic/ memory state of a  channel is encoded into the direction and value of polarization of ferroelectric or paraelectric insulating material embedded between source and drain electrodes. For example, a polarization pointing along the vertical direction of device heterostructure growth is a state “1” and opposite to that direction is a state “0”. This polarization in a channel material is modulated by applying electric field and/ or mechanical stress.
In  tetragonal ferroelectric Barium Titanium Oxide (BTO), which we utilize as a proposed ferroelectric channel material in this work, under applied vertical $\langle 001 \rangle$ mechanical stress remanent polarization rotates away from the $\langle 001 \rangle$ axis towards  $\langle 110 \rangle$ direction as simulated with Density-Functional Theory (DFT) in Fig. \ref{fig:deviceFig_dft} \cite{giannozzi2009quantum} 
and with phase field model of energy landscapes (Eq. \ref{eq:LKfree energy abstract}, \cite{wang2010temperature}) in Fig. \ref{fig:deviceFig1_energybarriers_a}. In bulk ferroelectric   BTO there are six equivalent energy minima along principal crystal axes at zero field and zero stress corresponding to the spontaneous polarization ($P_s$) of 0.25 C/m$^2$, which shift to $\langle 110 \rangle$ and equivalent directions under stress. In the device source-drain voltage and a third gate electrode control an electric field and a stress in the channel, respectively.  As in a MOSFET case,  the switch between different states is accomplished by setting both gate voltage and source-drain voltage high. For a ferroelectric channel applying only gate voltage or only source-drain voltage, the device does not switch during characteristic delay time  $\tau$ due to a finite energy barrier for switching between two polarization states.  The barrier is reduced by applied compressive stress and further reduced by field, as illustrated by energy landscapes changes for BTO  in Fig. \ref{fig:deviceFig1_energybarriers_b} and in Fig. \ref{fig:deviceFig1_energybarriers_c} and by the resultant barrier dependence on stress in Fig. \ref{fig:deviceFig1_energybarriers_d}. This mechanism is suitable both for logic and for memory applications. To describe device and material behavior we developed a multi-physics modeling flow in COMSOL \cite{comsol2025} summarized in Fig. \ref{fig:deviceFig_simulationflow}  in Section \ref{sec:ModelsEquations}. We use a phase field model for energy landscapes in Eq. \ref{eq:LKfree energy abstract}.  We solve Landau-Khalatnikov (LK) equation Eq. \ref{eq:LKfree energy dynamics}  for switching dynamics in the ferroelectric or paraelectric channel. As a stressor material, we use a linear piezoelectric from PZT to PMN-PT.  An individual transcapacitor device consists of a vertical heterostructure with metal gate electrode, piezoelectric stressor insulator, metal source electrode, ferroelectric or paraelectric channel insulator and metal drain electrode layers, as simulated in Fig.\ref{fig:deviceFig1_energybarriers_e} (see Table \ref{tab:material parameters} for specific material choices and their parameters). Our piezoelectrics are poled vertically up to induce the compressive vertical stress  under appropriate orientation directions of gate fields applied across them as we discuss further. We solve coupled time-dependent Poisson and equilibrium mechanics equations across all layers. Time scale response of strain is limited by the speed of sound in a material. 
It is shorter than a polarization response time, for example in BTO it is in the sub-picosecond range, smaller than the Merz switching time of ferroelectric by an order of magnitude in Eq. \ref{eq:MerzLaw},
justifying the use of equilibrium mechanics in simulations presented in this paper. To facilitate stress transfer the device stack is pinned (i.e., the mechanical displacement u = 0) at the inner gate/stressor and  drain/channel interfaces in our two-dimensional device simulations. This can be thought of being achieved by pinning large area devices by the substrate on the bottom, and by wafer bonding process on the top of the device stack. A general stress transfer structure vehicle  for finite size three dimensional devices is proposed and discussed further in this work.  A source electrode between a stressor and a channel has two functions: it screens the gate electric fields across the stressor, and it transfers to the channel the  stress created in the stressor through the converse piezoelectric effect under application of gate voltage. This requires the source electrode to have sufficient density of free carriers which defines its   screening length being at least less than half of its thickness, and be an elastic material with the high Young's Modulus to enable an efficient stress transfer across its thickness. 
For our considered materials and source electrode thicknesses of 2 - 6 nm, the Young's Modulus of greater than 50 GPa provides the target level of stress in the channel. The target level of stress $\sigma$ created by a gate stressor capacitor is set by the stress sensitivity of the channel charge, which is measured by $dQ_{D}/d\sigma$, with $Q_{D}$ being a charge of the channel drain electrode. Electric potential, polarization  and stress distribution in the channel change in response to the applied gate voltage, as simulated in Figs. \ref{fig:deviceFig1_energybarriers_e}-\ref{fig:deviceFig1_energybarriers_f}.  In ferroelectric channel transcapacitor, gate action through stress transfer into the channel and the electrostrictive coupling between channel stress and polarization leads to switching barrier and  vertical component of remanent polarization lowering.  
Increasing gate voltage results in larger created vertical compressive stress in Fig. \ref{fig:deviceFig1_energybarriers_g}, which yields a hysteresis shrink and drain charge reduction with the corresponding positive drain-gate transcapacitance $dQ_{D}/dV_{GS} > 0$ in Fig.\ref{fig:deviceFig1_energybarriers_h} or coercive voltage reduction in Fig.\ref{fig:deviceFig1_energybarriers_i}. 
This dependence stems from coercive voltage $V_c$ being proportional to a switching barrier magnitude, and  being inversely proportional to the polarization component of remanent polarization along the switching field direction. 
In ferroelectric BTO the switching barrier collapses faster than $P_s$ (Figs.\ref{fig:deviceFig1_energybarriers_a}, \ref{fig:deviceFig1_energybarriers_b},  \ref{fig:deviceFig_dft_c} and \ref{fig:deviceFig_dft_d}) leading to the coercive voltage going to zero at high stress. 
The reduction of coercive voltage with stress $dV_c/d\sigma$ is an important metric for the ferroelectric channel material. 
The coercive voltage reduction and positive drain gate transcapacitance are key device metrics for ferroelectric channel devices that enable their proposed here application for memory and logic. 
In ferroelectric channel transcapacitors,  a device level charge gain (Eq. \ref{eq:CG_Device})  $\Delta Q_{D}/\Delta Q_{G}$ $>$ 1 is achieved,  a key requirement for driving circuits. 
In Fig. \ref{fig:deviceFig1_energybarriers_j} drain charge under gate voltage pulse undergoes full $2P_s$ channel polarization switching, exceeds the change of charge on the gate $Q_{G}$ for drain voltage range from  0.1 V  to the coercive voltage of 0.22 V. 
As V$_{DS}$ drain voltage is approaching the coercive voltage V$_{c}$, a smaller stress and a smaller gate voltage V$_{GS}$ change are sufficient to overcome the channel switching barrier and nominal charge gains asymptotically reach high values, more than 3000 in Fig. \ref{fig:deviceFig1_energybarriers_j}. 
For a stable logic device operation in a transcapacitor with a ferroelectric channel, the applied V$_{DS}$ needs to be below the coercive voltage to withstand thermal fluctuations. 
For V$_{DS}$=V$_{GS}$=0.15 V, 
70 mV below V$_c$, we achieve a charge gain of 10 times as shown in the inset in Fig. \ref{fig:deviceFig1_energybarriers_j}.
A positive transcapacitance required for logic operation is also achieved with a paraelectric channel which has zero remanent polarization at zero field. Under compressive $\langle 001 \rangle$ stress, the phase space for a $\langle 001 \rangle$ field-induced polarization  reduces as simulated with a paraelectric phase field model in  Figs. \ref{fig:deviceFig1_energybarriers_k}-\ref{fig:deviceFig1_energybarriers_l}. In a transcapacitor device in Fig. \ref{fig:deviceFig1_energybarriers_m} drain charge reduces  under higher gate voltage and transferred vertical stress, the paraelectric response becomes a linear dielectric response in Fig. \ref{fig:deviceFig1_energybarriers_n}. This positive drain-gate transcapacitance  is a necessary requirement for making a logic transcapacitor element. This reduction of polarization drain charge with increasing compressive $\langle 001 \rangle$ stress has been confirmed in our nanoindenter experiments on paraelectric BTO in Section \ref{sec:experim_demo}.

\subsection{Piezo-Transcapacitor Memory devices and memory circuits operation}
\label{memory_circuits_details}

The piezo-transcapacitor device with a ferroelectric channel is a non-volatile (NV) memory element with a built-in selector. The schematic of the device, its corresponding hysteresis behavior  are depicted in Fig. \ref{fig:piezoConcept}, Figs. \ref{fig:deviceFig1_energybarriers_e} -  \ref{fig:deviceFig1_energybarriers_h}.  A  1-bit memory element is obtained by using drain as a bit line (BL), source  as  a plate line (PL) and gate terminal as a word line (WL).  BL is applied across the BTO ferroelectric and determines its hysteresis loop. WL is applied across the PZT stressor and controls the strain transferred to the BTO, resulting in coercive voltage change in the ferroelectric BTO channel. 

 We ground the source terminals of all the memory elements, hence the voltage applied at BL and WL, become $V_{DS}$ and $V_{GS}$ respectively. The idea is to apply WL to select a given memory bit and read/write that using the BL signal. 
Using as an example the hysteresis loops depicted in Fig. \ref{fig:deviceFig1_energybarriers_h}, it is relatively straightforward to understand how the device works as a memory element with a built-in selector using the two-assist  mechanism. Let's say, the initial critical voltage $V_c$ of the device is $V_{ci}$ at WL $=0$ and the final $V_c$ with WL $=1.25V$ is $V_{cf}$. For the WL $= 0 V$ case, applying a BL pulse with amplitude $|BL| < V_{CI}$ would not result in a polarization reversal of the BTO. On the other hand, if WL is positively biased, a BL pulse with amplitude $V_{CF} < |BL| < V_{CI}$ will traverse the coercive voltage $V_{CF}$, enabling a polarization reversal of the BTO. Here we fully utilize the two-assist polarization switching mechanism - by field and by stress.

A good choice in the simulated device in Fig. \ref{fig:deviceFig1_energybarriers_e} for the BL amplitude is $0.14 V$, assuming a WL amplitude of $1.25$ V, to operate between the coercive voltages $V_{CI}$ and $V_{CF}$. The WL amplitude of $1.25$ V is selected to bring $V_{CF}$ close to 0 V. The bit line charge $Q_{BL}$ can be used to assign the memory states as 0 or 1. For example, the state could be assigned to be 0 when $Q_{BL} < 0$, and $1$ when $Q_{BL} > 0$. The WL is always pulsed positively, while the BL amplitude is set to positive to write a 1 and negative to write a 0. This enables writing of the memory bit using BL signal when it is selected by WL signal. The figure of merit ON/OFF ratio important for the write operation bit selectivity in the simulated memory device is 
\begin{equation}
Q_{ON/OFF}=
\frac{(Q_{BL}(WL = 1.25 V, BL = 0.14 V) - Q_{BL}(WL = 0 V, BL = 0.0 V))}
{(Q_{BL}(WL = 0 V, BL = 0.14 V) - Q_{BL}(WL = 0 V, BL = 0.0 V))} = 26.4.
\label{eq:QonQoff_deviceSim}
\end{equation}

\begin{figure}[ht!]
  \centering
  \begin{subfigure}{0.48\linewidth}
    \centering
    \includegraphics[width=\linewidth]{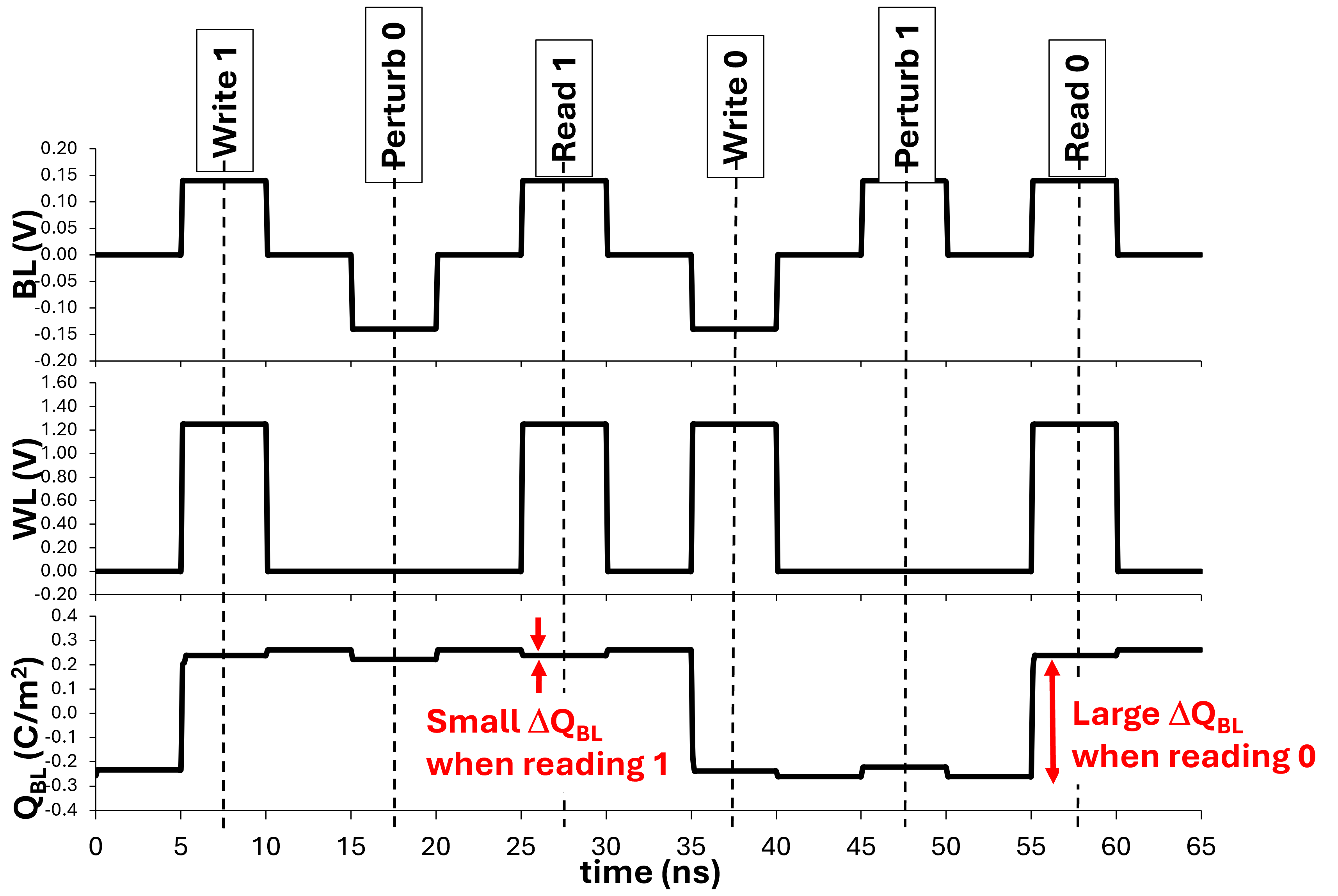}   
    \caption{}\label{fig:readDestructive_suppl}
  \end{subfigure}  \hfill
    \begin{subfigure}{0.48\linewidth}
    \centering
    \includegraphics[width=\linewidth]{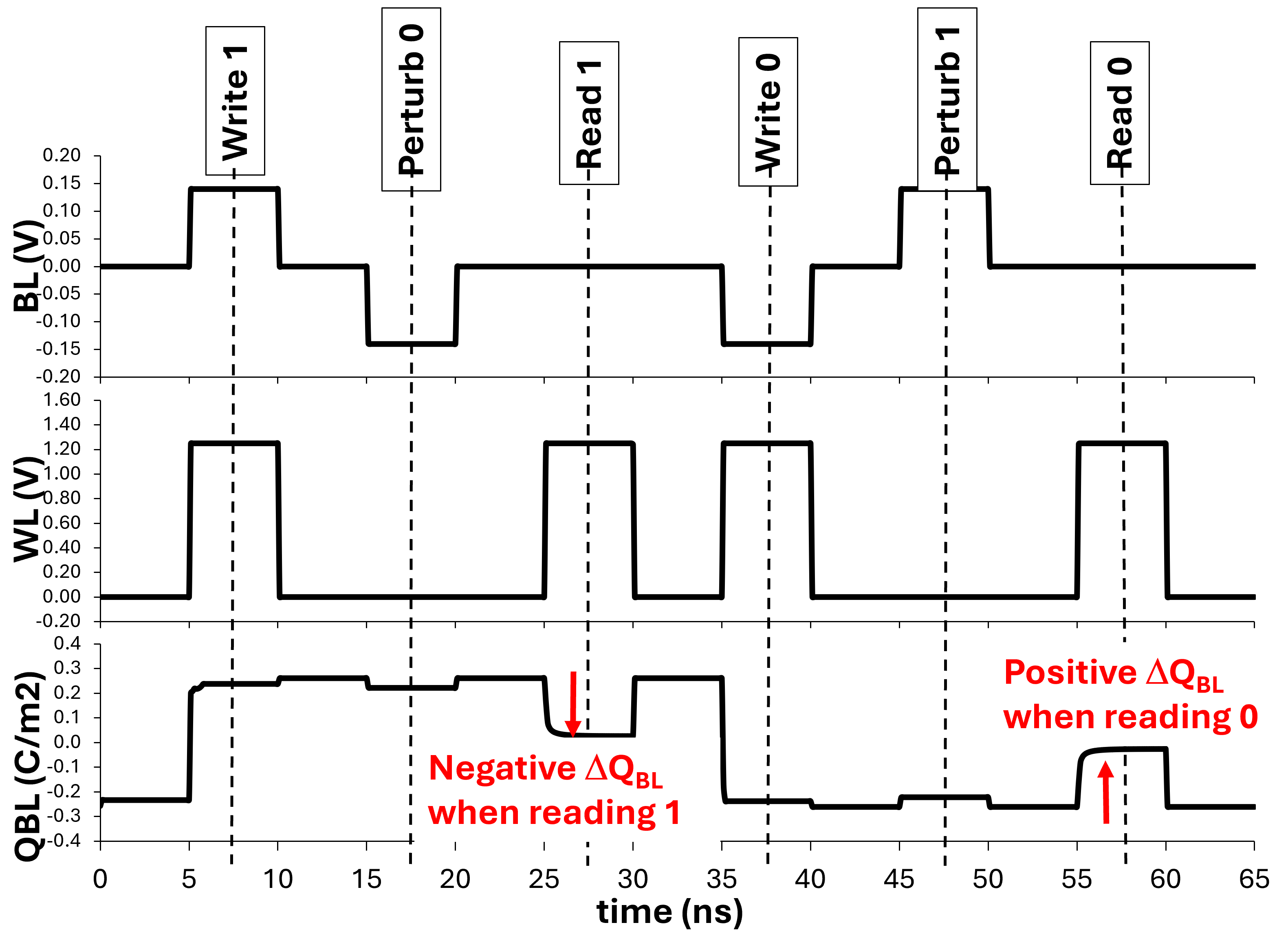}   
    \caption{}\label{fig:nonreadDestructive_suppl}
  \end{subfigure}
  \begin{subfigure}{0.48\linewidth}
    \centering
     \includegraphics[width=\linewidth]{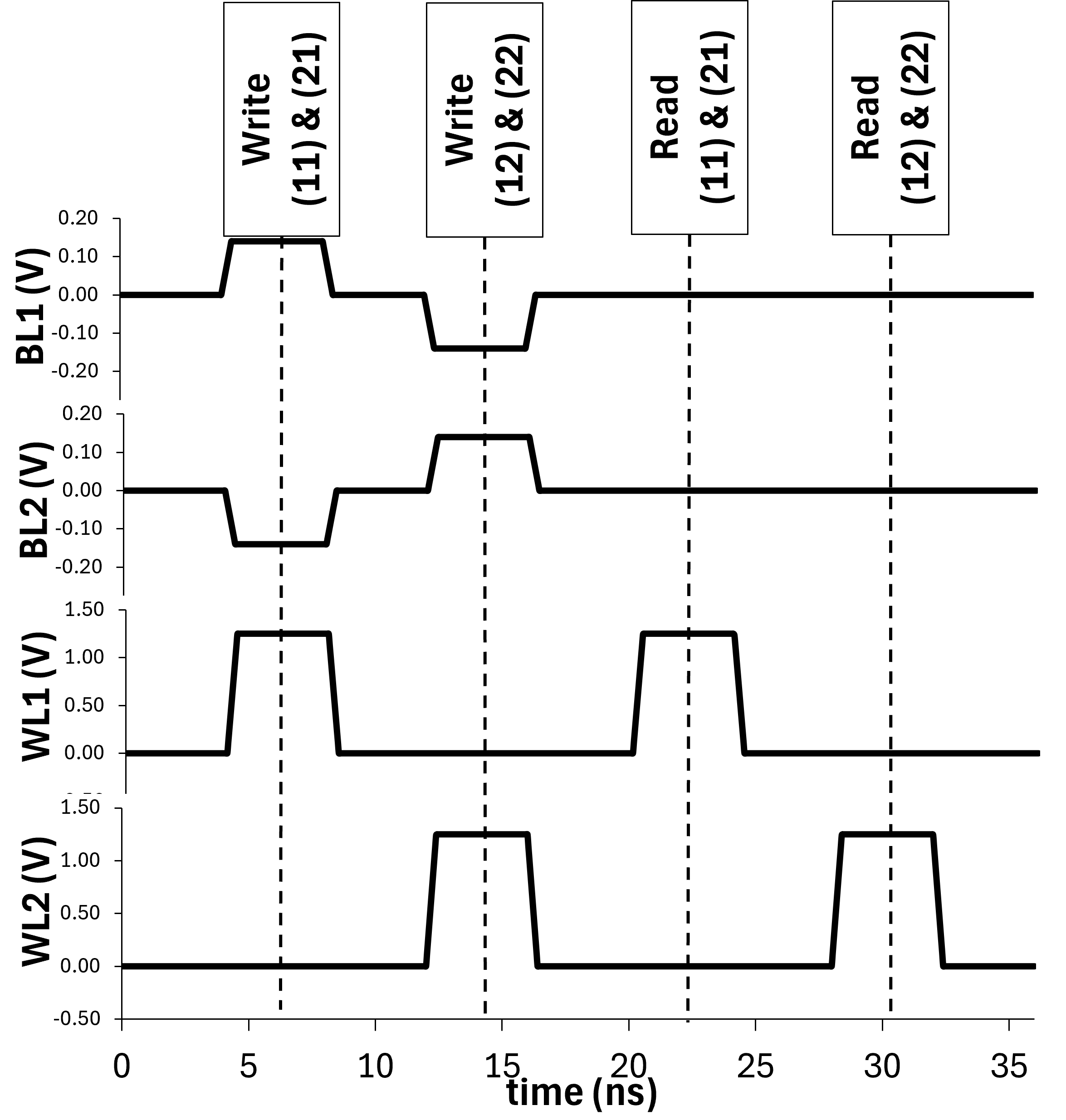}   
    \caption{}\label{fig:mem2x2VoltageSignals}
  \end{subfigure}
  \hfill
  \begin{subfigure}{0.48\linewidth}
    \centering
    \includegraphics[width=\linewidth]{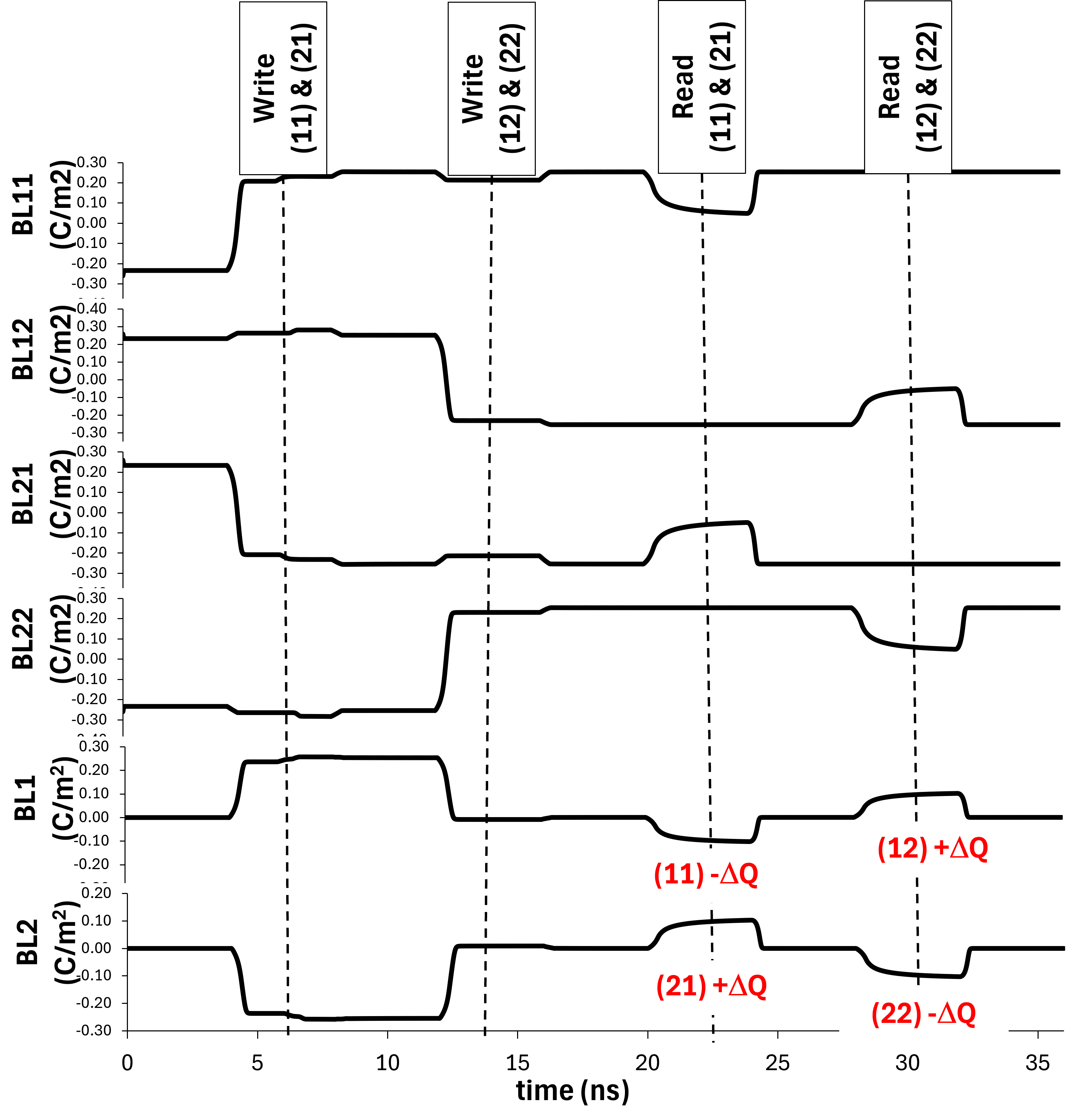}   
\caption{}\
\label{fig:Mem2x2Response}
\end{subfigure}
  \begin{subfigure}{0.48\linewidth}
    \centering
    \includegraphics[width=\linewidth]{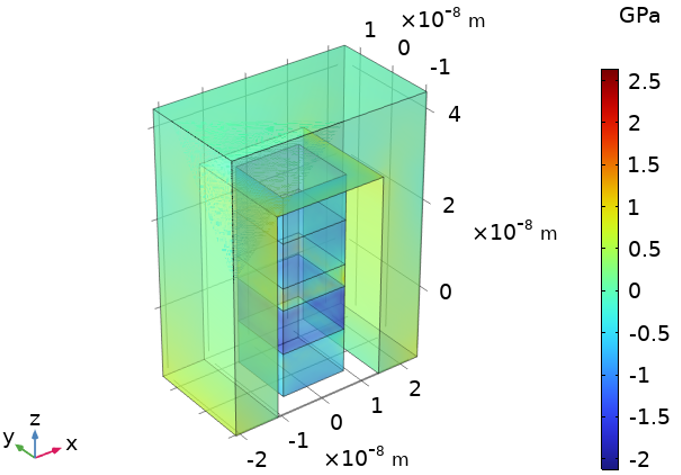}   
    \caption{}\label{fig:3ddesign_main}
  \end{subfigure}\hfill
  \begin{subfigure}{0.48\linewidth}
    \centering
    \includegraphics[width=\linewidth]{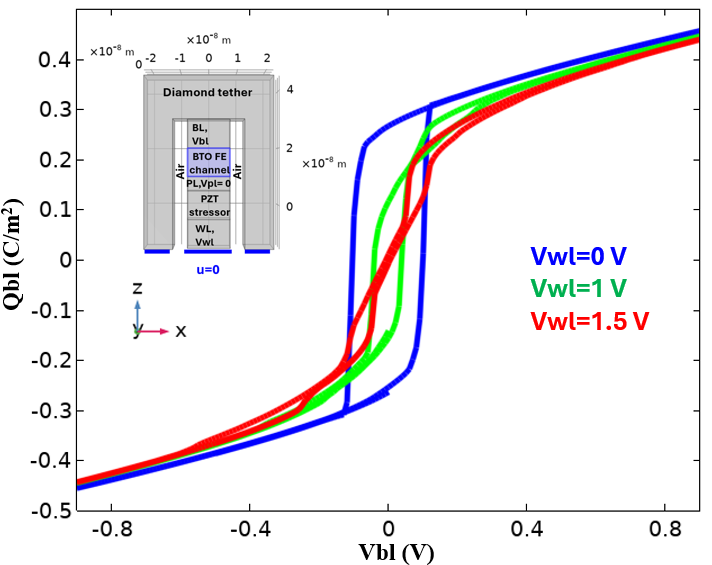}   
    \caption{}\label{fig:3ddesign_hysteresis_main}
  \end{subfigure}
  \caption{\textbf{Circuit operations results for memory.}   \textbf{a-b} Read destructive and non-read destructive memory operations for a single bit.
  \textbf{c} Voltage signals applied to the 2x2 array. \textbf{d} Charge response of the 2x2 array. \textbf{e} Vertical stress distribution at $WL = 1.5V$ and f) a hysteresis vs voltage at BL (Vbl) at various WL voltages (Vwl), for a memory bit 15x15 nm$^2$ area nanopillar design with a tether and side airgaps. }
  \label{fig:mem_circuit_results}
\end{figure}

For reading this memory bit, one can do it in both read destructive (similar to 1T1C FERAM) and non-read destructive ways. Reading the memory non-destructively has a huge advantage since the bit does not need to be written back after reading. It enables faster and denser memory arrays. 
The non-destructive memory operation of a single transcapacitor memory bit is simulated in Fig.~\ref{fig:nonreadDestructive_suppl}. Using a judicious $WL$ amplitude, it is possible to induce close to half of the polarization reversal charge without altering the state. For reading, the bit is selected with WL as for the write operation, and BL line is kept at zero. Contrast this to the destructive read operation in Fig~\ref{fig:readDestructive_suppl} where the BL is pulsed as well during read. A 1 is read when $\Delta Q_{BL} < 0$ (detected as a negative current pulse) and a 0 is read when $\Delta Q_{BL} > 0$ (detected as a positive current pulse). Another advantage of this reading approach is that there is no interference from other bits sharing the same BL, since only the $WL$ pulse is applied to the bit being read. Consequently, the non-destructive read option may be able to support larger array sizes. In addition to the write and read operations, Fig.~\ref{fig:nonreadDestructive_suppl} shows that perturbed pulses on the shared BL —when other bits on different wordlines are written or read— do not alter the state. When only WL is pulsed, the remaining switching barrier exceeds thermal energy to suppress disturbs of a bit state. Disturb dynamics will be further investigated experimentally in our future work.

A destructive read application of the piezo-transcapacitor memory is described by the sequence of points in Fig~\ref{fig:readDestructive_suppl}. The read operation can be arbitrarily set equal to the write 1 operation, such that a small change in $Q_{BL}$ (detected as a small negative current pulse) happens when the state is already a 1. On the other hand, a relatively large $Q_{BL}$ change equal to the total polarization reversal (detected as a larger positive current pulse) happens if the state was 0. Therefore, the read 0 operation is destructive and the bit must be rewritten back.

Fig.~\ref{fig:Figure1_2x2_memory} shows a circuit model of a 2x2 memory array built from 1-bit memory elements. For each row, we tie BL nodes of the devices together and call it a bitline (BL) signal. For each column, we tie their WL lines together and call it a wordline (WL) signal. Here, the idea is to select a column with the WL, similar to the 1T1C FERAM array, and read and write the memory words using the BL signal. 
The bit notation used is (nm), where n represents the bit location and m represents the word location. The memory operation of a 2x2 array in the read non-destructive way is simulated in the Fig.~\ref{fig:Mem2x2Response}. The corresponding voltage signals are shown in Fig.~\ref{fig:mem2x2VoltageSignals}.

Similarly to the 1-bit simulations, the bits are assigned to be 0 when $\Delta Q_{BL} < 0$, and 1 when $\Delta Q_{BL} > 0$. The non-destructive reads produce a positive bit-line charge change when the state is 0, and a negative charge change when the state is 1.
For the Fig~\ref{fig:Mem2x2Response} example, the initial bit states and write waveforms are designed to force a flip for all the bits in the array. To aid the eye, the Table~\ref{tab:2x2ArrayStates} illustrates the intended results. As illustrated and verified by the COMSOL simulations, the write operations produce the intended flipping of the individual bit charges, as shown by the top four graphs of Fig.~\ref{fig:Mem2x2Response}. The bottom 2 graphs in Fig.~\ref{fig:Mem2x2Response} show the total charge in each bit line, normalized by area, demonstrating the expected direction of charge changes in accordance with the written bit state. As intended, each bit-line shows a change in charge associated only with the individual bit of the word-line being pulsed, with no interference from the other wordline. By extension, this should be true for arrays of any size.

Individual transcapacitor memory bits and their arrays can be realized using nanopillar with side airgaps design simulated in Fig. ~\ref{fig:3ddesign_main}. It has been shown experimentally that nanopillar structures can exhibit a giant piezoelectricity \cite{liu2020giant}. Here an efficient stress transfer from PZT stressor to BTO ferroelectric channel across the PL electrode metallic material is facilitated both by the finite size 15 $\times$ 15 nm$^2$ device area, 5 nm airgaps on each side along the width (x-direction in the figure), and a pi-shaped tether with 15 nm on top of BL electrode and 10 nm on airgap sides (as illustrated in the inset in Fig. \ref{fig:3ddesign_hysteresis_main}). A diamond material with Young's
Modulus of  1050 GPa is used for the tether. The otherwise free standing structure is fixed to the substrate at tether and a bottom WL metallic electrode. The substrate is simulated to be lattice matched to channel BTO. Other simulation details and parameters are included in Section \ref{device_3ddev}. A stress transfer is made efficient through stress exchange by edge forces on airgap sides of the nanopillar, as well as by fixed top boundary of the device by the tether. We demonstrate  the hysteresis shrink of a bit line charge Qbl vs Vbl voltage applied across the channel at the three values of Vwl voltages  applied across the PZT stressor. We obtain coercive voltage reduction in the channel from 0.1 V to 0 when Vwl increases from 0 to 1.5 V in Fig. \ref{fig:3ddesign_hysteresis_main}. The corresponding vertical transferred stress of -1 GPa in BTO is obtained in simulated stress distribution in the nanopillar for Vwl = 1.5 V in Fig. \ref{fig:3ddesign_main}. Note that the finite size geometry  reduces the coercive voltage and stress transfer from their planar limits, yet yields sufficient device parameter space for the device operation. This full transcapacitor device functionality for finite size devices will allow to build full integrated memory arrays circuits as we discuss further in  Section \ref{sec:future}.

\subsection{Piezo-Transcapacitor logic}
\label{logic_device_details}

\begin{figure}[ht!]
  \centering
  \begin{subfigure}[t]{0.33\linewidth}
    \includegraphics[width=\linewidth]{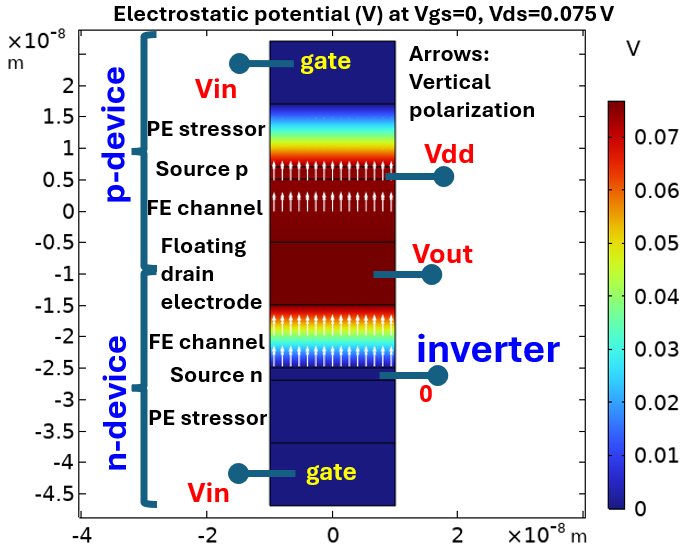}
    \caption{}\label{fig:deviceFig_logic_a}
  \end{subfigure}\hfill
  \begin{subfigure}[t]{0.33\linewidth}
    \includegraphics[width=\linewidth]{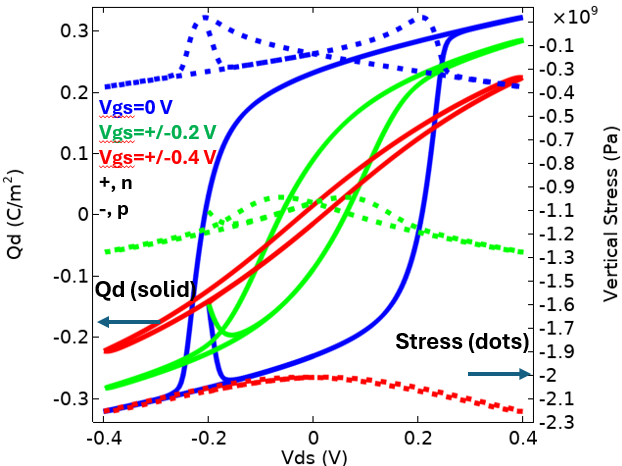}
    \caption{}\label{fig:deviceFig_logic_b}
  \end{subfigure}\hfill
  \begin{subfigure}[t]{0.33\linewidth}
    \includegraphics[width=\linewidth]{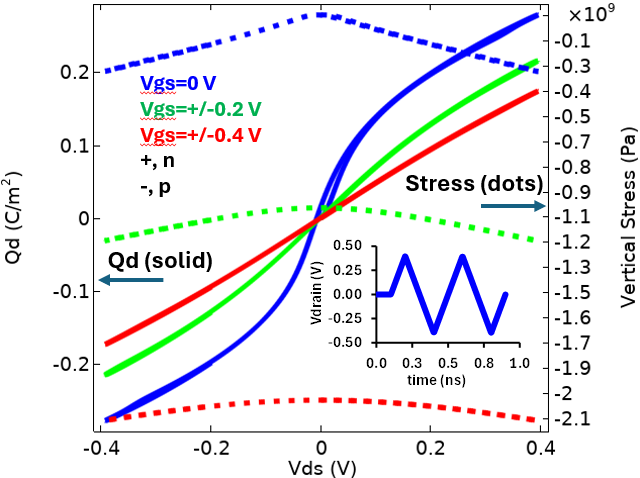}
    \caption{}\label{fig:deviceFig_logic_c}
  \end{subfigure}
  \\
  \begin{subfigure}[t]{0.33\linewidth}
    \includegraphics[width=\linewidth]{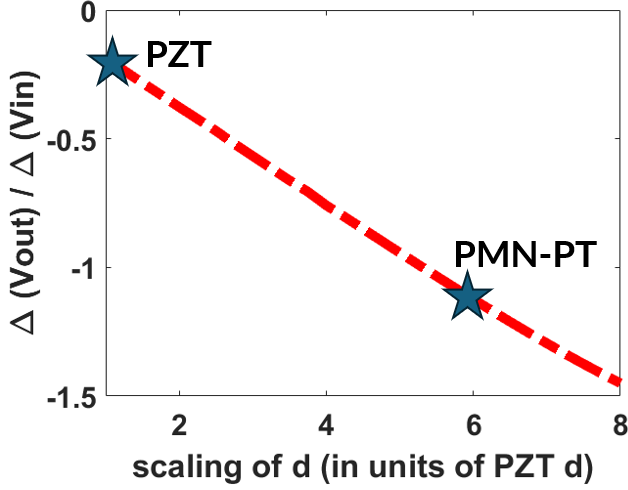}
    \caption{}\label{fig:deviceFig_logic_d}
  \end{subfigure}\hfill
   \begin{subfigure}[t]{0.33\linewidth}
    \includegraphics[width=\linewidth]{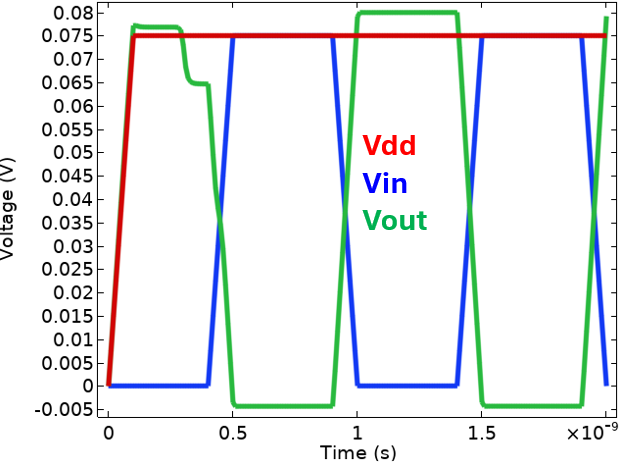}
    \caption{}\label{fig:deviceFig_logic_e}
  \end{subfigure}\hfill
  \begin{subfigure}[t]{0.33\linewidth}
    \includegraphics[width=\linewidth]{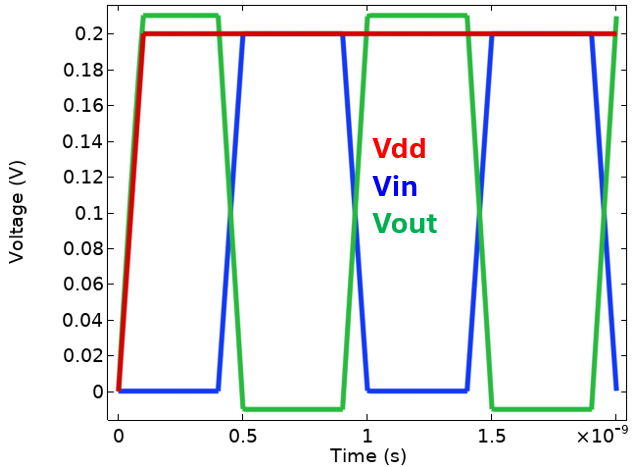}
    \caption{}\label{fig:deviceFig_logic_f}
  \end{subfigure}
  \\
  \begin{subfigure}[t]{0.33\linewidth}
    \includegraphics[width=\linewidth]{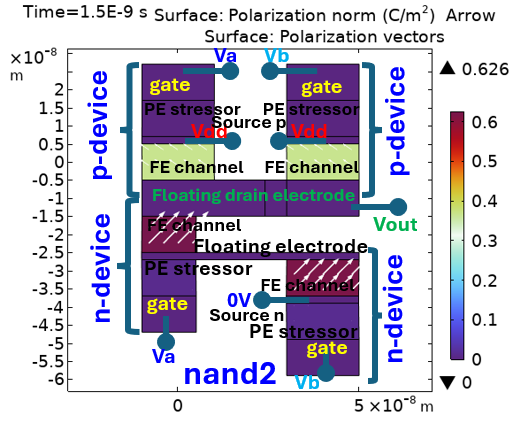}
    \caption{}\label{fig:deviceFig_logic_g}
  \end{subfigure}\hfill
  \begin{subfigure}[t]{0.33\linewidth}
    \includegraphics[width=\linewidth]{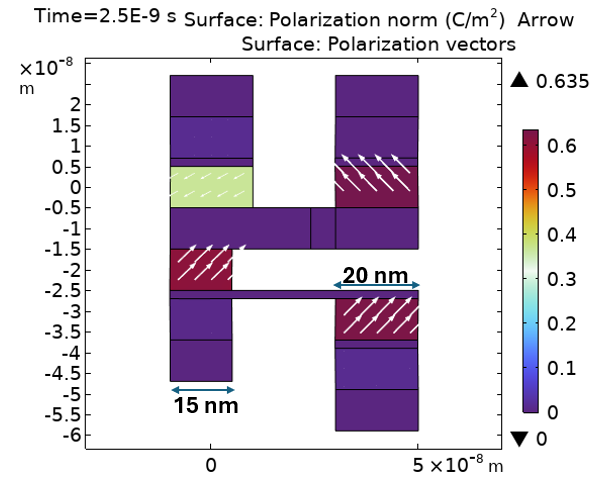}
    \caption{}\label{fig:deviceFig_logic_h}
  \end{subfigure}\hfill
   \begin{subfigure}[t]{0.33\linewidth}
    \includegraphics[width=\linewidth]{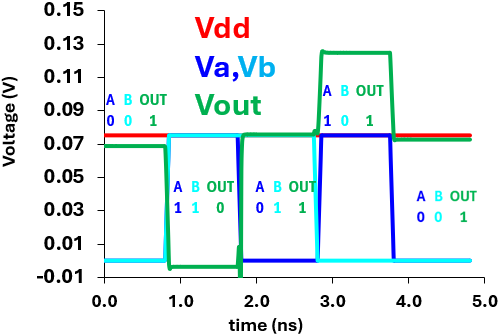}
    \caption{}\label{fig:deviceFig_logic_i}
  \end{subfigure}
  \caption{\textbf{Logic device operation,  inverter and NAND demonstration.}
Electrically connecting a transcapacitor n-type device  with its vertically flipped image p-device  results in an inverter device (\textbf{a}) with floating electrode used as an output node Vout.  Electrostatic potential Vout follows the charge re-distribution determined by the polarization state in each device shown for ferroelectric channel inverter. \textbf{b}-\textbf{c}: individual device responses to applied voltages.    Drain charge (solid) and vertical stress (dots) in response to Vds signal  (inset in \textbf{c}) vs Vgs for ferroelectric channel  (\textbf{b}) and for paraelectric channel  (\textbf{c}). Drain charge hysteresis shrink or charge reduction are obtained for Vgs $>$ 0 in n-type and for Vgs $<$ 0 in p-type devices.
Voltage inversion margins in an inverter increase   as a function of stressor piezo-coefficient $d$  (\textbf{d}). Logic devices are simulated with BTO channel, and PMN-PT stressor. Work functions of source of n-device and of source of p-device electrodes are shifted by +0.2 eV and -0.2 eV for paraelectric channel inverter.  Utilizing coercive voltage $V_c$ reduction and  polarization  reduction under stress and field yield voltage inversion in a ferroelectric channel (\textbf{e}) and in a paraelectric channel  (\textbf{f}). NAND logic gate is demonstrated using two p-devices and two n-devices. Polarization distribution for two logic stages at 1.5 ns  (\textbf{g}) at 2.5 ns (\textbf{h}) corresponding to time trace (\textbf{i}) which reproduces the full logic truth table of the NAND gate.}
\label{fig:deviceFig_logic}
\end{figure}

Fig.~\ref{fig:deviceFig_logic_a} shows a simulated inverter circuit with two complementary transcapacitor devices, called N-transcapacitor and P-transcapacitor. All insulator thicknesses are 10 nm, and device widths are  20 nm simulated with periodic boundary conditions along width direction. Channel and stressor thicknesses determine  operating voltages, scaling them commensurately  allows to obtain an equivalent device operation at higher thicknesses and biases as verified with the simulation for up to 20 nm insulator thicknesses. Additional device details and parameters are provided in 
Table~\ref{tab:material parameters}.  
Gates of both devices are tied to form an input terminal Vin and drains tied together to form the output terminal Vout, similar to CMOS. The source terminals of the devices are tied to different supply voltages. The source of p-device receives a supply voltage V$_{DD}$, while the source of n-device is kept at 0 V in  Fig. \ref{fig:deviceFig_logic_a}. This setup enables the inverter to receive input Vin signal at the gate terminals and produce output Vout at the floating drain electrode terminal. For example, the output node has the high potential close to  V$_{DD}$ in the logic state 0 of the inverter (V$_{DD}$) in the simulated electrostatic potential distribution in the device  in Fig \ref{fig:deviceFig_logic_a}. Both the input and output nodes are floating in extended circuits, since they are connected to the previous and next stages of the logic.   

 Complementary n-type and p-type devices are vertically flipped mirror images of each other as shown in Fig. \ref{fig:deviceFig_logic_a}. 
 In FE channels, for an inverter initialization, drain electrodes of N and P devices are initialized to $-P_s$ and $+P_s$ respectively, making the initial charge at the output node zero. For paraelectric channel devices, 
 the work function of the source electrodes of N and P devices are shifted by 0.2 eV and -0.2 eV respectively, to achieve initializing N device to $-P_s$ and P device to $+P_s$ at $V_{DS}=0$ voltage. Note that an inverter functionality can be achieved with both ferroelectric and paraelectric channels using the electrode workfunction engineering to  operate with supply voltage V$_{DD}$ above their corresponding coercive voltages V$_c$.  Using supply voltages  V$_{DD}$  $<$  V$_c$ in transcapacitors with ferroelectric channels does not require reset cycles in logic operations as we discuss further.
 Individual device characteristics of drain charge and vertical stress in response to V$_{DS}$ signal (inset in  \ref{fig:deviceFig_logic_c} for n-type, negative of this is for p-type) at the three values of gate voltage are simulated for a ferroelectric channel in Fig. \ref{fig:deviceFig_logic_b} and  for a paraelectric channel in \ref{fig:deviceFig_logic_c} for both n and p types of devices shown with no source workfunction offsets.  For these device terminal characteristics V$_{DS}$ signal is applied to drains, and sources are kept at 0 V. For large gate voltage magnitudes, drain charge reduces through the hysteresis shrink in Fig. \ref{fig:deviceFig_logic_b} and transition to linear dielectric response is obtained in Fig. \ref{fig:deviceFig_logic_c}. The P-transcapacitor  has a similar device operation as the N-transcapacitor for opposite signs of V$_{DS}$ across the channel and of V$_{GS}$ across the stressor.  

A transcapacitor circuit is dominated by capacitances as their circuit components and hence contains floating nodes. The circuits work on two core principles. First, at the floating nodes charges are always conserved. Second, voltage of the floating nodes is determined via charge balance by capacitance based voltage division among all elements which influence the charge in a channel. 
In contrast to a transistor circuit, where a continuous current flows from the drain, charges up the gate capacitor of the next stage; for a novel paradigm transcapacitor circuit, the displacement charge to the gate of the next stage is provided by the displacement charge from the drain of the previous stage. For an unloaded inverter in Fig. \ref{fig:deviceFig_logic_a}, a circuit model  consists of p-device and n-device  source-drain capacitances $C_{DS,p}$, and  $C_{DS,n}$ of channel capacitors connected in series (see Fig. \ref{fig:deviceFig_circuitinv}). The effect of stressors on their electro-mechanically coupled channel charges can be expressed through the effective capacitances $C_{DG,p}$, and  $C_{DG,n}$ of two capacitors connected in parallel.
The figure of merit of individual devices, positive drain-gate transcapacitance $TC = -  \mathrm{
 (effective)} ~C_{DG}$ controls the magnitude of the released charge and the degree of inversion. This can be understood from a charge balance on a floating drain electrode in an unloaded inverter: 
\begin{equation}
\begin{split}
C_{DS,p}  (Vout - V_{DD}) + C_{DS,n} Vout  + (C_{DG,p} +C_{DG,n}) (Vout-Vinp) = 0, \\
Vout = \frac{C_{DS,p}  V_{DD} +(C_{DG,p} +C_{DG,n}) Vinp }{C_{DS,p} +C_{DS,n} + C_{DG,p} +C_{DG,n}}. 
\label{eq:chargebalanceinv}
\end{split}
\end{equation}
For equivalent n and p devices, with $C_{DS,p}  = C_{DS,n} = C_{DS}$ and $C_{DG,p}  = C_{DG,n} = C_{DG}$,  
 the necessary condition to get voltage inversion simplifies to having positive transcapacitance, or $C_{DG} < 0$ which is achieved in reduction of polarization charge under stress in ferroelectric or paraelectric channel device. For $C_{DG}=-C_{DS}/2$ the full rail-to-rail inversion is obtained. For  $C_{DG} < -C_{DS}/2$, a voltage margin gain is obtained, for  $C_{DG} > -C_{DS}/2$ inversion margins are degraded.

For driving successive logic stages, V$_{GS}$ needs to be comparable to the supply voltage V$_{DD}$ in magnitude. This requires a strong stressor material to obtain the target vertical stress $\sigma$ at a lower gate field. For logic application in Fig. \ref{fig:deviceFig_logic} we use PMN-PT material which has six to seven times larger piezo-coefficient than the PZT material which meets memory  requirements. Figures of merit for a stressor material for logic, $d \sigma/d V_{GS}$, and $d \sigma/d Q_{G}$  (where $Q_{G}$ is a gate charge), need to be maximized for an optimal fanout as discussed in Section \ref{sec:future}.

The operations of an inverter can be understood as follows, somewhat similar to CMOS: as the input voltage goes to logic 1, $V_{GS}$ for the N and P devices are at logic 1 and 0 respectively, turning off the P device and turning on the N-device. N device goes into reduced charge state (lower energy barrier state, shrinking its hysteresis for FE channels, and linear dielectric state for PE channels), 
while P-transcapacitor remains in the original high charge state. 
This way, N-transcapacitor releases the corresponding charge due to its positive drain-gate transcapacitance
and develops a negative potential on a floating node leading to low or negative $V_{DS}$ across it, while the P-transcapacitor develops larger negative $V_{DS}$ across it. This released charge from the N-transcapacitor serves to drive the gates of the next stage transcapacitor devices, and the dielectric component of the P-transcapacitor. Hence, input of logic level 1, leads to the output of zero (or negative), since the N-transcapacitor drops a voltage across it, while the P-transcapacitor has large voltage across it.  The circuit becomes an inverter circuit reaching voltage inversion margins which increase with the magnitude of piezo-coefficient of the stressor as simulated in Fig. \ref{fig:deviceFig_logic_d}.

Since the degree of inversion depends on the amount of the released charges and final capacitance values, the output voltage is not limited by V$_{DD}$, and a rail-to-rail voltage gain $\Delta Vout/ \Delta Vin > 1$ in inversion is obtained both for ferroelectric channel in Fig. \ref{fig:deviceFig_logic_e} and for paraelectric channel in Fig. \ref{fig:deviceFig_logic_f}.  Unloaded inverters for both types of channels are simulated for a limit of low load. Both types of devices achieve an inverter functionality at low supply voltages of V$_{DD}$ = 75 mV  and V$_{DD}$ = 200 mV for ferroelectric and paraelectric channels, respectively. Operating at V$_{DD}$ = 75 mV supply under the original V$_c$ of the FE channel device (in this case 0.2 V) allows the FE channel to not switch fully and  still reach inversion.  We note here the importance of a key function of source electrode in our proposed novel device: it screens gate fields, removing the detrimental gate voltage follow effect and allows achieving voltage inversion.
 
For both types of channels with released charges being less than $P_s*A$, the channel recovers its original state when Vin is 0. This partial switched charge lowers the magnitude of required stress created by stressor to achieve charge gains.  Paraelectric channel based devices are particularly desirable for low energy consumption, due to either none or very small critical voltages. We call energy consumption in hysteresis as damping losses. For FE channel based devices as well, damping losses do not happen if the desired polarization state is maintained throughout the inverter operation and the full hysteresis traversal does not happen.
If the polarization states are flipped during the operation for FE channel based devices, they may require reset. Energy comparison is described in detail later in the paper in Section \ref{sec:EdCompar}.

A universal logic gate is defined as a multi-input gate which can make a complete set of logic, such as a NAND2 gate. A transcapacitor based NAND2 gate can be created with two N-transcapacitors stacked in series and two P-transcapacitors in parallel as shown in the Fig.~\ref{fig:NandGateCircuit}. NAND2 realization for ferroelectric channel transcapacitors is demonstrated in the device simulation in Figs. \ref{fig:deviceFig_logic_e} -\ref{fig:deviceFig_logic_i}. The working principle of this gate remains the same as for the inverter, and is based on charge balance in the device.  The NAND2 circuit contains two p-devices ('p1' and 'p2') connected in parallel sharing a floating drain output Vout electrode. This electrode is also shared with the first n-device ('n1'), which is connected in series via a floating electrode at the source to a drain of another n-device ('n2').
As in an inverter case, supply voltage V$_{DD}$ is applied to sources of p-devices. The source of the n2 device is kept at 0 V.   Gates of p1 and n1 devices  are tied to form an input terminal Va, gates of p2 and n2 devices  are tied to form a second input terminal Vb.  This setup enables the NAND2 gate to receive two input Va and Vb signals at the corresponding gate terminals and produce output Vout at the floating drain electrode terminal. The thickness of all insulators here was kept the same as for the inverter at 10 nm. As indicated in Fig. \ref{fig:deviceFig_logic_h}, the width of n1 was reduced by 5 nm to optimize the NAND2 performance. All other device widths are kept the same as for the inverter at 20 nm. The full NAND2 logic truth table was tested with a time trace in Fig. \ref{fig:deviceFig_logic_i}. For example, at time 1.5 ns for logic state 1 on Va, and for logic state 1 on Vb, the output potential on a floating Vout node goes to zero (or negative). The corresponding distribution of polarization in Fig. \ref{fig:deviceFig_logic_g}  shows that two p-devices  and two n-devices  have the same direction polarization states, correspondingly. At time 2.5 ns for logic state 0 on Va, and for logic state 1 on Vb, the output potential on a floating Vout node goes high (logic 1). The corresponding distribution of polarization in Fig. \ref{fig:deviceFig_logic_h}  shows that two p-devices have opposite direction of  polarization, with p2 device polarization rotating due to transferred stress under action of Vb gate, therefore lowering the total  surface charges on each side of a floating drain electrode and hence raising the floating potential Vout. Similarly, logic state 1 is obtained on Vout for all other combinations of Va and Vb where at least one of them is at logic 0.

 Fundamental elements of logic operation  inverters and NAND2 gates at low voltage of operation can be realized with novel computing platform transcapacitor devices.

\section{Energy and Delay Gains with Transcapacitors}
\label{sec:EdCompar}

Conventional CMOS logic circuits work in a resistive switching regime. In this case, the energy to switch is estimated as
\begin{equation}
E_{rs}=C V_{DD}^2
\label{eq:nonadiabE}
\end{equation}
where $C$ is the total capacitance of the logic stage. It includes the capacitance of the driving sub-circuit, the interconnect, the load capacitance, and the parasitic capacitances. The switching time is dominated by the switching delay $RC$. The dissipated power is proportional to the effective resistance $R$. The effective capacitance includes the capacitances of the driving devices, of the interconnecting wires, and of the load in the next stage of the circuit. Therefore, the resistance factor is canceled in the result.

Reactive logic is switched at a slower time, $\tau$. Typically slower switching is achieved by ramping the supply voltage over that time. This results in a smaller dissipated energy \cite{galisultanov2017capacitive}
\begin{equation}
E_{re}=C V_{DD}^2 \frac{RC}{\tau}
\label{eq:adiabE}
\end{equation}
Switching of transistor logic over time scales of the circuit’s $RC$ time and thus will not be reactive. 
For transistor logic, reactive
regime can be achieved by switching the power supply at a rate slower than the RC time. 

In our case of transcapacitor logic, the 
reactive regime can be achieved even at a constant supply voltage. The slowness of switching arises from the stochastic thermal nature of switching and is described by the Merz law.
According to it, the expected time of the thermal switching $T \rightarrow \tau_M$ is:
\begin{equation}
\tau_M = \tau_0 \exp{ \left[ \left( \frac{V_a}{V} \right) ^{\alpha_M} \right]}
\label{eq:MerzLaw}
\end{equation}
Here $\tau_0$ is the switching time at a very high voltage $>>V_a$, $V$ is the applied voltage, $V_a$ is the activation voltage (proportional to the coercive voltage $V_c$), and ${\alpha_M}$ is the Merz exponent. 
This time can be in the range from tens of picoseconds to nanoseconds.
For ferroelectric logic, the clocking time is limited by and needs to be set to longer than $\tau_M$.
Compare this with the CMOS or wire delays, which is in the range of picoseconds. 
This difference in scales ensures the ability of ferroelectric circuits to function in the reactive regime. 
The slowdown of the circuit operation into the 
reactive regime permits a trade-off of lower switching energy for slower speed. It is a very favorable trade-off. 
Oftentimes, chip architects optimize the throughput per power (TOPS/W), which amounts to the inverse energy of switching. 
The slowness of switching can be compensated by placing multiple cores on a chip, for which the chip area is typically available.

We made a side-by-side comparison of the delay and energy consumption in simple circuits based on CMOS transistors and on transcapacitors, respectively. 
This comparison follows the method developed for the benchmarking of exploratory beyond-CMOS devices \cite{nikonov2013overview,nikonov2015benchmarking}. 
In doing so, we rely on the analytical expressions in the sections above. These estimates are first obtained for simpler logic gates and then aggregated for more complicated subcircuits. More details on the methods and parameters used for the estimates are given in 
Sections~\ref{sec:EdCMOS} and~\ref{sec:EdTpol}.
The parameters of the CMOS transistor correspond to the LP transistors of the N5 generation technology from TSMC
\cite{yeap20195nm}.

\begin{figure}[ht!]
\centering
\begin{subfigure}[t]{0.48\linewidth}
\includegraphics[width=\linewidth]{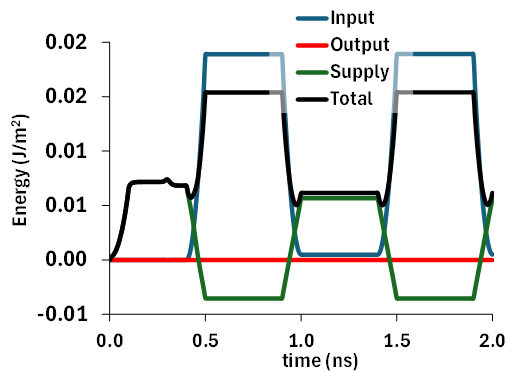}
\caption{}
\label{fig:EnergyUnloadedInv}
\end{subfigure}\hfill
\begin{subfigure}[t]{0.48\linewidth}
\includegraphics[width=\linewidth]{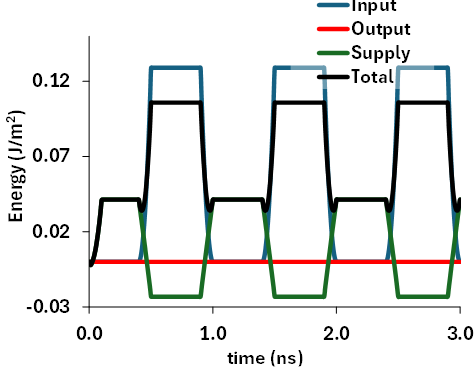}
\caption{}\label{fig:EnergyFO1Inv}
\end{subfigure}
\centering
\begin{subfigure}[t]{0.48\linewidth}
\includegraphics[width=\textwidth]{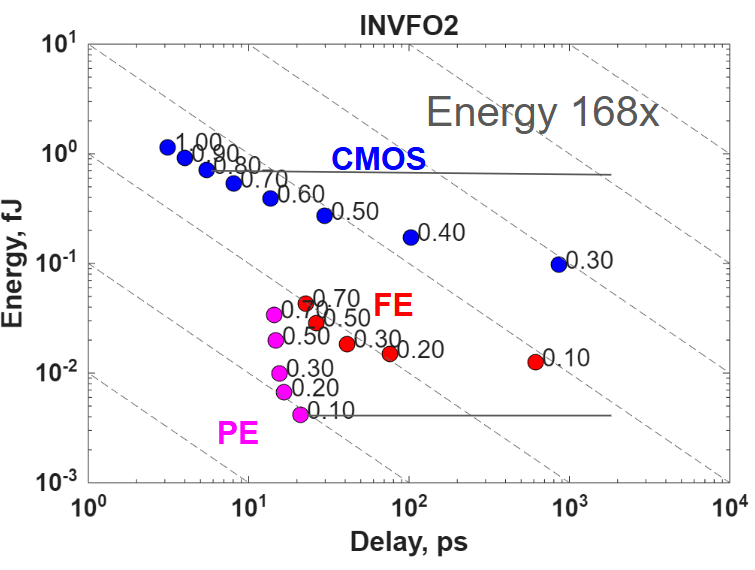}
\caption{}
\end{subfigure}
\begin{subfigure}[t]{0.48\linewidth}
\includegraphics[width=\textwidth]{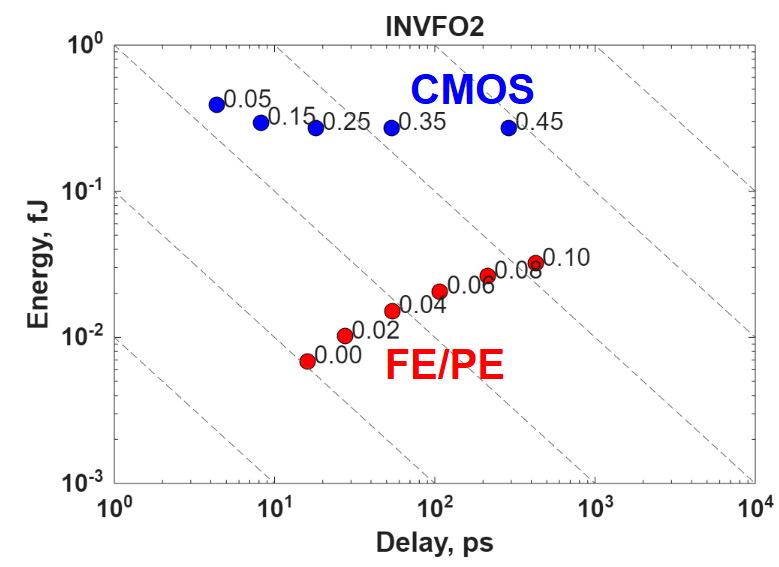}
\caption{}
\end{subfigure}
\caption{
Energy-efficient switching with transcapacitors. Energy flows (through the input, output, supply, and total) vs. time for operation of unloaded inverter with ferroelectric channel (a) and paraelectric channel (b). Benchmarking of energy and delay (following the standard methodology
\cite{nikonov2013overview,nikonov2015benchmarking}) of inverters with fanout of 2 based on transistor (CMOS) and ferroelectric (FE) and paraelectric (PE) transcapacitor devices (c) for various values of supply voltage (labeled next to markers) and (d) for various values of threshold voltage for CMOS and coercive voltage for FE (labeled next to markers).}
\label{fig:energyFig_inverter}
\end{figure}

The cumulative flows of energy (per unit area) into and out of an unloaded transcapacitor inverter are shown in Fig.~\ref{fig:EnergyUnloadedInv} for a ferroelectric channel and in Fig.~\ref{fig:EnergyFO1Inv} for a paraelectric channel.
Note that these results are obtained for the unloaded inverters (i.e., ones driving a zero load in the next logic stage). Also in these plots (unlike the following ones) the Joule energy dissipation in resistances is neglected.
These results underscore the feature of transcapacitors that they have a relatively high value of energy flows in parts of the switching cycle and similar opposite flows in opposite parts of the switching cycle.
They approximately cancel, and thus the total dissipated energy is much smaller. This approximate recovery of energy is the result of 
the reactive switching and of the absence of the direct current path from the supply to the ground.

We provide estimates for the energy vs. delay of 
an inverter with a fanout of 2 (FO2) and of a 32-bit adder,
see Fig.~\ref{fig:energyFig_inverter}.
They are made over a range of supply voltages,
while the threshold voltage for the transistors 
is $V_t=0.3V$ for CMOS transistors, the coercive voltage $V_c=40mV$ (labeled 'FE') and the coercive voltage $V_c=4mV$ (labeled 'PE') for transcapacitors
in Fig.~\ref{fig:energyFig_inverter}(c).
The dependence on the threshold voltage for CMOS transistors (at $V_{DD}=0.7V$) and on the coercive voltage for transcapacitors (at $V_{DD}=0.2V$) is shown in Fig.~\ref{fig:energyFig_inverter}(d).
For CMOS, we use the Kogge-Stone architecture of the adder.
For ferroelectric circuits, the adder is synthesized of majority gates with fanin from 3 to 7 as per~\cite{mathuriya2024majority}.

Our estimates for the CMOS switching energy are different from some published 
values, e.g. $\approx 10 aJ$ for the 7nm node
\cite{datta2022toward}. Those estimates seem to only account for only the
intrinsic capacitance of the gate, and taking the area of the gate to be the square of the process node parameter, $F$.
In contrast, our estimates include the parasitic capacitance, the capacitance and resistance of wires to the next logic gate. Also our benchmarking uses the realistic gate length and the perimeter of the fin as well as the 2 fins in a transistor.
Note that our estimates are in agreement with other published results, in the range of hundreds of attojoules for the 7nm node \cite{stillmaker2017scaling}.

In general, a lower voltage result results in slower switching and smaller switching energy. 
In CMOS, the delay is mainly determined by the on-current. As the supply voltage approaches the threshold voltage, the on-current drops off dramatically, and thus the delay of CMOS circuits exceeds even that of ferroelectric ones. 
This trend makes lowering the supply voltage in CMOS very problematic. 
This problem cannot be resolved by lowering the threshold voltage, since in that case the off-current and the corresponding dissipation become untenable.
Transcapacitor circuits have an advantage in this respect. 
They do not have a continuous conduction path from the supply to the ground, and therefore they are not limited by the off-current of a device. 
Moreover, ferroelectric layers can be made thick enough to make the leakage current negligible. 
So, the standby power dissipation (not computed here) is expected to be better for transcapacitors. 
In addition, these circuits do not have a dramatic current drop-off at lower supply voltage. 
Even though at higher supply voltage they have a longer delay dominated by the intrinsic (Merz) switching time, they experience a smaller delay increase as the supply voltage decreases.

\begin{table}[ht]
\caption{Energy win factors in transcapacitors relative to CMOS transistors due to various attributes.}
\label{tab:energyWins}
\begin{tabular*}{\textwidth}{@{\extracolsep\fill}lc}
\toprule%
Attribute & Energy Win \\
\midrule
FE smaller capacitance & 1.8 \\
Reactive win  at iso-voltage & 9.2 \\
Supply voltage 0.8V→0.1V & 10.1 \\
Majority gates micro-architecture &  6.2 \\
\midrule
Total & 1040 \\
\botrule
\end{tabular*}
\end{table}

Several factors contribute to making transcapacitors an attractive option for lower-energy logic; see Table~\ref{tab:energyWins}. 
The factors in the table are in agreement with Fig.~\ref{fig:energyFig_inverter}.
1) They require a smaller capacitance area than transistors (which typically have several tall fins in a device). 2) They can be operated in the adiabatic regime. 3) They can operate at a lower voltage and still have a tolerable delay. The win factor of the supply voltage is determined from the ratio of energies for the 
paraelectric ('PE') cases in 
Fig.~\ref{fig:energyFig_inverter}(c).
Note that this energy is not proportional to the square of voltage, but instead has 
a complicated dependence including the intrinsic damping and the Joule heating types of energy dissipation.
4) They are capable of creating majority gates (unlike CMOS), and thus provide savings in the area and number of switched devices to perform a given logic function. Majority gates with non-linear capacitors are based on our work in~\cite{mathuriya2024majority} technology and circuits similar to~\cite{pudi2017majority} extended to  include MAJ-3,5,7 gates. Thus, in comparing a higher voltage CMOS and low-voltage ferroelectric circuits, for the inverter with fanout of 2 we predict a win of ~168x in energy at the expense of ~1.1x longer delay. 
For the adder, the energy win is ~1040x while the delay is ~0.43x that of CMOS.

\section{Experimental Demonstration}
\label{sec:experim_demo}

\begin{figure}[ht!]
\centering
\includegraphics[width=13cm]{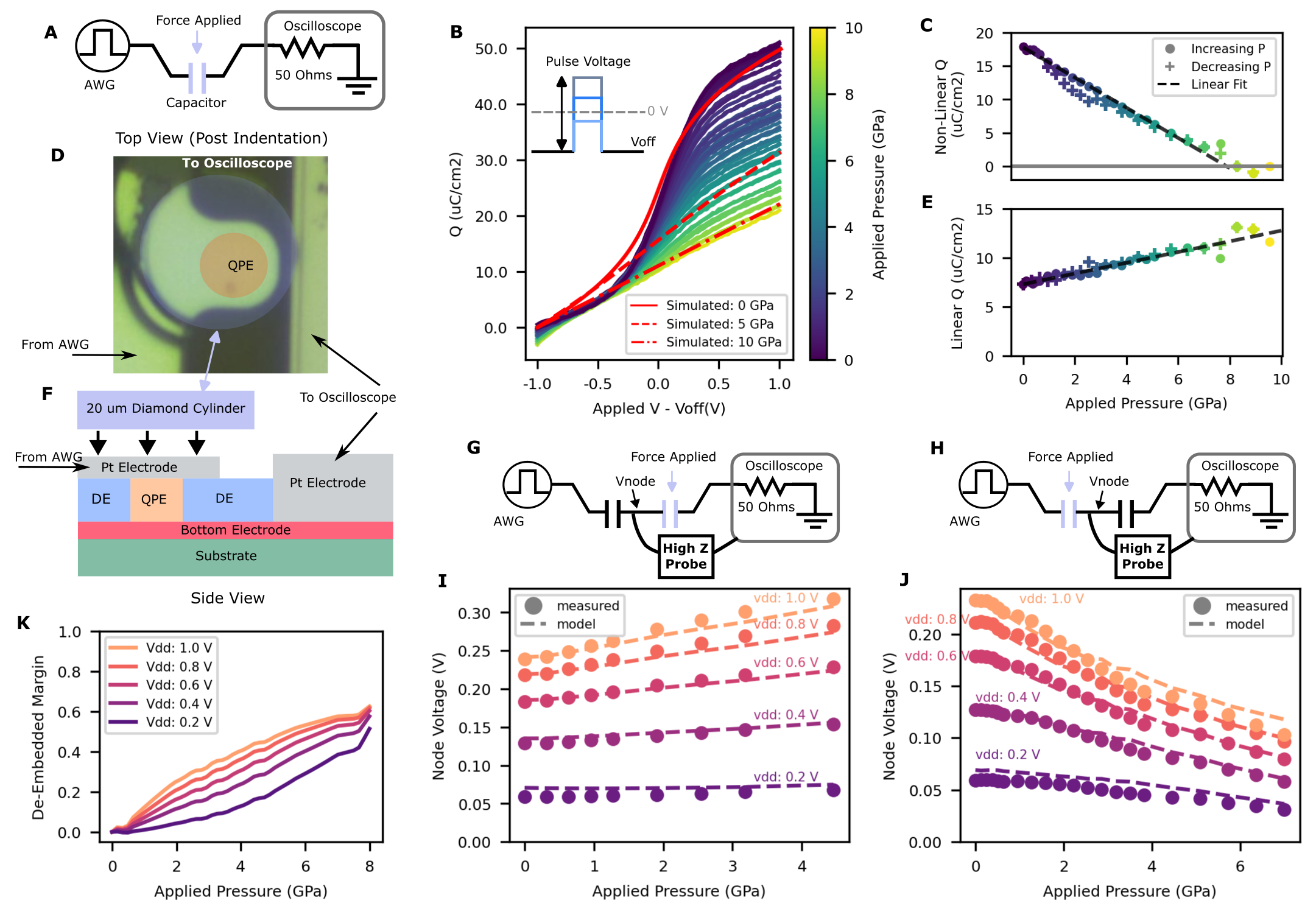}
\caption{\textbf{Experimental Demonstration of Half TCAP Inverter} (\textbf{A}) Pressure is applied to a capacitor under test while its charge response is measured with an AWG and Oscilloscope. (\textbf{B}) The integrated charge response from a first order reversal curve (FORC) measurement shows an evolution from non-linear to linear with increasing pressure and the sample response is well capture by multiphysics simulation (red lines). Fitting the charge response to a $\tanh$ + linear model shows an approximate linear dependence of both the non-linear response (\textbf{C}) and linear response (\textbf{E}) on pressure with a complete extinction of the non-linear response by 8 GPa. There is a minor hysteresis observed in the non-linear response from 3.5 GPa to 1.5 GPa as seen in the deviation of the plus symbols in (\textbf{C}) from the linear fit. (\textbf{D}) Image of device after stress cycle applied by 20 $\mu$m diamond cylinder tip with false color added. (\textbf{F}) Side view of device shown in (\textbf{E}) where the quasi-paraelectric (QPE) BTO stack is surrounded by a back-filled dielectric (DE) AlOx layer. (\textbf{G}) and (\textbf{H}) show the circuit implementation to test a half-inverter where only on capacitor of two capacitors is modulated by measuring the node voltage using a high impedance (``high Z'') probe. The resulting response is shown in \textbf{(I)} and \textbf{(J)}, respectively. A model constructed using the data from \textbf{B} for both half inverters shows excellent agreement with the measure data (dashed lines in \textbf{(I)} and \textbf{(J)} as well). The node voltage is less than
$V_{DD}/2$
in all cases because the High Z probe adds a substantial linear capacitance to the floating node. \textbf{(K)} Computed distance from $V_{DD}/2$
normalized by $V_{DD}/2$
using the model in \textbf{(I)} and \textbf{(J)} without the probe capacitance loading. }
\label{fig:demo_stress_effect}
\end{figure}

To validate the working principle of the strain-based transcapacitor in Fig.~\ref{fig:piezoConcept}, we perform an analogous experiment where we replace the stressor material with an external method to apply force, as shown in Figs.~\ref{fig:demo_stress_effect}A,E,H. We deposit a stack of 30 nm strontium ruthenate (SRO), 20 nm barium titanate (BTO), and 15 nm SRO using pulsed laser deposition (PLD) on a dysprosium scandate (DSO) substrate. Growth conditions are optimized to produce a BTO with almost no remanent polarization leading to a quasi-paraelectric film. The SRO functions as an electrode compatible with the growth of the BTO film. Standard lithography techniques are then used to fabricate 10 um diameter capacitors with large Pt leads suitable for wirebonding. Details of the growth and fabrication are discussed in the supplementary information. By wirebonding the device leads to a custom printed circuit board, we are able to put the sample into a nano-indenter system and directly characterize the charge response of the sample at static pressures applied by the nano-indenter tip. 

We use a 20 um diameter cylindrical diamond tip to apply the force to ensure that the 10 um diameter capacitor is fully contained within the footprint of the tip and that the force is applied uniformly owing to the flat surface of the tip. The use of diamond also prevents any shorting or charge leaking through the nano-indenter system. To characterize the charge response, we take first order reversal curves (FORC's) by applying a static offset to fully polarize the device in one direction and then recording the charge response to pulses away from that offset until the device is fully polarized in the opposite direction (Fig.~\ref{fig:demo_stress_effect}B inset).  

Over the application of pressure from 0 Pa to 10 GPa and back down to 0 Pa, we observe the complete suppression of the nonlinear quasi-paraelectric behavior of the FORC to a purely linear response (purple to yellow curves in Fig.~\ref{fig:demo_stress_effect}B). By fitting the FORC's to a model consisting of tanh plus a linear component we can understand the form of the nonlinear suppression with pressure,  Fig.~\ref{fig:demo_stress_effect}C. The response is largely linear with pressure (black dashed line) with a slope of $-2.26 uC/cm^2$ GPa with complete suppression occurring close to 8 GPa, and minor hysteresis observed on the return to 0 Pa from max pressure (plus symbols). 
This modulation of polarization with pressure is expected \cite{uchino1981electrostrictive}, but this is the first time a direct measurement of polarization response to parallel uniaxial pressure has been performed. 
The nonlinear suppression can be understood as the forcing of the lattice distortion that supports the polarization rotation from out-of-plane  to the polarization into the plane, as predicted by polarization dynamics driven by energy landscapes gradients of paraelectric in Figs. \ref{fig:deviceFig1_energybarriers_k} - \ref{fig:deviceFig1_energybarriers_n}, reducing the out-of-plane response that we measure electrically. 

We also find that the linear charge response component, effectively the relative permittivity of the BTO film, almost doubles from 0 Pa to 10 GPa (Fig.~\ref{fig:demo_stress_effect}D).  Effectively,   the  suppression of the nonlinear parts results in a greater linear charge response. We reproduced the experimental charge response behavior by performing simulations of the paraelectric stack under vertical compressive stress applied by the nanoindenter (overlayed lines in Fig.~\ref{fig:demo_stress_effect}B). We considered a heterostructure stack  consisting of the four layers of source electrode,  BTO paraelectric,  drain electrode and Diamond nanoindenter as shown in Fig. \ref{fig:deviceFig_NI_a}.  The whole stack is pinned to the bottom DSO substrate. See Section \ref{device_nidetails} for other details and parameters of this simulation. Displacing nanoindenter vertically down, exerts vertical force on the stack and results in the uniform  compressive vertical stress. For example, displacing nanoindenter by 2.4 nm results in -10 GPa as simulated in \ref{fig:deviceFig_NI_b}. We simulated Forc characteristics by applying source-drain voltage pulses in Fig. \ref{fig:deviceFig_NI_c} and collecting the integrated charge for various nanoindenter displacements and voltage pulse magnitudes. In agreement with experiments, FORC charge for a given magnitude of a voltage pulse reduces for larger displacements or stresses and saturates to linear dielectric response below -8 GPa of vertical pressure exerted by the nanoindenter, as shown in Fig. \ref{fig:deviceFig_NI_d}. 

While the creation of a full-inverter is precluded by the geometry of the experiment, our devices are fabricated such that multiple capacitor share a common bottom electrode. By connecting two capacitors in series by their common bottom electrode, we can then create a half inverter by applying pressure to one capacitor in the chain at a time as shown in Figs.~\ref{fig:demo_stress_effect}F,G. We probe what would be the inverter output, Vnode, using a passive, high-impedance oscilloscope probe with 10 MOhm resistance in parallel with $\approx$40 pF linear capacitance. At 1 V, the effective capacitance of a single QPE 10 um diameter BTO capacitor is $\approx$23 pF. Thus the presence of the oscilloscope probe adds a substantial load to the Vnode voltage, bringing the zero pressure value closer to $V_{DD}/4$
instead of the expected 
$V_{DD}/2$
if the probe loading was negligible. The large capacitive loading also contributes to a substantial asymmetry in the pressure responses of the two configurations. Despite this however, we still observe NTCAP and PTCAP-like responses, with Vnode increasing and decreasing with pressure, respectively. Furthermore, a simple model using the data from Figs.~\ref{fig:demo_stress_effect}B-D combined with a simple linear capacitor representing the probe load shows excellent agreement with measured data over the full range of voltages, 
$V_{DD}$,
and pressures (gray dashed lines, Figs.~\ref{fig:demo_stress_effect}I,J). 

Given the excellent agreement of our model with the data, we can de-embed the effect of the probe and extract the expected behavior of the system when not loading by an experimental probe. We set the extra linear capacitance to zero and compute the expected de-embedded relative voltage margin
$VM = \frac{2 V_{node}}{V_{DD}} - 1$, see
Fig.~\ref{fig:demo_stress_effect}K. Without the extra linear capacitance, the margin is symmetric for both configuration and reaches a peak value 
of  0.62 at 8 GPa and 
$V_{DD}=1V$. More details about the model and its computation can be found in the supplementary information. Ultimately, the observed near-recoverable transition in polarization response under applied external vertical stress and its straightforward extension to working PTCAP and NTCAP structure confirms the proposed mechanism for operation of future transcapacitor devices.

The 2-terminal capacitors used in these experiments have demonstrated the specs desired to build logic class devices. In our unpublished work, we measure low voltage ~100 mV $V_c$ as the critical voltage with thresholding at 150 mV. For switching time, at 200 nm device diameter, it reaches the instrument limit of 29 ps, and begins to asymptote to the measurement system risetime limit of 20 ps. Also exemplar fatigue measurement are demonstrated where the device is cycled at 20x the switching voltage, +/-4 V at 500 MHz. The device is directly cycled $>10^{15}$ times without any degradation in properties.

\section{Future directions}
\label{sec:future}

Ferroelectric materials have rich physics, including domain formation, domain boundary movements, phase transition, depolarization with the modulation of carrier concentration in the attached electrodes and coupling to magnetism, stress, temperature, etc. Utilizing them, we can build various incarnations~\cite{mathuriya2024barrier}~\cite{mathuriya2025channel}~\cite{mathuriya2026electromechanical}~\cite{manipatruni2025transpolarizer} of a transcapacitor.
Extensive research,
similar in scope to that conducted on each known type of a transistor,
is required to explore the geometry and the structure for each type of a transcapacitor, to better understand its mechanism of operation, requirements for the material, methods of its fabrication and its incorporation into and integrated circuit.

We list the types of transcapacitors that are envisioned. Among them are
A) the MEMS-based transcapacitor has been proposed \cite{galisultanov2017capacitive,pillonnet2017adiabatic};
B) the floating island transcapacitor consists of two polarized layers and a floating electrode between them;
C) the field assist transcapacitor operates on the basis of the influence of an additional field, created by the gate, on the coercive value of the switching field, applied between the source and the drain;
D) the field effect (metal-insulator transition) transcapacitor relies on the gate voltage to change the conductance of a semiconducting layer adjacent to a polarizer layer;
E) the piezoelectric transcapacitor relies on the transfer of stress between two adjacent capacitors;
F) the phase transition transcapacitor, which operates by switching the phases of polarization, e.g. from a state with vortices to a state without vortices \cite{kavle2024highly}.
The principles of operation, advantages, disadvantages, breakthroughs required, and research directions for their implementation are summarized in Table~\ref{tab:tpol_types}.

\begin{table}[ht!] 
\centering 
\caption{Types of transcapacitors, their principles of operation, advantages and disadvantages, and required breakthroughs.}
\label{tab:tpol_types}
\begin{tabularx}{\textwidth}{
    | >{\hsize=0.5\hsize}X  
    | >{\hsize=1.75\hsize}X 
    | >{\hsize=0.9\hsize}X 
    | >{\hsize=1\hsize}X
    | >{\hsize=0.85\hsize}X |}
\hline
\textbf{Type} & 
\textbf{Principle of operation} & 
\textbf{Advantages} & 
\textbf{Disadvantages} & 
\textbf{Breakthroughs required} \\
\hline
\textbf{Floating island} &
Gate voltage causes the shift of FE hysteresis &
Demonstrated transcapacitance &
Charge gain ~1. Voltage inversion not shown. &
N/A \\
\hline
\textbf{Field assist} &
Electric field lowers the barrier for FE switching. Shifts or narrows the hysteresis. &
Transcapacitance measured. Charge gain 3 predicted. &
Voltage inversion not shown &
Operation with multiple FE layers \\
\hline
\textbf{Field-effect (MIT)} &
Gate voltage changes carrier density.Shifts or narrows the hysteresis. &
Predicted gain $>10$ and voltage inversion &
No demo so far &
Integration of the field effect into a FE stack \\
\hline
\textbf{Piezo-electric} &
Piezo stress from one capacitor changes polarization in another capacitor. &
Predicted gain 4.9 and voltage inversion &
Substrate clamping can be a limitation &
Implement in stronger piezo materials \\
\hline
\textbf{Phase transition} &
Vortices are formed in a thin FE layer. A gate field destroys them and varies the polarizability. &
Potentially large transcapacitance &
Vortex state difficult to observe &
Understanding of the operation \\
\hline
\textbf{MEMS} &
Applied voltage moves a cantilever, changes distance between cap electrodes. &
Predicted gain and voltage inversion &
Moving parts &
Scale to nanometer sizes \\
\hline
\end{tabularx}
\end{table}

Furthermore, the determination needs to be done for each type of the transcapacitor on whether it can satisfy the essential requirements for digital logic, see \cite{waser2012nanoelectronics}, listed above in 
Section~\ref{tcap_as_a_concept}.
We have performed a set of simulations that indicate that all these requirements can be satisfied by transcapacitors. However a more thorough and conclusive determination is required in the context of realistic circuits.

In particular, for the piezoelectric transcapacitor, emphasized in this article, the identification of more realistic options for the stressor and the channel of the device is required. We also need a better understanding of the contributions of in-plane and out-of-plane stress. 
For the stressor (gate) capacitor, materials providing a higher stress while collecting smaller change of charge at a given applied voltage are preferred.  
For the channel (drain) capacitor, materials are needed that create a greater charge change and a greater change of the coercive voltage for a given applied stress at low voltages.
This would obviously improve the figures of merit discussed in Section~\ref{sec:FOMs} and enable low energy computing.
Research is required to improve the interfaces and 
to provide materials with higher stiffness for electrodes; stiffer encapsulation materials to ensure that the top and bottom boundaries are close to fixed;
high efficiency of the stress transfer and good charge screening in the source electrode layer. Process integration with air gaps on the sides, while fixing the top and bottom boundary conditions (similar to the one shown in fig.~\ref{fig:3ddesign_main}) would ensure a high efficiency stress transfer.

\section{Conclusions}
\label{sec:conclusions}

We presented a novel computing device, 
the transcapacitor, as well as a demonstration of 
a key effect underlying its operation.
The transcapacitor is a three-terminal 
device that 
employs ferroelectric or paraelectric materials and
produces the output charge from one terminal (drain) in response to a voltage applied between two other terminals (source and gate).
It presents a radical departure from ubiquitous transistors.
Logic and memory circuits based on transcapacitors 
rely on the voltage determined by the charge re-distribution instead of the resistive voltage division.
Conventional MOSFETs suffer from Boltzmann-limited voltage scaling, resistive heating, and rush-through currents. 
The transcapacitors operate by charge transfer 
and decrease or eliminate these parts of energy dissipation.
This paradigm shift enables operation at an order-of-magnitude lower supply voltage.
The nature of switching of reactive circuits provides additional energy efficiency.
This breakthrough addresses three fundamental limitations of CMOS: voltage scaling, channel dissipation, and transition currents, potentially enabling more than 100× device-level  improvements for next-generation computing.

\section{Acknowledgments}
\label{sec:ack}

The project is in part supported by Government support under DARPA Spark Tank grant agreement \#HR0011-25-3-0173, led by Todd Bauer (PM).
The work was also done with Government support under the Fast and Curious program Agreement No.
\#HR0011269E206, awarded by DARPA. We extend our gratitude to Prof. Sunil Bhave.

\FloatBarrier
\bibliography{paper}

@article{weinstein2010resonant,
  title={The resonant body transistor},
  author={Weinstein, Dana and Bhave, Sunil A},
  journal={Nano letters},
  volume={10},
  number={4},
  pages={1234--1237},
  year={2010},
  publisher={ACS Publications}
}

@article{Gupta2026transswitching,
  author    = {Gupta, Pushpendra and Puebla, Sergio and Gomez-Ortiz, Fernando and Li, Xinyan and Husain, Sajid and Liu, Ting-Ran and Meisenheimer, Peter and Srikrishna, Vishantak and Nikonov, Dmitri and Chen, Matthew and Kumar, Yogesh and Omar, Ashish and Achinuq, Barat and Roy, Sujoy and Shao, Yu-Tsun and Han, Yimo and Mathuriya, Amrita and Manipatruni, Sasikanth and Ghosez, Philippe and Junquera, Javier and Ramesh, Ramamoorthy},
  title     = {Trans-switching in layered ferroelectrics},
  journal   = {Research Square},
  year      = {2026},
  month     = {January},
  doi       = {10.21203/rs.3.rs-8704297/v1},
  url       = {https://www.researchsquare.com/article/rs-8704297/v1},
  note      = {Preprint, under review at Nature Portfolio}
}

@misc{mathuriya2024majority,
  author       = {Mathuriya, Amrita and Rios, Rafael and Odinaka, Ikenna and Doshi, Darshak and Dokania, Rajeev Kumar and Manipatruni, Sasikanth},
  title        = {Majority or Minority Logic Gate with Non-Linear Input Capacitors without Reset},
  howpublished = {U.S. Patent 11,967,954, Kepler Computing},
  year         = {2024},
  month        = apr,
  note         = {Granted 2024-04-23. Application 17/659,994, filed 2022-04-20}
}

@misc{mathuriya2024barrier,
  author       = {Mathuriya, Amrita and Rios, Rafael and Nikonov, Dmitri E. and Guha, Biswajeet and Dokania, Rajeev Kumar and Manipatruni, Sasikanth},
  title        = {Barrier Controlled Non-Linear Polar Material Based Capacitor and Associated Circuits},
  howpublished = {U.S. Patent Application No. 18/582,437, patent pending},
  year         = {2024}
}

@misc{mathuriya2025channel,
  author       = {Mathuriya, Amrita and Rios, Rafael and Nikonov, Dmitri E. and Guha, Biswajeet and Dokania, Rajeev Kumar and Manipatruni, Sasikanth},
  title        = {Channel Based Barrier Controlled Capacitor and Associated Circuits},
  howpublished = {U.S. Patent Application No. 19/067,701},
  year         = {2025},
  note         = {Patent pending}
}

@misc{manipatruni2025transpolarizer,
  author       = {Manipatruni, Sasikanth and Nikonov, Dmitri E. and Mathuriya, Amrita and Guha, Biswajeet and Dokania, Rajeev Kumar},
  title        = {Transpolarizer with Electrode Material Control},
  howpublished = {U.S. Patent Application No. 19/067,700},
  year         = {2025},
  note         = {Patent pending}
}

@misc{mathuriya2026electromechanical,
  author       = {Mathuriya, Amrita and Kotlyar, Roza and Nikonov, Dmitri E. and Rios, Rafael and Dokania, Rajeev Kumar and Manipatruni, Sasikanth},
  title        = {Electro-Mechanically Controlled Memory and Logic Devices},
  howpublished = {U.S. Patent Application No. 19/648,966},
  year         = {2026},
  note         = {Patent pending}
}

@article{pudi2017majority,
  title={Majority logic formulations for parallel adder designs at reduced delay and circuit complexity},
  author={Pudi, Vikramkumar and Sridharan, K and Lombardi, Fabrizio},
  journal={IEEE transactions on computers},
  volume={66},
  number={10},
  pages={1824--1830},
  year={2017},
  publisher={IEEE}
}

@article{horowitz2005scaling,
  title={Scaling, power, and the future of CMOS},
  author={Horowitz, Mark and Alon, Elad and Patil, Dinesh and Naffziger, Samuel and Kumar, Rajesh and Bernstein, Kerry},
  journal={IEEE InternationalElectron Devices Meeting, 2005. IEDM Technical Digest.},
  pages={7--pp},
  year={2005},
  publisher={IEEE}
}

@article{dallyhbm,
  title={Fine-grained DRAM: Energy-efficient DRAM for extreme bandwidth systems},
  author={O'Connor, Mike and Chatterjee, Niladrish and Lee, Donghyuk and Wilson, John and Agrawal, Aditya and Keckler, Stephen W and Dally, William J},
  journal={Proceedings of the 50th Annual IEEE/ACM International Symposium on Microarchitecture},
  pages={41--54},
  year={2017}
}

@article{adhinarayanan2025folded,
  title={Folded Banks: 3D-Stacked HBM Design for Fine-Grained Random-Access Bandwidth},
  author={Adhinarayanan, Vignesh and Beckmann, Bradford M and Li, Wantong and Seyedzadeh, Mohammad and Blagodurov, Sergey and Aguren, Derrick and Lee, Hayden Hyungdong},
  journal={Proceedings of the 52nd Annual International Symposium on Computer Architecture},
  pages={1819--1833},
  year={2025}
}

@phdthesis{kim2010equalized,
  title={Equalized on-chip interconnect: Modeling, analysis, and design},
  author={Kim, Byungsub},
  year={2010},
  school={Massachusetts Institute of Technology}
}

@article{li2022interconnect,
  title={Interconnect in the Era of 3DIC},
  author={Li, Shenggao and Lin, Mu-Shan and Chen, Wei-Chih and Tsai, Chien-Chun},
  journal={2022 IEEE Custom Integrated Circuits Conference (CICC)},
  pages={1--5},
  year={2022},
  publisher={IEEE}
}

@article{bennett1985fundamental,
  title={The fundamental physical limits of computation},
  author={Bennett, Charles H and Landauer, Rolf},
  journal={Scientific American},
  volume={253},
  number={1},
  pages={48--57},
  year={1985},
  publisher={JSTOR}
}

@article{landauer1976,
  title={Fundamental limitations in the computational process},
  author={Landauer, Rolf},
  journal={Berichte der Bunsengesellschaft f{\"u}r physikalische Chemie},
  volume={80},
  number={11},
  pages={1048--1059},
  year={1976},
  publisher={Wiley Online Library}
}

@article{cao2023future,
  title={The future transistors},
  author={Cao, Wei and Bu, Huiming and Vinet, Maud and Cao, Min and Takagi, Shinichi and Hwang, Sungwoo and Ghani, Tahir and Banerjee, Kaustav},
  journal={Nature},
  volume={620},
  number={7974},
  pages={501--515},
  year={2023},
  publisher={Nature Publishing Group UK London}
}

@article{yang2025multi,
  title={Multi-node scaling potential of monolithic CFET},
  author={Yang, S and Verschueren, L and B{\"o}mmels, J and K{\"u}kner, H and Lin, JY and Bufler, FM and Sankatali, V and Farokhnejad, A and Van de Put, M and Hellings, G},
  journal={2025 IEEE International Electron Devices Meeting (IEDM)},
  pages={1--4},
  year={2025},
  publisher={IEEE}
}

@article{guptatrans,
  title={Trans-switching in layered ferroelectrics},
  author={Gupta, Pushpendra and Puebla, Sergio and G{\'o}mez-Ortiz, Fernando and Li, Xinyan and Husain, Sajid and Liu, Ting-Ran and Meisenheimer, Peter and Srikrishna, Vishantak and Nikonov, Dmitri and Chen, Matthew and others},
  journal={in press},
  pages={},
  year={2026}
}

@article{UygarAvci,
  title={Process integration and future outlook of 2D transistors},
  author={O’Brien, Kevin P and Naylor, Carl H and Dorow, Chelsey and Maxey, Kirby and Penumatcha, Ashish Verma and Vyatskikh, Andrey and Zhong, Ting and Kitamura, Ande and Lee, Sudarat and Rogan, Carly and others},
  journal={nature communications},
  volume={14},
  number={1},
  pages={6400},
  year={2023},
  publisher={Nature Publishing Group UK London}
}

@article{appenzeller2004band,
  title={Band-to-band tunneling in carbon nanotube field-effect transistors},
  author={Appenzeller, J and Lin, Y-M and Knoch, J and Avouris, Ph},
  journal={Physical review letters},
  volume={93},
  number={19},
  pages={196805},
  year={2004},
  publisher={APS}
}

@article{shalf2020future,
  title={The future of computing beyond Moore’s Law},
  author={Shalf, John},
  journal={Philosophical Transactions of the Royal Society A: Mathematical, Physical and Engineering Sciences},
  volume={378},
  number={2166},
  year={2020},
  publisher={The Royal Society}
}

@article{sakurai2002alpha,
  title={Alpha-power law MOSFET model and its applications to CMOS inverter delay and other formulas},
  author={Sakurai, Takayasu and Newton, A Richard},
  journal={IEEE Journal of solid-state circuits},
  volume={25},
  number={2},
  pages={584--594},
  year={2002},
  publisher={IEEE}
}

@article{pawashe2013scaling,
  title={Scaling limits of electrostatic nanorelays},
  author={Pawashe, Chytra and Lin, Kevin and Kuhn, Kelin J},
  journal={IEEE Transactions on Electron Devices},
  volume={60},
  number={9},
  pages={2936--2942},
  year={2013},
  publisher={IEEE}
}

@article{he2020tunable,
  title={A tunable ferroelectric based unreleased RF resonator},
  author={He, Yanbo and Bahr, Bichoy and Si, Mengwei and Ye, Peide and Weinstein, Dana},
  journal={Microsystems \& nanoengineering},
  volume={6},
  number={1},
  pages={8},
  year={2020},
  publisher={Nature Publishing Group UK London}
}

@article{fiebig2016evolution,
  title={The evolution of multiferroics},
  author={Fiebig, Manfred and Lottermoser, Thomas and Meier, Dennis and Trassin, Morgan},
  journal={Nature Reviews Materials},
  volume={1},
  number={8},
  pages={1--14},
  year={2016},
  publisher={Nature Publishing Group}
}

@article{singh2009static,
  title={Static synchronous compensators (STATCOM): a review},
  author={Singh, Bhim and Saha, R and Chandra, Ambrish and Al-Haddad, Kamal},
  journal={IET power electronics},
  volume={2},
  number={4},
  pages={297--324},
  year={2009},
  publisher={IET}
}

@article{fukuzumi2007optimal,
  title={Optimal integration and characteristics of vertical array devices for ultra-high density, bit-cost scalable flash memory},
  author={Fukuzumi, Yoshiaki and Katsumata, Ryota and Kito, Masaru and Kido, Masaru and Sato, Mitsuru and Tanaka, Hiroyasu and Nagata, Yuzo and Matsuoka, Yasuyuki and Iwata, Yoshihisa and Aochi, Hideaki and others},
  journal={2007 IEEE International Electron Devices Meeting},
  pages={449--452},
  year={2007},
  publisher={IEEE}
}

@article{scott1989ferroelectric,
  title={Ferroelectric memories},
  author={Scott, James F and Paz de Araujo, Carlos A},
  journal={Science},
  volume={246},
  number={4936},
  pages={1400--1405},
  year={1989},
  publisher={American Association for the Advancement of Science}
}

@book{waser2012nanoelectronics,
  title={Nanoelectronics and information technology: advanced electronic materials and novel devices},
  author={Waser, Rainer},
  year={2012},
  publisher={John Wiley \& Sons}
}

@article{datta2022toward,
  title={Toward attojoule switching energy in logic transistors},
  author={Datta, Suman and Chakraborty, Wriddhi and Radosavljevic, Marko},
  journal={Science},
  volume={378},
  number={6621},
  pages={733--740},
  year={2022},
  publisher={American Association for the Advancement of Science}
}

@article{galisultanov2017capacitive,
  title={Capacitive-based adiabatic logic},
  author={Galisultanov, Ayrat and Perrin, Yann and Fanet, Herv{\'e} and Pillonnet, Ga{\"e}l},
  journal={International Conference on Reversible Computation},
  pages={52--65},
  year={2017},
  address={Kolkata, India},
  doi={10.1007/978-3-319-59936-6},
  note={Conference proceedings}
}

@article{pillonnet2017adiabatic,
  title={Adiabatic capacitive logic: a paradigm for low-power logic},
  author={Pillonnet, Ga{\"e}l and Fanet, Herv{\'e} and Houri, Samer},
  journal={2017 IEEE International Symposium on Circuits and Systems (ISCAS)},
  pages={1--4},
  year={2017},
  address={Baltimore, MD, USA},
  doi={10.1109/ISCAS.2017.8050996},
  note={Conference proceedings}
}

@article{nikonov2015benchmarking,
  title={Benchmarking of beyond-CMOS exploratory devices for logic integrated circuits},
  author={Nikonov, Dmitri E and Young, Ian A},
  journal={IEEE Journal on Exploratory Solid-State Computational Devices and Circuits},
  volume={1},
  pages={3--11},
  year={2015},
  publisher={IEEE}
}

@article{kavle2024highly,
  title={Highly Responsive Polar Vortices in All-Ferroelectric Heterostructures},
  author={Kavle, Pravin and Ross, Aiden M and Kp, Harikrishnan and Meisenheimer, Peter and Dasgupta, Arvind and Yang, Jiyuan and Lin, Ching-Che and Pan, Hao and Behera, Piush and Parsonnet, Eric and others},
  journal={Advanced Materials},
  volume={36},
  number={50},
  pages={2410146},
  year={2024},
  publisher={Wiley Online Library}
}

@book{hu2010modern,
  title={Modern semiconductor devices for integrated circuits},
  author={Hu, Chenming Calvin},
  year={2009},
  publisher = "Pearson",
  edition   = "1st",
  address   = "London",
  isbn      = "978-0136085256",
}

@article{nikonov2013overview,
  title={Overview of beyond-CMOS devices and a uniform methodology for their benchmarking},
  author={Nikonov, Dmitri E and Young, Ian A},
  journal={Proceedings of the IEEE},
  volume={101},
  number={12},
  pages={2498--2533},
  year={2013},
  publisher={IEEE}
}

@article{behera2025anisotropic,
  title={Anisotropic Ferroelectricity in Polar Vortices},
  author={Behera, Piush and Ross, Aiden M and Shanker, Nirmaan and Meisenheimer, Peter and Manna, Mahir and Lin, Ching-Che and Hsu, Shang-Lin and Harris, Isaac and Kavle, Pravin and Husain, Sajid and others},
  journal={Advanced Materials},
  volume={37},
  number={1},
  pages={2410149},
  year={2025},
  publisher={Wiley Online Library}
}

@article{wang2010temperature,
  title={Temperature-pressure phase diagram and ferroelectric properties of BaTiO3 single crystal based on a modified Landau potential},
  author={Wang, JJ and Wu, PP and Ma, XQ and Chen, LQ},
  journal={Journal of Applied Physics},
  volume={108},
  number={11},
  year={2010},
  publisher={AIP Publishing}
}

@article{hu1998three,
  title={Three-dimensional computer simulation of ferroelectric domain formation},
  author={Hu, Hong-Liang and Chen, Long-Qing},
  journal={Journal of the American Ceramic Society},
  volume={81},
  number={3},
  pages={492--500},
  year={1998},
  publisher={Wiley Online Library}
}

@book{landau2013course,
  title={Course of theoretical physics},
  author={Landau, Lev Davidovich and Lifshitz, Evgenii Mikhailovich},
  year={2013},
  publisher={Elsevier}
}

@article{giannozzi2009quantum,
  title={QUANTUM ESPRESSO: a modular and open-source software project for quantum simulations of materials},
  author={Giannozzi, Paolo and Baroni, Stefano and Bonini, Nicola and Calandra, Matteo and Car, Roberto and Cavazzoni, Carlo and Ceresoli, Davide and Chiarotti, Guido L and Cococcioni, Matteo and Dabo, Ismaila and others},
  journal={Journal of physics: Condensed matter},
  volume={21},
  number={39},
  pages={395502},
  year={2009},
  publisher={IOP Publishing}
}

@article{rappe1990optimized,
  title={Optimized pseudopotentials},
  author={Rappe, Andrew M and Rabe, Karin M and Kaxiras, Efthimios and Joannopoulos, JD},
  journal={Physical Review B},
  volume={41},
  number={2},
  pages={1227},
  year={1990},
  publisher={APS}
}

@misc{comsol2025,
  title        = {COMSOL Multiphysics® Reference Manual, version 6.2},
  author       = {{COMSOL AB}},
  organization = {COMSOL AB},
  address      = {Stockholm, Sweden},
  year         = {2025},
  url          = {https://www.comsol.com}
}

@article{horowitz20141,
  title={1.1 computing's energy problem (and what we can do about it)},
  author={Horowitz, Mark},
  journal={2014 IEEE international solid-state circuits conference digest of technical papers (ISSCC)},
  pages={10--14},
  year={2014},
  publisher={IEEE},
  note={Conference proceedings}
}

@article{yeap20195nm,
  title={5nm cmos production technology platform featuring full-fledged euv, and high mobility channel finfets with densest 0.021 $\mu$m 2 sram cells for mobile soc and high performance computing applications},
  author={Yeap, Geoffrey and Lin, SS and Chen, YM and Shang, HL and Wang, PW and Lin, HC and Peng, YC and Sheu, JY and Wang, M and Chen, X and others},
  journal={2019 IEEE International Electron Devices Meeting (IEDM)},
  pages={36--7},
  year={2019},
  publisher={IEEE},
  note={Conference proceedings}
}

@book{weste2015cmos,
  title={CMOS VLSI design: a circuits and systems perspective},
  author={Weste, Neil HE and Harris, David},
  year={2015},
  publisher={Pearson Education India}
}

@article{uchino1981electrostrictive,
  title={Electrostrictive effect in perovskites and its transducer applications},
  author={Uchino, Kenji and Nomura, Shoichiro and Cross, Leslie E and Newnham, Robert E and Jang, Sei J},
  journal={Journal of Materials science},
  volume={16},
  number={3},
  pages={569--578},
  year={1981},
  publisher={Springer}
}

@article{choi2004enhancement,
  title={Enhancement of ferroelectricity in strained BaTiO3 thin films},
  author={Choi, Kyoung Jin and Biegalski, Michael and Li, YL and Sharan, A and Schubert, J and Uecker, Reinhard and Reiche, P and Chen, YB and Pan, XQ and Gopalan, Venkatraman and others},
  journal={Science},
  volume={306},
  number={5698},
  pages={1005--1009},
  year={2004},
  publisher={American Association for the Advancement of Science}
}

@article{pan2024clamping,
  title={Clamping enables enhanced electromechanical responses in antiferroelectric thin films},
  author={Pan, Hao and Zhu, Menglin and Banyas, Ella and Alaerts, Louis and Acharya, Megha and Zhang, Hongrui and Kim, Jiyeob and Chen, Xianzhe and Huang, Xiaoxi and Xu, Michael and others},
  journal={Nature materials},
  volume={23},
  number={7},
  pages={944--950},
  year={2024},
  publisher={Nature Publishing Group UK London}
}

@article{resta1992theory,
  title={Theory of the electric polarization in crystals},
  author={Resta, Raffaele},
  journal={Ferroelectrics},
  volume={136},
  number={1},
  pages={51--55},
  year={1992},
  publisher={Taylor \& Francis}
}

@article{liu2020giant,
  title={Giant piezoelectricity in oxide thin films with nanopillar structure},
  author={Liu, Huajun and Wu, Haijun and Ong, Khuong Phuong and Yang, Tiannan and Yang, Ping and Das, Pranab Kumar and Chi, Xiao and Zhang, Yang and Diao, Caozheng and Wong, Wai Kong Alaric and others},
  journal={Science},
  volume={369},
  number={6501},
  pages={292--297},
  year={2020},
  publisher={American Association for the Advancement of Science}
}

@article{ekpu2024characterising,
  title={Characterising the mechanical properties of a composite material comprising aluminium and silicon carbide},
  author={Ekpu, Mathias and Ikpeseni, Sunday C},
  journal={Silicon},
  volume={16},
  number={10},
  pages={4333--4342},
  year={2024},
  publisher={Springer}
}

@article{stillmaker2017scaling,
  title={Scaling equations for the accurate prediction of CMOS device performance from 180 nm to 7 nm},
  author={Stillmaker, Aaron and Baas, Bevan},
  journal={Integration},
  volume={58},
  pages={74--81},
  year={2017},
  publisher={Elsevier}
}

@article{solomon2015pathway,
  title={Pathway to the piezoelectronic transduction logic device},
  author={Solomon, PM and Bryce, Brian A and Kuroda, Marcelo A and Keech, Ryan and Shetty, Smitha and Shaw, Thomas M and Copel, Matthew and Hung, L-W and Schrott, Alejandro G and Armstrong, C and others},
  journal={Nano letters},
  volume={15},
  number={4},
  pages={2391--2395},
  year={2015},
  publisher={ACS Publications}
}

@article{newns2012low,
  title={A low-voltage high-speed electronic switch based on piezoelectric transduction},
  author={Newns, Dennis and Elmegreen, Bruce and Hu Liu, Xiao and Martyna, Glenn},
  journal={Journal of Applied Physics},
  volume={111},
  number={8},
  year={2012},
  publisher={AIP Publishing}
}

@article{newns2012piezoelectronic,
  title={The piezoelectronic transistor: A nanoactuator-based post-CMOS digital switch with high speed and low power},
  author={Newns, Dennis M and Elmegreen, BG and Liu, X-H and Martyna, Glenn J},
  journal={MRS bulletin},
  volume={37},
  number={11},
  pages={1071--1076},
  year={2012},
  publisher={Cambridge University Press}
}
\clearpage

\section*{Supplementary Materials}
\label{supplementary}
\setcounter{section}{0}
\setcounter{equation}{0}
\setcounter{table}{0}
\setcounter{figure}{0}
\renewcommand{\thesection}{S.\Roman{section}}
\renewcommand{\thesubsection}{S.\Roman{subsection}}
\renewcommand{\theequation}{S.\arabic{equation}}
\renewcommand{\thetable}{S.\arabic{table}}
\renewcommand{\thefigure}{S.\arabic{figure}}
\makeatletter
\makeatother

\section{Analytical Transcapacitor Model}
\label{sec:analyticalSpiceShrinkModel}

\begin{figure}[htbp]
    \centering
    \begin{subfigure}{0.70\textwidth}
        \includegraphics[width=\textwidth]{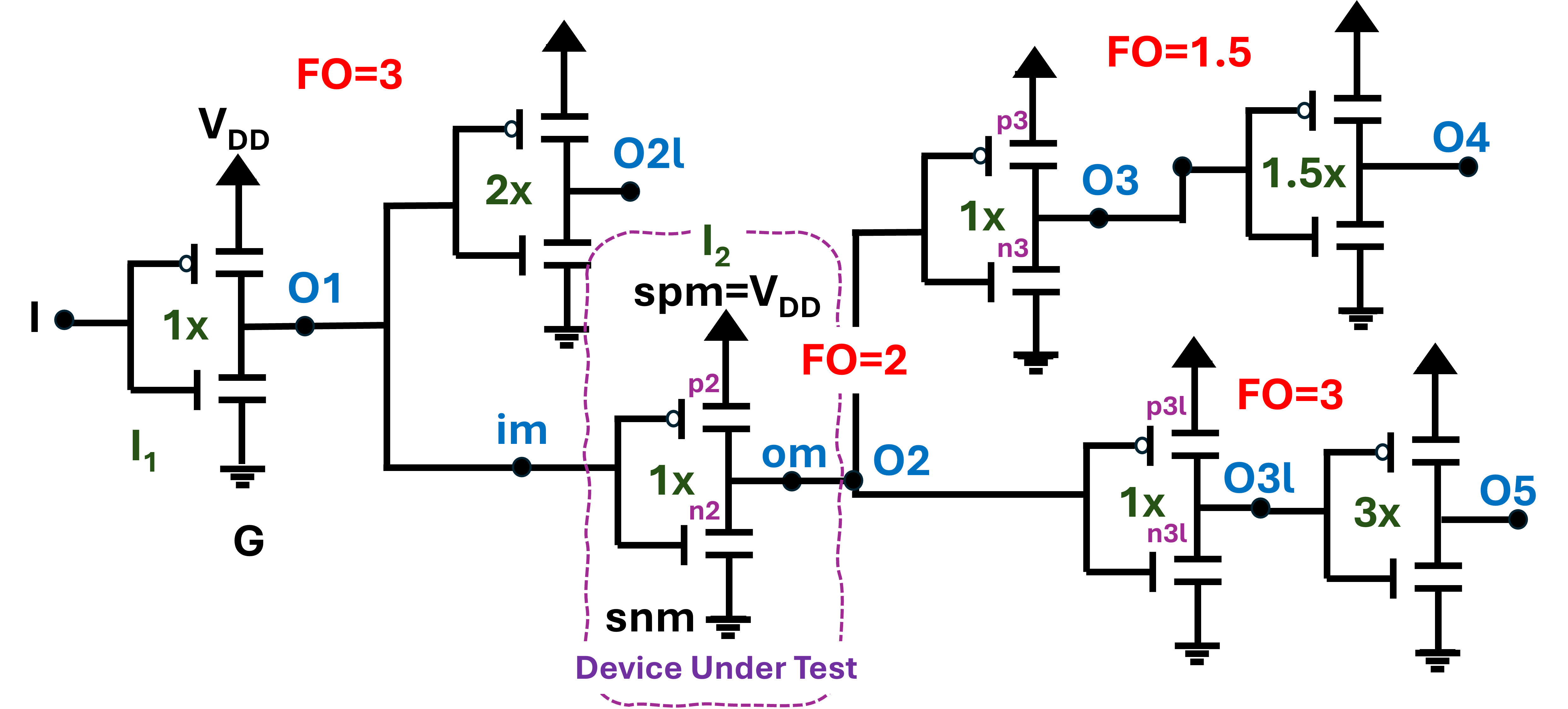}
        \caption{}
        \label{fig:SupplSpiceInv_InverterChain}
    \end{subfigure}%
    \begin{subfigure}{0.28\textwidth}
        \includegraphics[width=\textwidth]{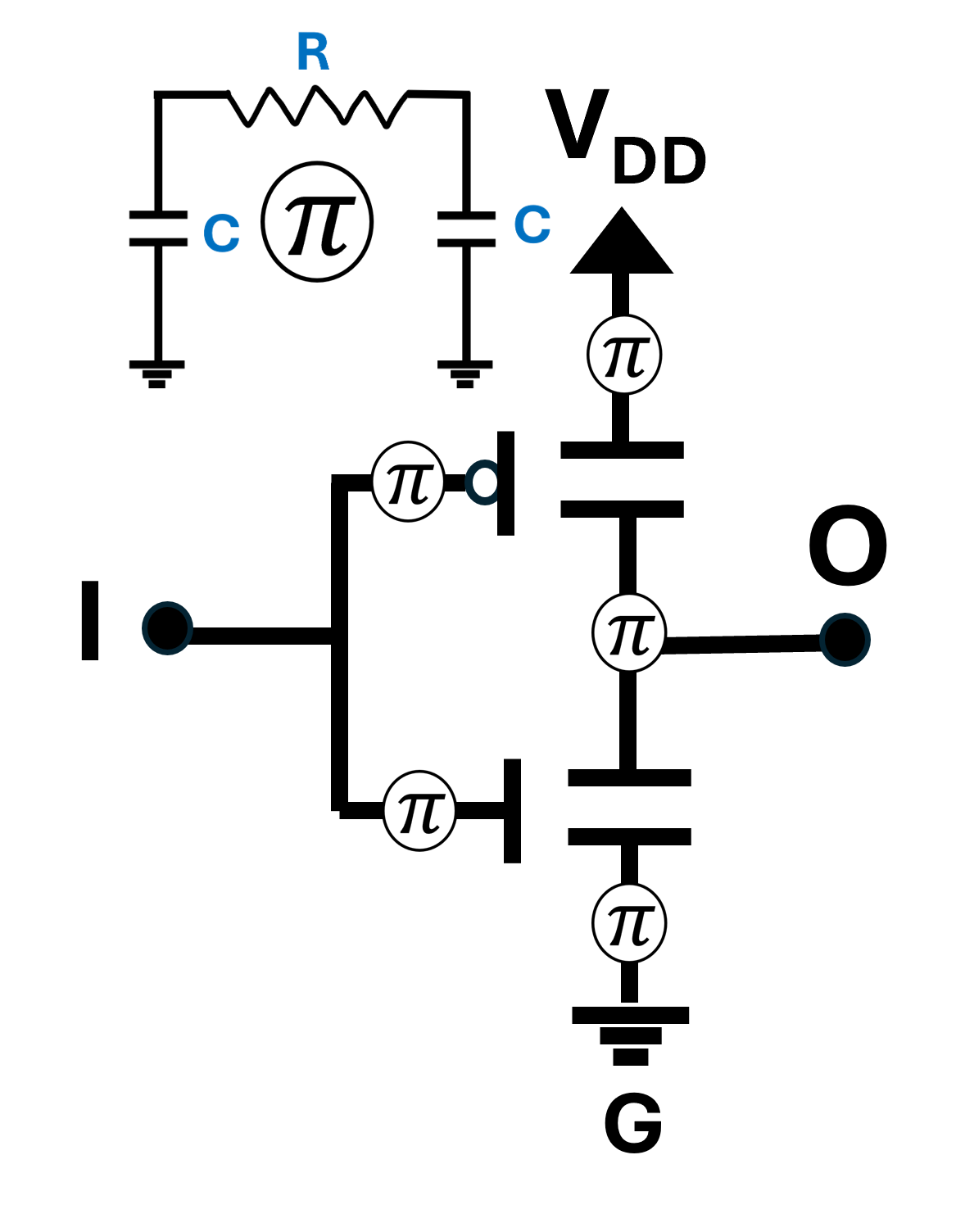}
        \caption{}
        \label{fig:SupplSpiceInv_PIRC}
    \end{subfigure}
    \caption{(a) Inverter chain circuit simulated with Spice, (b) the single inverter model for a) using $\Pi$ interconnect wire model for each wire.}
    \label{fig:InvChainCircuitSuppl}
\end{figure}

We build a Spice model with analytical equations to demonstrate circuit operations and figure of merits for a transcapacitor device which is similar to the piezo-TCAP behavior in operation. We build a chain of inverters with the model, as shown in fig.~\ref{fig:InvChainCircuitSuppl}, with a fan-out up to 3, and include wire parasitics with a PI interconnect model with RC components.  The model imitates the behavior of a FE material when its energy barrier is reduced through the application of stress or another mechanism through the gate field application, see \ref{fig:piezoStressCoupling}.
In this Spice model, we describe the channel with a ferroelectric, and gate as an anti-ferroelectric behavior. 
For simplicity, the model only captures the relevant branch of a hysteresis for both P (negative branch) and N (positive branch) devices.  

\begin{figure}[htbp]
    \centering
    \begin{subfigure}{0.49\textwidth}
        \includegraphics[width=\textwidth]{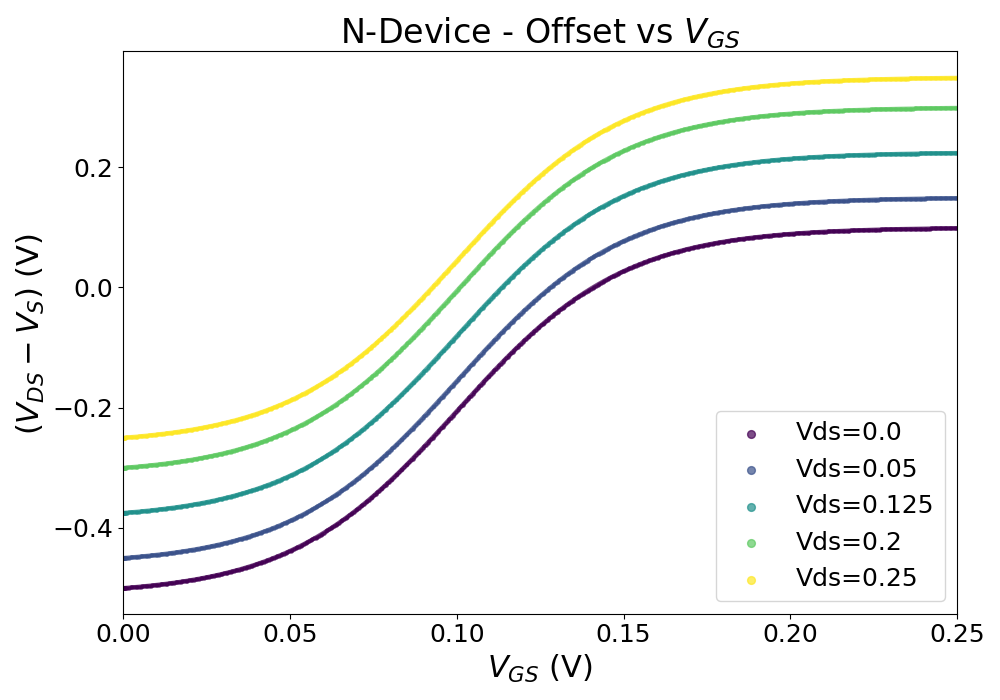}
        \caption{}
        \label{fig:SpiceSuppl_device_NOffset}
    \end{subfigure}%
    \begin{subfigure}{0.49\textwidth}
        \includegraphics[width=\textwidth]{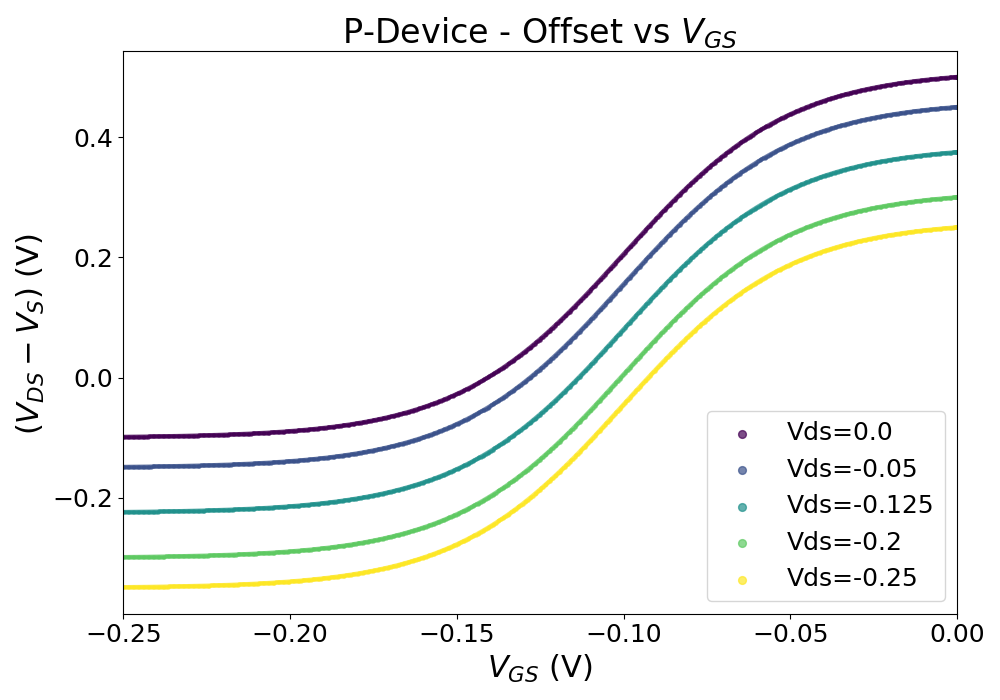}
        \caption{}
        \label{fig:SpiceSuppl_device_POffset}
    \end{subfigure}
    \vspace{0.3em}
    \begin{subfigure}{0.49\textwidth}
        \includegraphics[width=\textwidth]{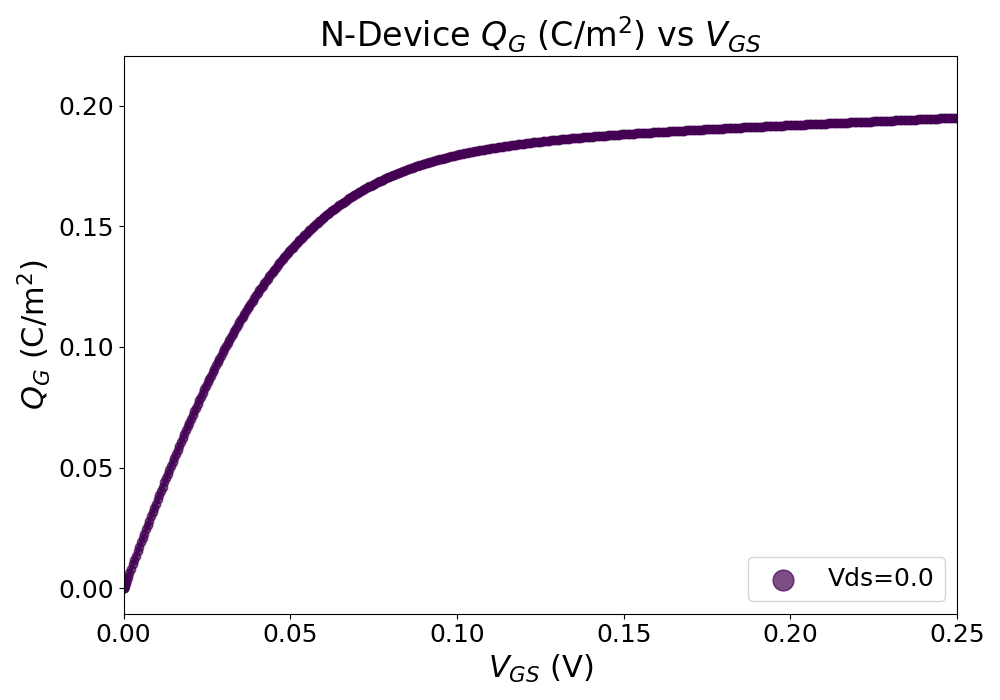}
        \caption{}
        \label{fig:SpiceSuppl_device_N_QG}
    \end{subfigure}%
    \begin{subfigure}{0.49\textwidth}
        \includegraphics[width=\textwidth]{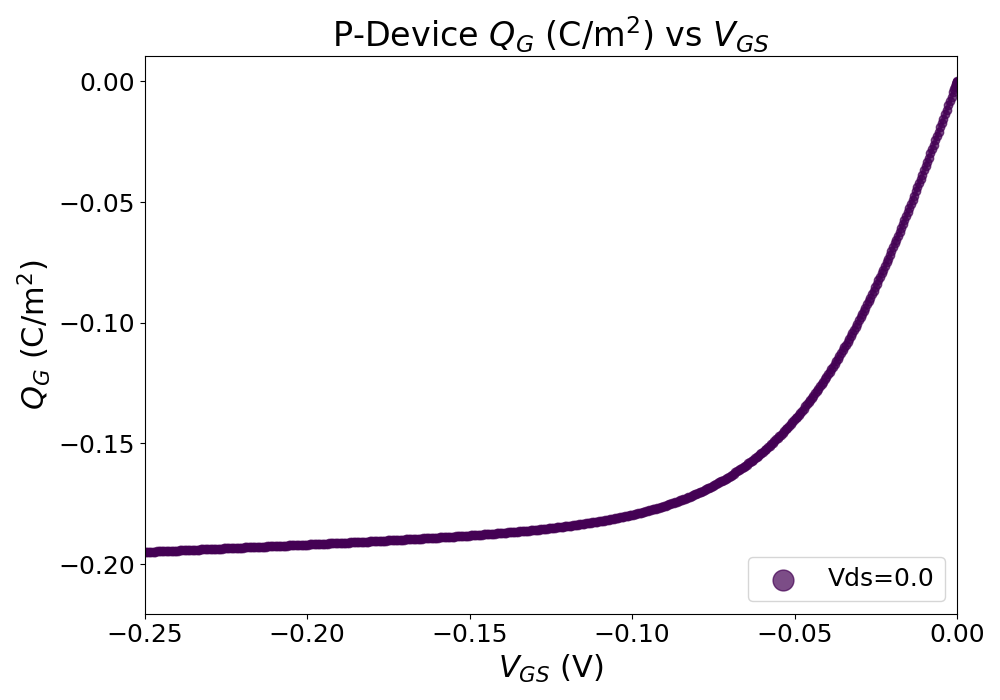}
        \caption{}
        \label{fig:SpiceSuppl_device_P_QG}
    \end{subfigure}
    \caption{(a-b) Offset voltage vs $V_{GS}$ for N and P devices, (c-d) $Q_G$ vs $V_{GS}$ for N and P devices.}
    \label{fig:SpiceSuppl_device}
\end{figure}

We describe the model taking an example of an N-TCAP, and the model is similarly applicable for a P-TCAP. 
Please note that the same model is applicable for a paraelectric (PE) channel, where an initial work-function difference between electrodes can create the starting position of the channel response,
see Fig.~\ref{fig:SpiceSuppl_device}.
Together with the saturation polarization ($P_{S,CH}$ for channel and $P_{S,G}$ for gate), additional dielectric caps are added for both channel and gate with $C_D$ and $C_{GG}$ respectively. Initial offset of the channel $V_{offset}$ represents where the FE branch intersects the voltage axis at $V_{DS}=0$. For a FE channel, this is an equivalent of its original critical voltage with $V_{GS}=0$. For a paraelectric channel, this is an equivalent of the initial offset produced by the work-function differences of the electrodes. Charges on each terminal are denoted by $Q$ and the current with $I$. Area of the gate and channel (A) is assumed to be the same similar to the piezo-TCAP model. Parameter t=1 for a N-TCAP and t=-1 for P-TCAP. 

Here, the effect of gate on the channel is expressed by the gate voltage causing  horizontal and vertical inward shrinks of the hysteresis branch similar to what is predicted by our COMSOL modeling for BTO based channel, with an application of stress. The effect of $V_{GS}$ is non-linear on the channel, while enabling a full swing of $V_{DS}$ from 0 to $V_{DD}$. When $V_{GS}$ is between 0 and $V_{TH,G}$, the effect of gate voltage on the channel is minimal, and becomes higher and gets saturated later, controlled by $V_{X,G}$ parameter. This is accomplished by calculating $\alpha_{NL}$ between 0 and 1 and updating the shrink voltage $V_S$ accordingly from its initial value. The full horizontal shrink is a slight bit higher than $V_{offset}$, to enable full $2*P_{S,CH}$ charge release of the channel, as $V_{DS}$ hits close to zero. For a FE channel, this can represent the final critical voltage (with $V_{GS}=V_{DD}$) of the other branch of its hysteresis. 
\begin{align}
V_{GS} &= V(g,s), \qquad V_{DS} = V(d,s) - t\,V_{offset} \\[6pt]
\alpha_{NL} &= \frac{\tanh\!\left(\dfrac{|V_{GS}| - V_{TH,G}}{V_{X,G}}\right)
              - \tanh\!\left(\dfrac{-V_{TH,G}}{V_{X,G}}\right)}
             {1 - \tanh\!\left(\dfrac{-V_{TH,G}}{V_{X,G}}\right)} \\[6pt]
V_S &= -\,t\,\alpha_{NL}\,(V_{offset} + 2\,V_{X,CH}) \\[6pt]
P_{CH} &= \left(1 - t\,\beta\,V_{GS}\right) P_{S,CH}
     \tanh\!\left(\frac{V_{DS} - V_S}{V_{X,CH}}\right) + C_D\,V(d,s) \\[6pt]
P_G &= P_{S,G}\,\tanh\!\left(\frac{V_{GS}}{V_{X,G}}\right) + C_{GG}\,V_{GS} \\[6pt]
Q_G &= A\,P_G, \qquad Q_D = A\,P_{CH}, \qquad Q_S = -Q_D - Q_G \\[6pt]
I(d) &= \frac{dQ_D}{dt}, \qquad
       I(g) = \frac{dQ_G}{dt}, \qquad
       I(s) = \frac{dQ_S}{dt}
\end{align}
$P_{S,CH}=25~\mu\text{C/cm}^2$, $P_{S,G}=18~\mu\text{C/cm}^2$, $C_D=6~\mu\text{F/cm}^2$, $C_{GG}=6~\mu\text{F/cm}^2$, $V_{X,CH}=0.05$~V, $V_{X,G}=0.05$~V, $V_{TH,G}=0.1$~V, $\beta=1.5$~V$^{-1}$, $V_{offset}=0.5$~V, $A=(10~\text{nm})^2=1\times10^{-16}$~m$^2$, $t=\pm1$; $V_{DD}=0.25$~V; $f=10$~GHz, $T=0.1$~ns, $t_r=20$~ps;  PI interconnect model with $R=100~\Omega$, $C=12$~aF. 
Various characteristics of this device model are shown in Figs.~\ref{fig1:QDvsVDS_N}, \ref{fig1:QDvsVDS_P}, \ref{fig:fig2_tcap_loadline} and~\ref{fig:SpiceSuppl_device}.

\begin{figure}[htbp]
    \centering
    \begin{subfigure}{0.32\textwidth}
        \includegraphics[width=\textwidth]{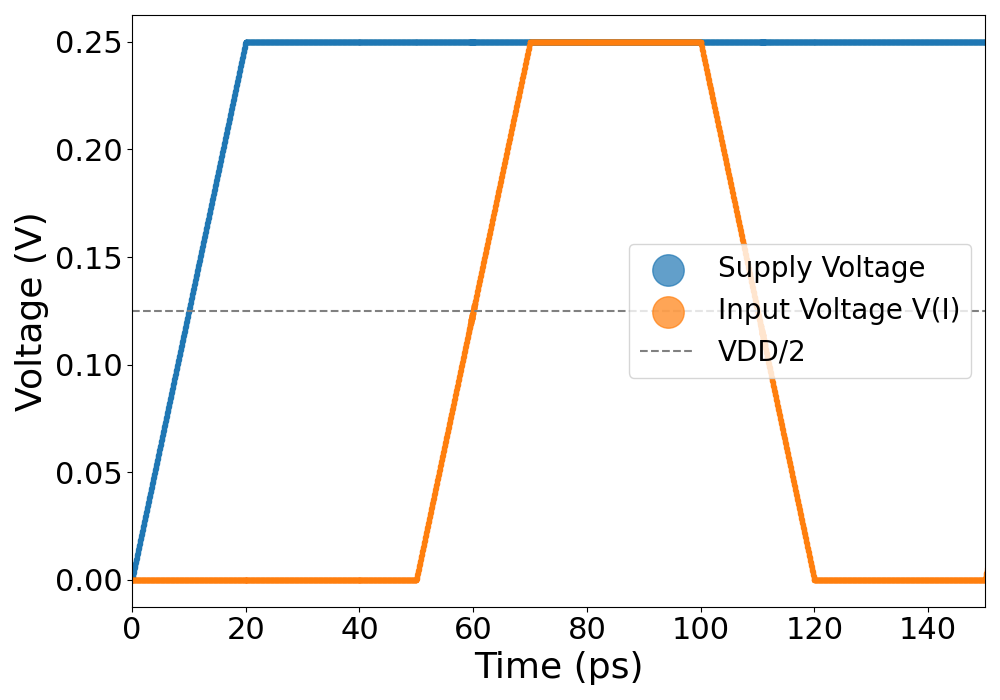}
        \caption{}
        \label{fig:SupplSpiceInv_Inputs}
    \end{subfigure}%
    \begin{subfigure}{0.32\textwidth}
        \includegraphics[width=\textwidth]{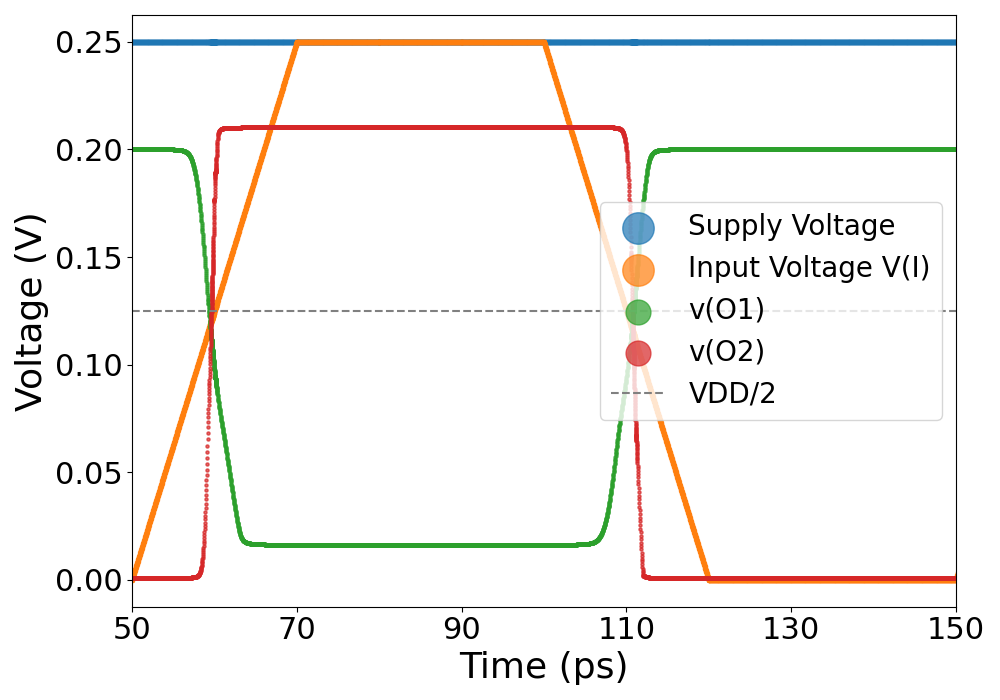}
        \caption{}
        \label{fig:SupplSpiceInv_Voltages}
    \end{subfigure}%
    \begin{subfigure}{0.32\textwidth}
        \includegraphics[width=\textwidth]{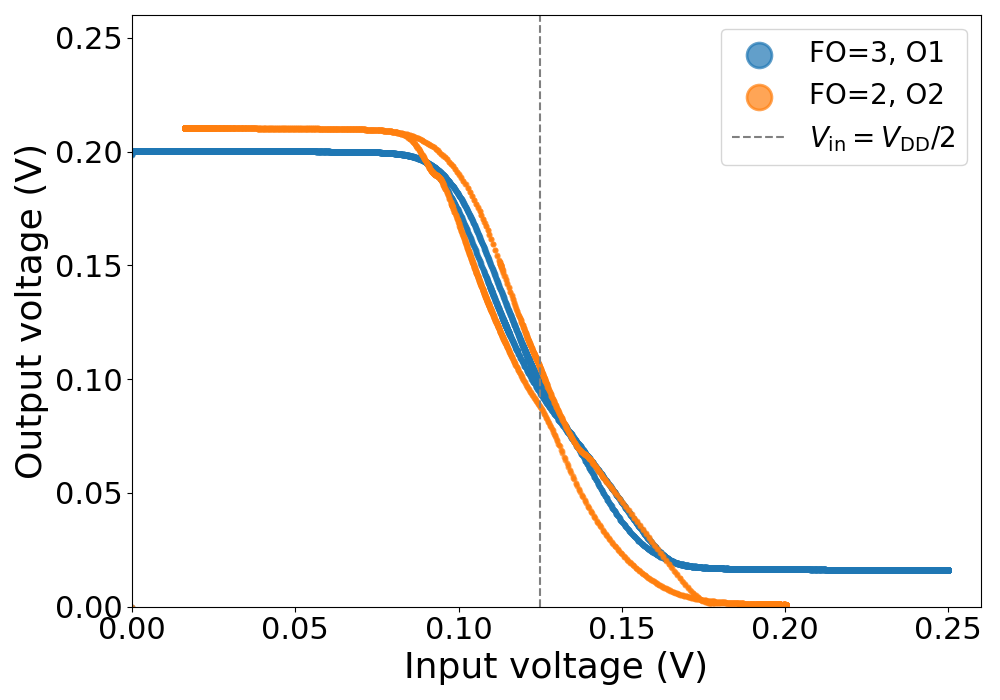}
        \caption{}
        \label{fig:SupplSpiceInv_Transfer}
    \end{subfigure}
    \vspace{0.5em}
    \begin{subfigure}{0.32\textwidth}
        \includegraphics[width=\textwidth]{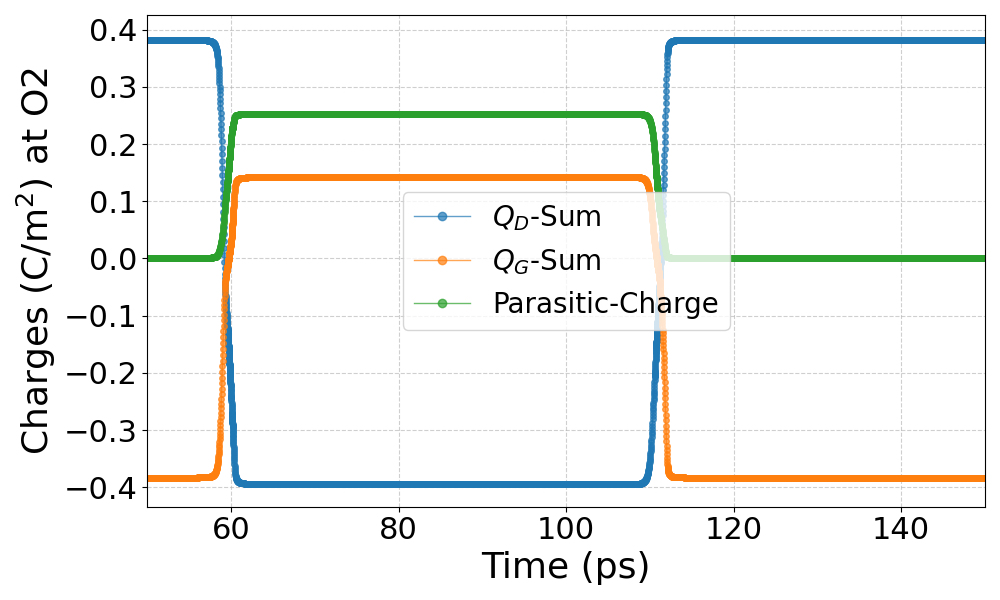}
        \caption{}
        \label{fig:SupplSpiceInv_ChargeConservation}
    \end{subfigure}%
    \begin{subfigure}{0.32\textwidth}
        \includegraphics[width=\textwidth]{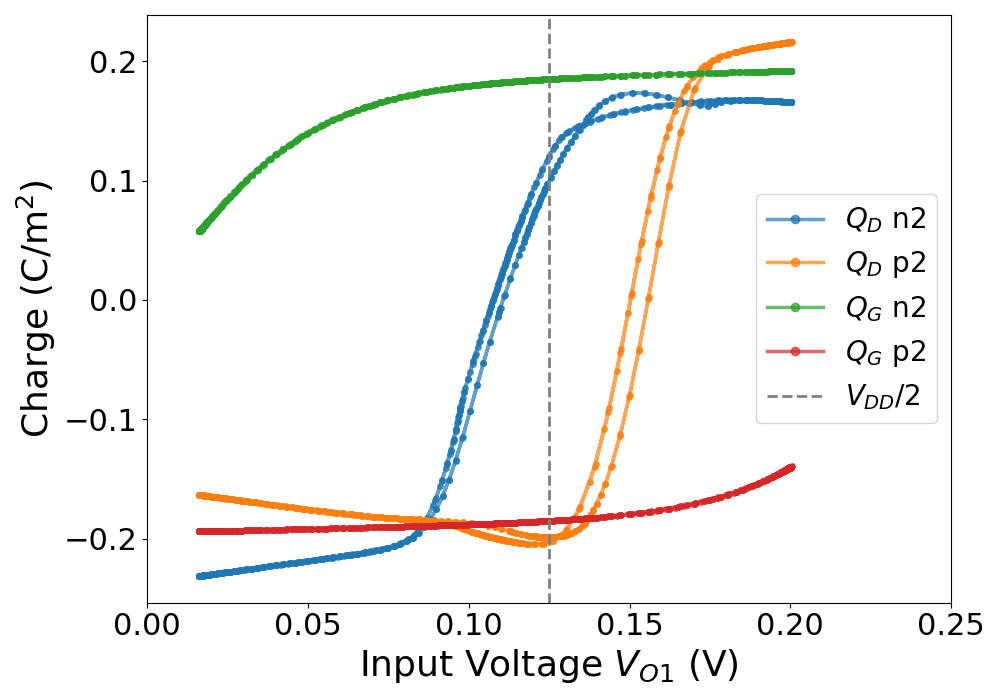}
        \caption{}
        \label{fig:SupplSpiceInv_QD}
    \end{subfigure}%
    \begin{subfigure}{0.32\textwidth}
        \includegraphics[width=\textwidth]{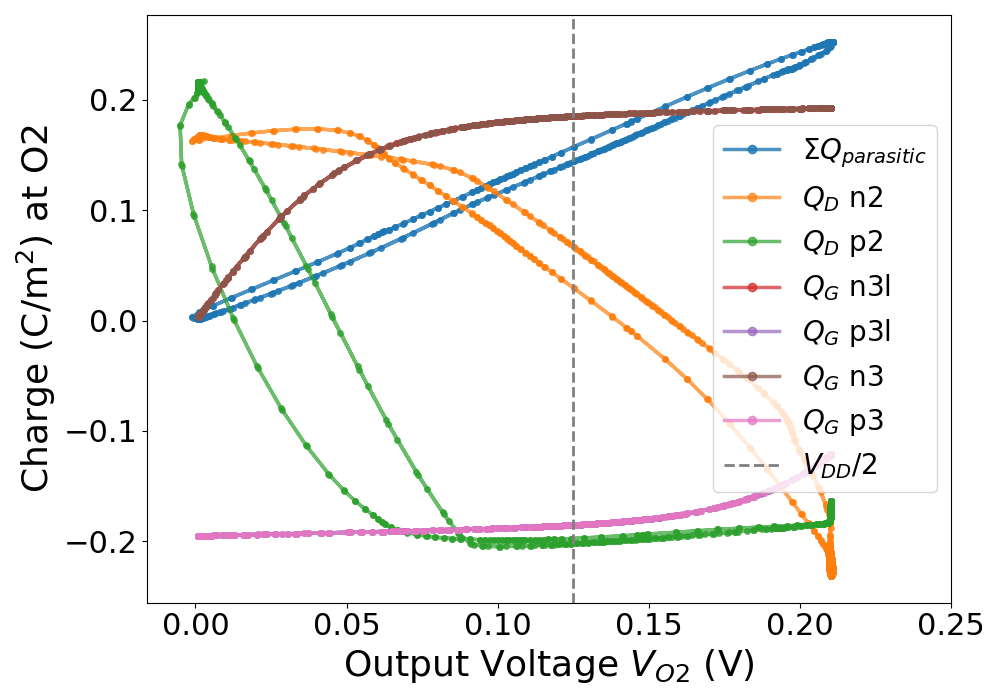}
        \caption{}
        \label{fig:SupplSpiceInv_QG}
    \end{subfigure}
    \caption{(a) Input and supply voltages for the inverter under test I2, (b) Input and output voltages for I2, (c) Inverter transfer curve for I2,
    (d) Charge conservation at the output node O2, (e) Drain and gate charges of I2 inverter vs I2's input voltage (f) Drain and gate charges at node O2 wrt. I2's output voltage.}
    \label{fig:SupplSpiceInv_voltageCharges}
\end{figure}

\begin{figure}[htbp]
    \centering
    \begin{subfigure}{0.32\textwidth}
        \includegraphics[width=\textwidth]{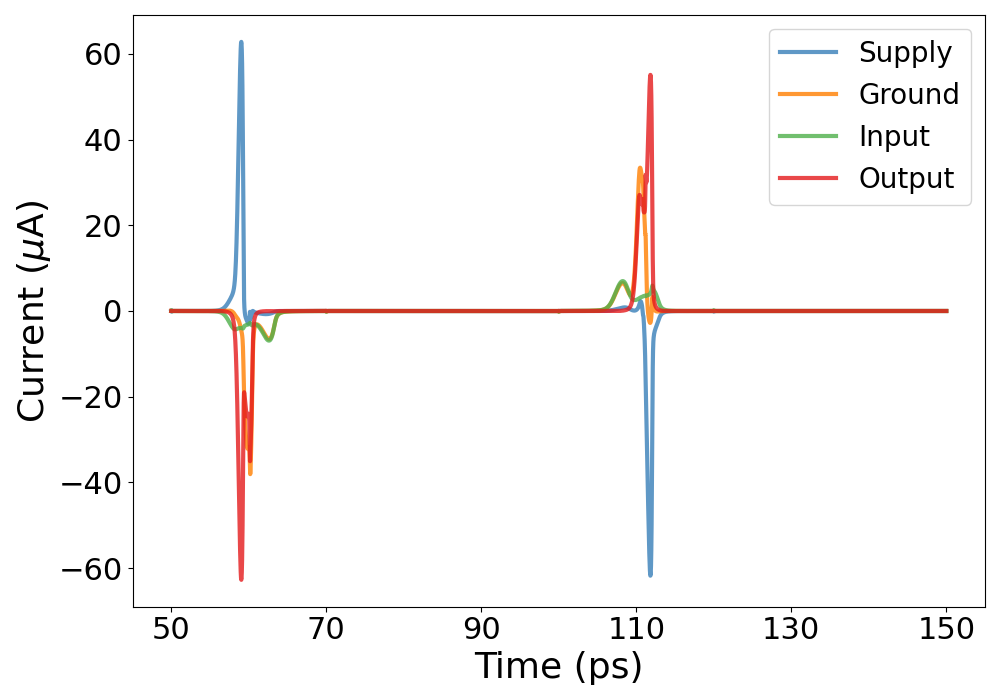}
        \caption{}
        \label{fig:SpiceEnergy_sub1}
    \end{subfigure}%
    \begin{subfigure}{0.32\textwidth}
        \includegraphics[width=\textwidth]{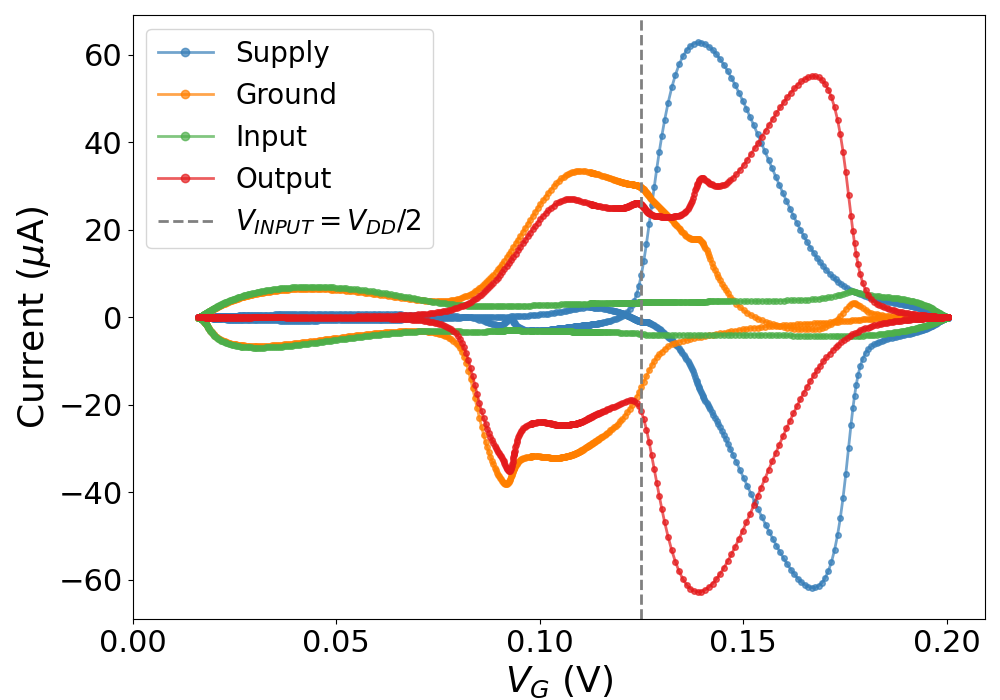}
        \caption{}
        \label{fig:SpiceEnergy_sub2}
    \end{subfigure}%
    \begin{subfigure}{0.32\textwidth}
        \includegraphics[width=\textwidth]{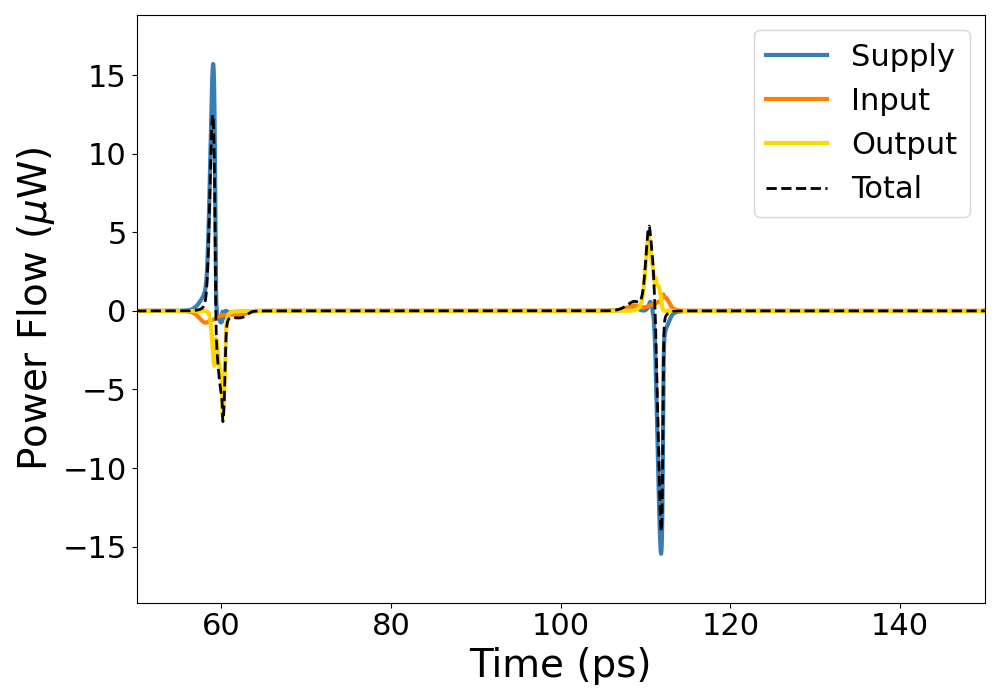}
        \caption{}
        \label{fig:SpiceEnergy_sub3}
    \end{subfigure}
    \vspace{0.3em}
    \begin{subfigure}{0.32\textwidth}
        \includegraphics[width=\textwidth]{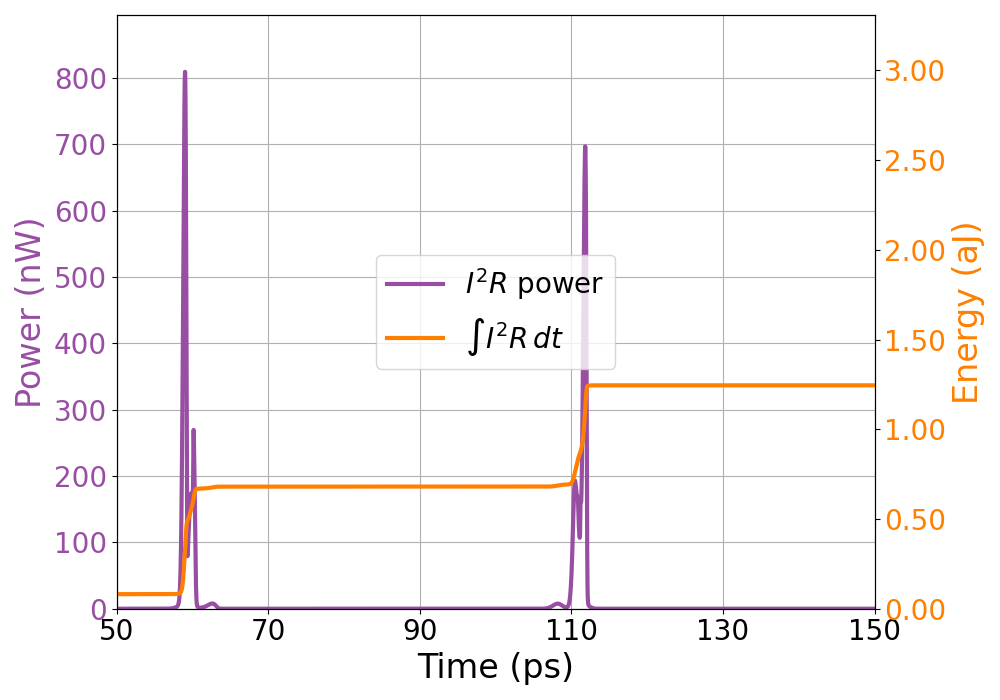}
        \caption{}
        \label{fig:SpiceEnergy_sub4}
    \end{subfigure}%
    \begin{subfigure}{0.32\textwidth}
        \includegraphics[width=\textwidth]{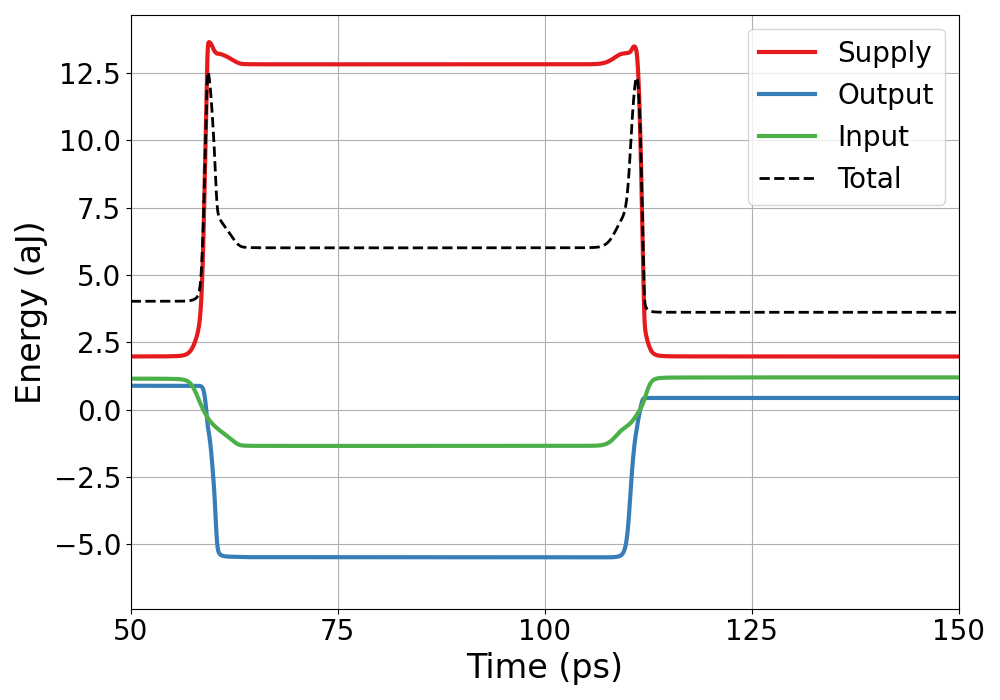}
        \caption{}
        \label{fig:SpiceEnergy_sub5}
    \end{subfigure}%
    \begin{subfigure}{0.32\textwidth}
        \includegraphics[width=\textwidth]{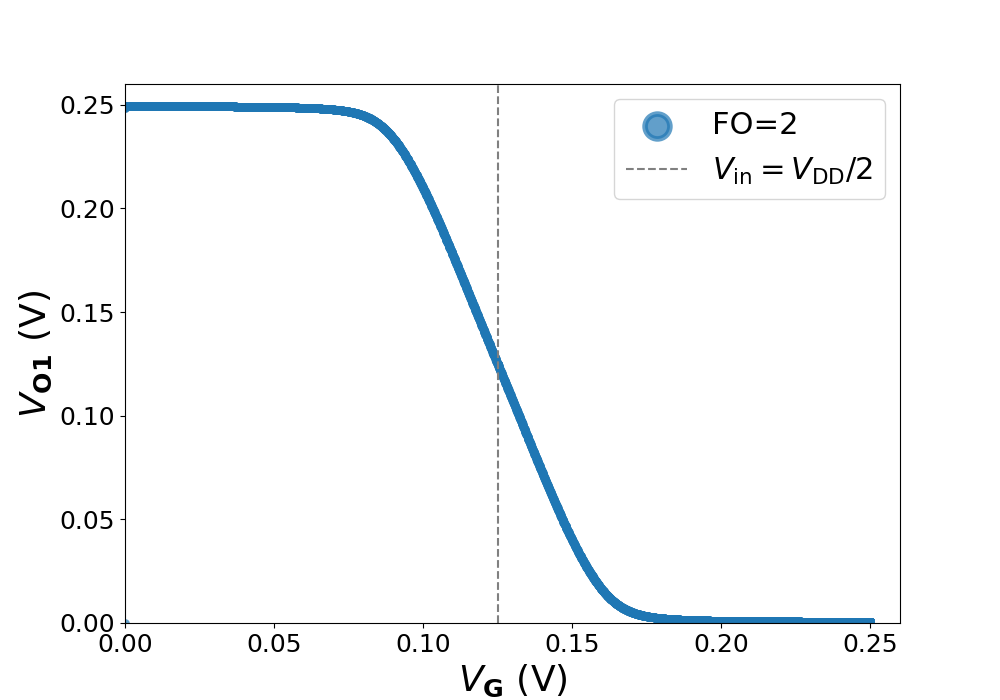}
        \caption{}
        \label{fig:INV_unloaded_transfer}
    \end{subfigure}
    \caption{(a) Current vs time for I2. (b) Current vs input voltage for I2 (c) Power flow through I2, (d) Dissipated power and energy for I2, (e) Energy flow for I2, (f) Unloaded inverter transfer characteristics.}
    \label{fig:SpiceEnergy}
\end{figure}

\begin{figure}[htbp]
    \centering
    \includegraphics[width=0.7\textwidth]{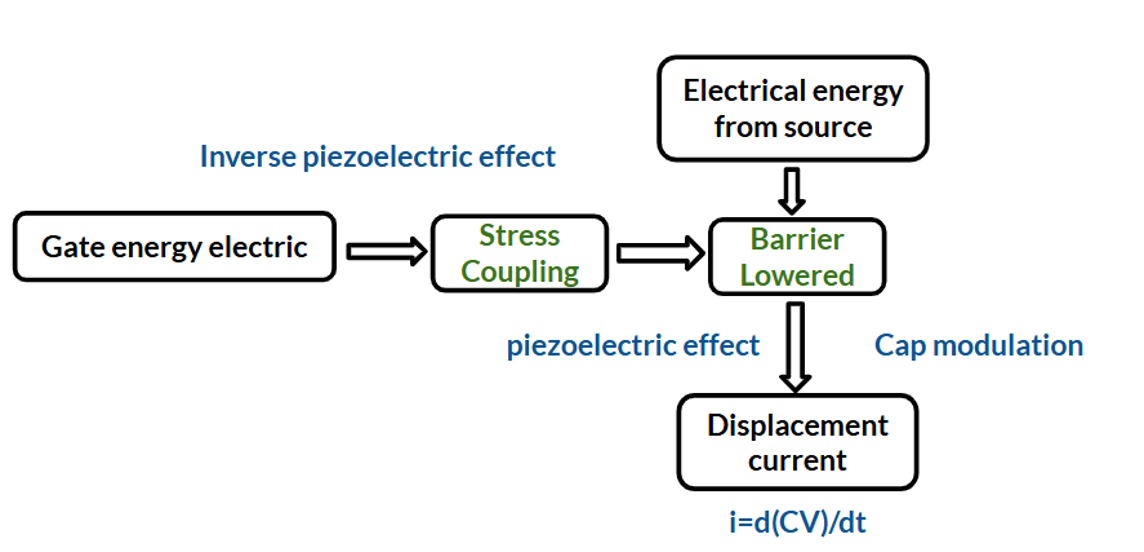}
    \caption{Stress coupling effect for piezo-TCAP, leading to energy barrier lowering for a ferroelectric.}
    \label{fig:piezoStressCoupling}
\end{figure}

The simulated inverter chain is shown in fig.~\ref{fig:SupplSpiceInv_InverterChain}. We use $\pi$ interconnect model for wires, as shown in fig~\ref{fig:SupplSpiceInv_PIRC}. The inverter chain contain various fanouts, up to 3. The first inverter drives a fanout of 3, and produces a voltage signal of lower margins compared to full rail to rail voltage signal. Output of the first inverter drives an inverter named I2, which is our device under test with fanout of 2, for studying its behavior. 
Although, the input signal of I2 (which is O1) has slightly lower voltage margins, output signal O2 is able to get to larger range voltage signal compared to its input and can drive a fanout of 2 further. 
Input voltages and output characteristics including charges at various nodes are shown in fig.~\ref{fig:SupplSpiceInv_voltageCharges} and current, power flow, dissipated power, dissipated energy and energy usage are shown in fig.~\ref{fig:SpiceEnergy}.
Charge on the drains of the inverter I2 does a full swing of ~$2*P_{S,CH}$ to be able to charge the next stage gate load. During a full pull-up and pull-down cycle, most of the energy is recovered, other than the dissipation losses in the parasitics of the wire circuitry, which is roughly 1.2 aJ. 
Since TCAP circuits are mostly fully capacitive circuits, the $\pi$ model used here with parasitic caps grounded, may not be fully suitable, since it works asymmetrically for N and P devices push pull mechanism to get the charges redistributed. In reality, parasitic capacitances are distributed in the circuit and reference to both the supply and ground voltages. However, for demonstration purposes with analytical model, the $\pi$ interconnect model serves the purpose of showing power, energy flow and dissipation losses. 

\section{Reactive Circuits Energy 
Recovery and Dissipation}
\label{sec:reactive} 

The energy dissipation in switching of resistive circuits is well 
explained \cite{hu2010modern}. 
This consideration is applicable to circuits with a direct conduction path from the voltage supply to the ground terminals, including CMOS transistor circuits.
Even though the resistances are the cause of this dissipation, 
the resulting energy is independent of them and instead is determined by the transferred charge $\Delta Q$ and the supply voltage $V_{DD}$, or in other words, by the capacitance $C$, as follows 
\begin{equation}
W_{diss}(resist) \approx \Delta Q V_{DD} = CV_{DD}^2.
\end{equation}
This energy is provided by the supply terminal. One half of it is
stored in the energy of the capacitor. And all of this energy is 
dissipated within the full cycle of "up" and "down" voltage switching.

\begin{figure}[ht!]
\centering
\begin{subfigure}[t]{0.5\linewidth}
\includegraphics[width=\textwidth]{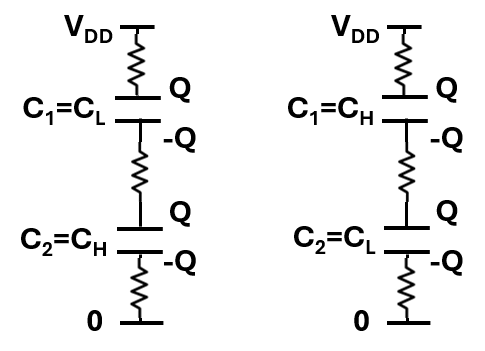}
\end{subfigure}
\caption{The example capacitive voltage divider circuit used for the estimate of the dissipated energy in the reactive energy recovery regime.
The circuit is switched between the left and right configurations.
The capacitances of the two capacitors are externally switched from $C_L$ to $C_H$ and vice versa.
Resistances of each wire is $R$.}
\label{fig:reactive}
\end{figure}

To elucidate the energy flows and
energy dissipation, we choose the 
capacitive voltage divider circuit
in Fig.~\ref{fig:reactive}.
The charge varying vs. time is $Q(t)$
is the same in the two capacitors.
The current in each of the three wires with resistances $R$ is 
\begin{equation}
I = \frac{dQ}{dt}
\end{equation}
The voltage drop across the circuit is
\begin{equation}
V_{DD}=V_1+V_2+IR_{tot}
\label{eq:voltDrop}
\end{equation}
where the total resistance $R_{tot}=3R$, $V_1$ and $V_2$ are voltages across the respective capacitors: 
\begin{eqnarray}
V_1(t)  =  Q / C_1, 
V_2(t) = Q/ C_2.
\end{eqnarray}
To mimic the behavior of transcapacitors,
we assume that the capacitance of the capacitors varies linearly with time over the time $\tau$ in the following manner:
\begin{eqnarray}
C_1(t) & = & C_L + (C_H - C_L) (t / \tau) \\
C_2(t) & = & C_H + (C_L - C_H) (t / \tau)
\end{eqnarray}
which constitutes a switch of voltage up.
The transition in which the capacitances switch down, i.e., in 
reverse, would complete the full cycle and shows a similar behavior.
We will focus on the switch up only.
The charge in capacitors 
\begin{equation}
Q(t) = V_{DD} 
\frac{C_1 C_2}{C_1+C_2}
\end{equation}
At the beginning of the switch, the charge is
\begin{equation}
Q(0) = V_{DD} 
\frac{C_H C_L}{C_H+C_L}
\end{equation}
and in the middle of the switch, the charge is
\begin{equation}
Q(\tau/2) = V_{DD} 
\frac{C_H + C_L}{4}
\end{equation}
Thus the change of charge is
\begin{equation}
\Delta Q  =
\frac{V_{DD} (C_H - C_L)^2}
{4 (C_L + C_H)}
\end{equation}
The current in the circuit is
\begin{equation}
I(t) = K (1 - 2t/\tau)
\end{equation}
where
$K = 4 \Delta Q / \tau$.

The power and flow of energy from the supply is 
\begin{equation}
P_{sup} = V_{DD} I , 
W_{sup} = V_{DD} Q_{sup}
\end{equation}
where $Q_{sup}$ is the net charge transferred out of the supply terminal.
In a half of the switch $Q_{sup} = \Delta Q$.
In the full switch
$Q_{sup}=0$, and therefore $W_{sup}=0$.

The corresponding flow from the ground is identically zero.
The dissipated power in resistors is
\begin{eqnarray}
P_{diss}(t) & = & I(t)^2 R_{tot}
= K^2 
\left(1 - 2t/ \tau
\right) ^2 R_{tot}
\end{eqnarray}
The total dissipated energy 
in one switch e.g. "up"
is obtained by integrating power over time 
\begin{equation}
W_{diss} = K^2 R \tau
\end{equation}
This is an example for a more general expression of dissipation in reactive energy recovery circuits.
\begin{equation}
W_{diss}(react) = (4 \Delta Q)^2 R  / \tau
\end{equation}

The energy stored in the capacitors is
\begin{equation}
U = \frac{C_1 V_1^2 + C_2 V_2^2}{2} =
\frac{QV_1 + QV_2}{2}
\end{equation}
Then the derivative of the capacitor energy is
\begin{equation}
\frac{dU}{dt} 
= (V_1+V_2) I
- \frac{V_1^2 dC_1}{2dt} 
- \frac{V_2^2 dC_2}{2dt} 
= P_{ele} + P_{ext}
\end{equation}
The first term on the right is $P_{ele}$, the work done 
by the electric circuit, and the second and third terms are $P_{ext}$, 
the work done by an external force to vary the capacitance.

The balance of powers is obtained by multiplying 
Eq.~(\ref{eq:voltDrop}) by the current:
\begin{equation}
V_{DD}I=V_1 I+V_2 I+I^2R_{tot}
\end{equation}
or in other words,
\begin{equation}
P_{sup}=P_{ext}+P_{diss}
\end{equation}
By substituting the above definitions
\begin{equation}
P_{sup}= \frac{dU}{dt} - P_{ext} +P_{diss}
\end{equation}
The total energies in switching are obtained by integrating powers over the time of switching. Since the charges are equal in the beginning and the end, the net energy flows from the supply is zero. 
Also the change of the capacitor energy is zero. Thus we arrive at the conclusion that the energy dissipated in resistances is supplied by the external force necessary to change the capacitances.
\begin{equation}
W_{ext} = W_{diss}
\end{equation}
If we extend this consideration to transcapacitors, the external work will in fact be performed by a flow of energy into the inputs of the logic gate, which are attached to the gates of transcapacitors.
The fact that the input energy flow is dissipated ensures the irreversibility of operation and this the isolation of the input state from perturbation at the output, see 
Section~\ref{sec:transcap_digital_switch}.

This derivation shows that the energy from the supply flows into the circuit and is then completely recovered back. 
The energy stored in the capacitors changes and then returns to the original value.
The above conclusion is true for a complete switch.
If we take the case of partial switching, then
the relationship between various terms in energy may be different.
If the resistance in the wires in the reactive circuit is sufficiently small, the net dissipated energy is much smaller than 
the the flow of energy back and forth from/into the supply and from/into the input and the output (and other terms in energy).
For example, for typical values, 
the supply energy can be
$W_{sup} \approx 2.5aJ$
but the dissipated energy can be 
$W_{diss} \approx 6.3zJ$
i.e. 400 times smaller.
Moreover, if the resistance in the wires of the reactive circuit 
is smaller than the resistance of the device (even in the "on" state),
then one concludes that the dissipated power in reactive circuits can be much smaller than that in the resistive circuits.
The charge gain and the energy gain are related to 
the energy exchange in a partial switching.
Therefore the supply energy is non-zero, and the output energy can be significantly greater
than the input one.
Note that the above conclusions are 
applicable to the constant supply voltage. In cases of the circuit reset by ramping 
the supply voltage down, potentially going negative and then up,
it is possible to have a non-zero net energy expenditure from the supply even though the charge state before and after are the same.
Also, if there is a non-zero leakage through capacitors or transcapacitors, it also leads to 
a net flow of energy from the supply terminal which 

\section{Energy and Delay for CMOS Circuits} \label{sec:EdCMOS}

In order to make a side-by-side comparison with the performance of a CMOS inverter, we adopt the analytical model of \cite{sakurai2002alpha}. 
There the drain current is approximated by a power law with the exponent, $\alpha$ , the velocity saturation index.
The on-current, $I_{d0}$, is the source-to-drain current at $V_{DS}=V_{GS}=V_{DD}$, and $V_{d0}$  is the drain saturation voltage.
The ratio of the threshold and supply voltages is
\begin{equation}
\nu = \frac{V_{th}}{V_{DD}}
\end{equation}
One of the key factors in the fanout of the logic gate, equal to the ratio of the load (output) and input capacitances:
\begin{equation}
FO = \frac{C_L}{C_{in}}
\end{equation}
The delay of a transistor logic stage is comprised of the ramp delay at the input and output delays
\begin{equation}
\tau_{tr}=\tau_{in} + \tau_{out}
\end{equation}
The output delay is 
\begin{equation}
\tau_{out} = \frac{C_L V_{DD}}{2 I_{d0}}
\end{equation}
And the input delay is
\begin{align}
\tau_{in} & = F_1 F_2\frac{C_L V_{DD}}{I_{d0}} \\
F_1 & = \frac{1}{2} - \frac{1-\nu}{1+\alpha} \\
F_2 & = \frac{0.9}{0.8} + \frac{V_{d0}}{V_{DD}} 
\ln
\left(
\frac{10V_{d0}}{e V_{DD}}
\right)
\end{align}
The switching energy of a transistor logic state can be separated into the charging energy of the load and the dissipation by the rush-through current (from the supply to the ground) as follows:
\begin{align}
E_{tr} & = E_{ch}+E_{ru} \\
E_{ch} & = C V_{DD}^2/2 \\
E_{ru} & = F_2 F_3 C V_{DD}^2 \\
F_3 & = \frac{1}{\alpha+1}
\frac{1}{2^{\alpha-1}}
\frac{(1-2\nu)^{\alpha+1}}
{(1-\nu)^\alpha}
\end{align}
Here is the exponent of the current dependence on the voltage, see \cite{sakurai2002alpha}.
In this analysis we are not considering the stand-by power dissipation. It is important and produces a significant contribution. However, its contribution depends on the activity factor. That factor in turn is dependent on the application being run by the chip.

\section{Energy and Delay for Transcapacitor Circuits}
\label{sec:EdTpol}

\begin{figure}[ht!]
\centering
\begin{subfigure}[t]{0.45\linewidth}
\includegraphics[width=\textwidth]{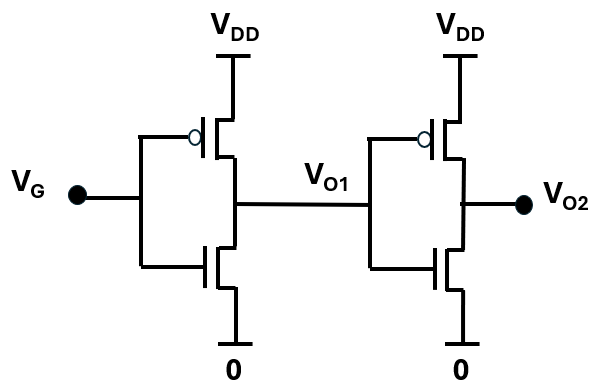}
\caption{}
\end{subfigure}
\begin{subfigure}[t]{0.48\linewidth}
\includegraphics[width=\textwidth]{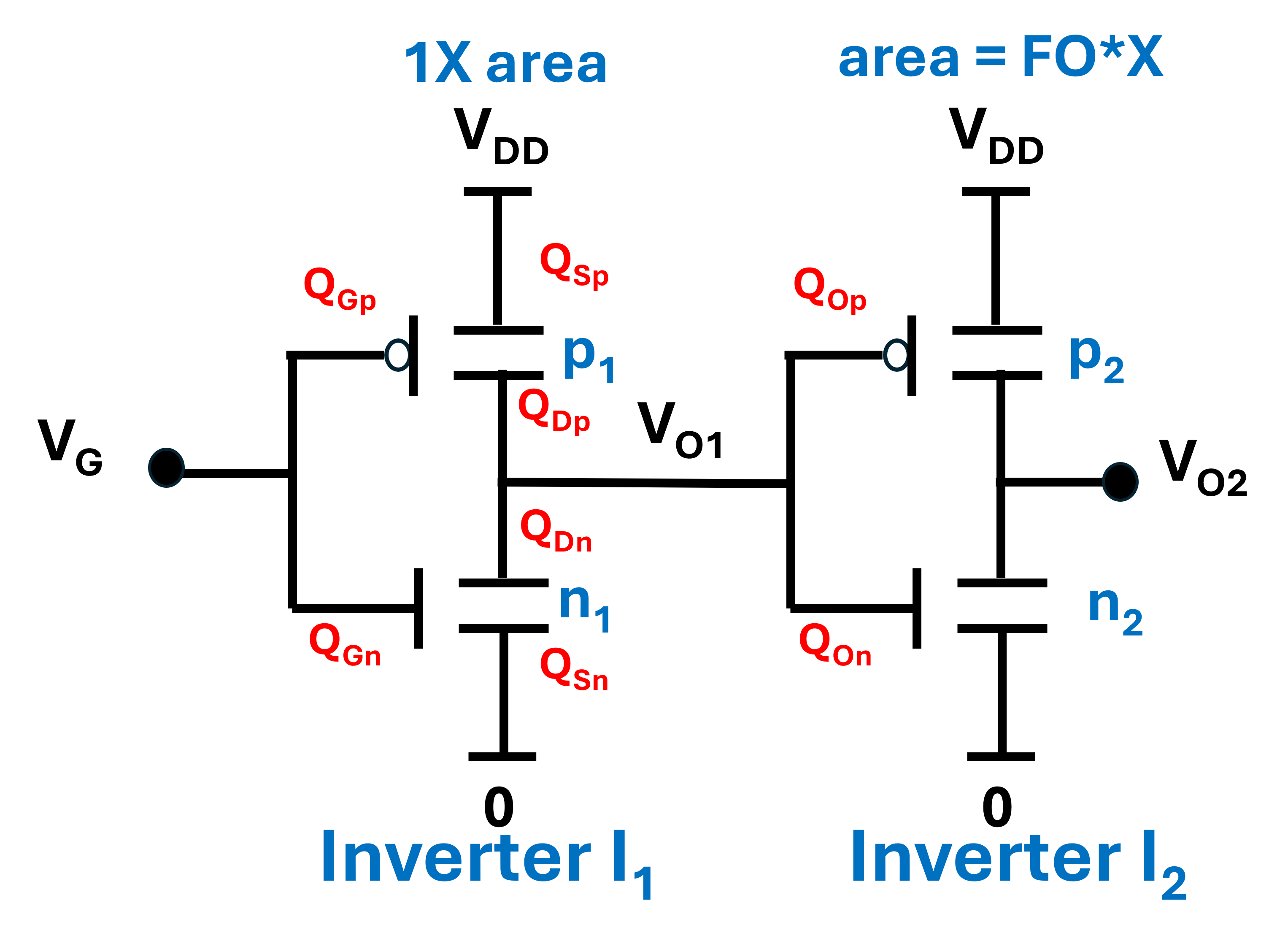}
\caption{}
\end{subfigure}
\caption{
Schematic for an inverter driving another inverter.
(a) CMOS-based. (b) Transcapacitor-based.
Input ($V_G$) and output ($V_{O1}$) in the signal nodes are shown. Supply terminals ($V_{DD}$) shown on the top, and ground terminals (0V) shown on the bottom. Charges at terminals of devices are designated and discussed in the text.}
\label{fig:tpol_inverter}
\end{figure}

An inverter based on transcapacitors is structured in a topology similar to that based on CMOS transistors, see Fig.~\ref{fig:tpol_inverter}. 

Here we consider two cascaded inverters such that the first is driving the second. 

To gain insight, we start with an analytical model for the transcapacitor inverter. 
It contains equations that express the net charge neutrality of each transcapacitor and the zero net charge at the floating node.
\begin{align}
& Q_{Gp}+Q_{Dp}+Q_{Sp}=0 \\
& Q_{Gn}+Q_{Dn}+Q_{Sn}=0 \\
& Q_{On}+Q_{Op}+Q_{Dp}+Q_{Dn}=0
\label{ch_conserv}
\end{align}
The capacitors in the next stages are greater by the factor of fanout (FO).
For simplicity, we are assuming linear capacitance for the gate, hence:
\begin{align}
Q_{Gn} & = C_{GS}V_G+C_{GD}(V_G-V_{O1}) \\
Q_{Gp} & = C_{GS}(V_G-V_{DD})+C_{GD}(V_G-V_{O1}) \\
Q_{On} & = FO \left( C_{GS}V_{O1}+C_{GD}(V_{O1}-V_G) \right) \\
Q_{Op} & = FO \left( C_{GS}(V_{O1}-V_{DD})+C_{GD}(V_{O1}-V_G) \right)
\end{align}
Polarization can be taken as an analytic S-shaped function of the applied voltage, where $V_{d0}$ is the saturated voltage for the polarized layer. Using the paraelectric polarization rather than the ferroelectric one, we neglect the coercive voltage compared to the supply voltage.
\begin{align}
P(V) & = P_s \tanh(V/V_x).
\end{align}
The polarizations in transcapacitors depends on both the source and drain voltages as follows.
\begin{align}
P_p & = -P(V_{O1}-V_{DD}+h(V_G-V_{DD}/2)) \\
P_n & = P(V_{O1}+h(V_G-V_{DD}/2))
\end{align}
The polarization of the layer contributes to the terminal charges.
To satisfy Eq.~(\ref{ch_conserv})
\begin{align}
Q_{Sp} & = - C_{GS}V_G -P_p \\
Q_{Dp} & = - C_{GD}(V_G-V_{O1}) + P_p\\
Q_{Dn} & = - C_{GD}(V_G-V_{O1}) + P_n\\
Q_{Sn} & = - C_{GS}(V_G-V_{DD}) - P_n
\end{align}
We designate empirical dimensionless factors corresponding to the horizontal ($h$) and vertical ($v$) shifts, and the slope ($sl$) of the charge curve:
\begin{align}
v & = -C_{GD}V_{DD}/(P_sA) \\
sl & = C_{GS}V_{DD}/(P_sA)
\end{align}
To be precise, both $sl$ and $v$ contribute to the slope. Here $h$ has the meaning of a dimensionless efficiency of a transcapacitor; and $v$ is the dimensionless transcapacitance. As discussed in
Section~\ref{logic_device_details},
the transcapacitance can be positive.

This simple model is derived from the gate having influence on the polarization by an electrostatic mechanism, i.e., creating the charges in the source and drain terminals. In that case, with positive $sl$ and $v$, the output voltage would increase with the gate voltage, i.e., one would not get a voltage inversion. In other types of transcapacitors (see sections below) additional effects (beside electrostatics) couple input voltage and polarization. This corresponds to negative voltage, and results in inversion of the output voltage relative to the input one.
The loadline analysis first relies on the total voltage drop through the two transcapacitors in the inverter. Secondly, it relies on the sum of charges at terminals connected to the output node as per the equations above; it is unlike the case of CMOS where the sum of currents is used. 

The above analysis of a transcapacitor inverter enables an estimate of energy and delay in simple logic gates. 
The following quantities are used: the linear part of the capacitance arising from the dielectric response of the polarized layer of thickness $t_{fe}$, having the dielectric constant
$\epsilon_{fe} $,
\begin{equation}
C_l = \frac{\epsilon_0 \epsilon_{fe} A} {t_{fe}}
\end{equation}
The range of variation during switching of the output voltage and a charge related to the remanent (nonlinear) part of polarization are set with the following empirical constants
\begin{align}
\Delta V & = 0.8 V_{DD} \\
\Delta Q_{fe} & = 2 P_s A
\end{align}
The total range of charge, where $C_{ic}$  is the capacitance of the interconnecting wire between this and the following logic gates,
\begin{equation}
\Delta Q = \Delta Q_{fe} +(C_l + C_{ic}) \Delta V
\end{equation}
The delay to charge the above capacitances is limited by the interconnect wire resistance $R_{ic}$,
\begin{equation}
\tau_{ch}=R_{ic} \Delta Q / \Delta V
\end{equation}		
And the total switching delay of the transcapacitor logic gate also includes the polarization layer (Merz) switching delay:
\begin{equation}
\tau_{Tp}=\tau_{ch}+\tau_{M}
\end{equation}
The dissipated energy comprises the Joule heat in the resistances of wires and losses in the material damping 
\begin{equation}
E_{Tp} = E_{da} + E_{re}
\end{equation}
The change in charge in the capacitors contributes to the current in the wires. The factor of 4 is intended to account for the current from the supply and ground networks and the current from each of the two capacitors to the next logic stage.
\begin{equation}
E_{re}=4 R_{ic} \Delta Q^2 / \tau_{tc}
\end{equation}
The damping dissipated energy in the material is proportional to the area of the hysteresis loop, related to the coercive voltage $V_c$.
It may be significantly narrower than the voltage range defined by the supply voltage $V_{DD}$:
\begin{equation}
E_{da}=2 V_c \Delta Q_{fe} 
\end{equation}
These equations are used below to estimate the energy and delay in transcapacitor circuits. For the following estimates, the size of the ferroelectric capacitor is taken to be $A=10*10 nm^2$.
The typical values used to get numerical values for the delay and energy
are in Table~\ref{tab:piezoFOM}.

\begin{table}[ht!] 
\centering
\caption{
Parameters to estimate the delay and energy ratios between CMOS and transcapacitors. }
\label{tab:compare_cmos_tcap}
\begin{tabularx}{\textwidth}{
| >{\hsize=1\hsize\raggedright\arraybackslash}X
| >{\hsize=0.5\hsize\raggedcenter\arraybackslash}X
| >{\hsize=0.5\hsize\raggedcenter\arraybackslash}X
| >{\hsize=0.5\hsize\raggedcenter\arraybackslash}X |
}
\hline
Quantity & Symbol & Value & Unit  \\
\hline
Supply voltage, CMOS & $V_{DD}$ & 0.8 & V  \\
\hline
Load capacitance & $C_L$ & 1 & fF  \\
\hline
Sakurai-Newton factors & $F_1,F_2,F_3$ & 0.29;1.43;0.0067 &   \\
\hline
Transistor on-current & $I_{d0}$ & 0.16 & mA  \\
\hline
Wire resistance & $R_{ic}$ & 400 &  $\Omega$ \\
\hline
Voltage swing, TCAP & $\Delta V$ & 100 & mV  \\
\hline
Charge change, TCAP & $\Delta Q$ & 80 &  aC \\
\hline
Intrinsic switching time & $\tau_{ferroic}$ & 0 ... 20 &  ps \\
\hline
Coercive voltage & $V_c$ & 0 ... 50 & mV  \\
\hline
\end{tabularx}
\end{table}

\section{Comparison of the Energy and Delay for adders}
\label{sec:Edadder}

We extend the comparison of the transistors and transcapacitors to more complex circuits.
For that we choose a $N=32$-bit adder circuit as an example.
One of the advantages of transcapacitors is an easier way to make majority gates. That permits us to design circuits with fewer gate and thus having smaller area and smaller dissipated energy. 
For example, the CMOS design of a 1-bit full adder cell 
comprises 26 transistors \cite{weste2015cmos}. 
The transcapacitor design only includes 3 majority gates with fanin of 3.
Each such majority gate has a transcapacitor at the output and at each input.

For the architecture of a CMOS transistor adder we choose the Kogge-Stone adder (KSA) scheme \cite{weste2015cmos}.
It includes 1-bit full adder (1bFA) cells as well as AOI cells.
Then the area of the adder is 
\begin{equation}
a(KSA,N) = 
N log_2(N)/2*a(1bFA) + N*a(AOI).
\end{equation}
A similar equation is true for the switching energy
$E(KSA,N)$, except it has an additional overall activity factor (which we approximately set to $1/3$).
The delay of such an adder is quite optimized and is
\begin{equation}
\tau(KSA,N) = 
log_2 (N)* \tau(AOI)+\tau(XOR)+\tau(NAND).
\end{equation}

For transcapacitors, we design the adder (MA) out of majority (MAJ) gates. Then its area  
\begin{equation}
a(MA,N) \approx 3N*a(MAJ).
\end{equation}
The same relationship between the energy and the area as above is true here.
The switching delay is estimated to be
\begin{equation}
\tau(MA,N) = 
(log_2 (N) +1)* \tau(MAJ).
\end{equation}

\begin{figure}[ht!]
\centering
\begin{subfigure}[t]{0.7\linewidth}
\includegraphics[width=\textwidth]{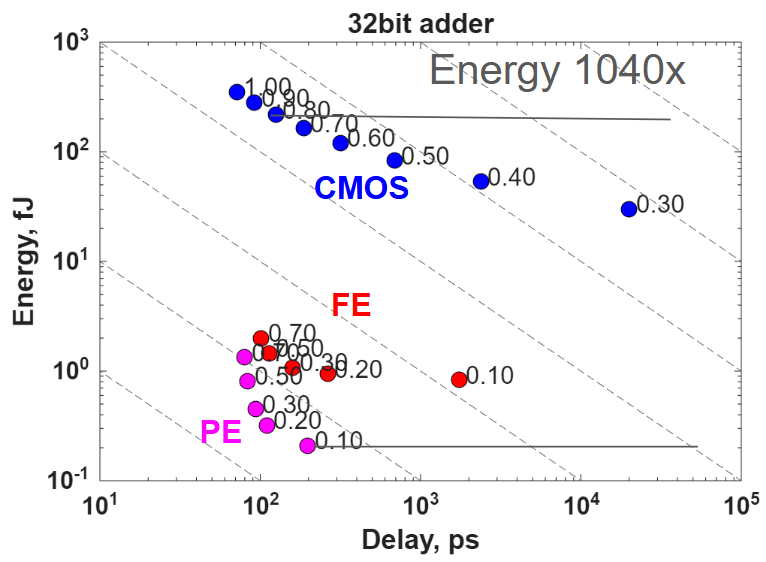}
\end{subfigure}
\caption{
Benchmarking of energy and delay (following the standard methodology \cite{nikonov2015benchmarking}) of 32-bit adders based on transistor (CMOS) and ferroelectric (FE) and paraelectric (PE) transcapacitor devices for various values of supply voltage (labeled next to markers).}
\label{fig:Ed_32adder}
\end{figure}

One can see that the advantage of transcapacitor circuits 
over transistor one for adders is 1040x, see Fig. ~\ref{fig:Ed_32adder}.
Meanwhile for inverters this advantage is 168x,
see Fig.~\ref{fig:energyFig_inverter}.
The ratio of these factors is the additional advantage 
conferred by a more efficient, majority gate based micro-architecture. This factor is listed in 
Table~\ref{tab:energyWins}.

\section{Analytical model Spice run on 5 nm node}
\label{sec:analyticalCLCH}

In this section, we use a simple analytical model for a TCAP device, based on the concept of modulating $C_{DS}$ (drain to source capacitance) with the application of gate voltage $V_{GS}$. We perform Spice simulations with this model in 5nm process node environment. 
Our CMOS simulations are based on a layout based inverter cell, with extracted RC parasitics in 5nm process node. For TCAP simulations, we keep the parasitic, while replacing the device with the analytical TCAP model described here. 

Our purpose here is to prove the advantage of a TCAP circuit element over CMOS, in 5nm process node environment. 
This model is similar to the one proposed in~\cite{galisultanov2017capacitive} for MEMS based capacitive devices. 
This ideal device model has a limitation that it ignores the impact of $C_{DG}$ (drain to gate capacitance) on the circuit operation, together with the impact of the drain voltage $V_{DS}$ on $C_{DS}$ which is necessary to make a physical device producing inversion and gain. 
However, this device implementation specific limitation doesn't hinder with our purpose stated above. It provides a support that a real physical device model of TCAP can provide significant advantage over CMOS even with the real parasitic present in the system. 

\begin{flalign}
&& C_{DS} &= \frac{C_H - C_L}{1 + e^{\frac{V_o - t V_{GS}}{V_x}}} + C_L & \\
&& Q_d &= \left( C_{DS} + \frac{1}{2} C_{GG} \right) V_{DS} - \frac{1}{2} C_{GG} V_{GS} & \\
&& Q_G &= C_{GG} V_{GS} - \frac{1}{2} C_{GG} V_{DS} & \\
&& Q_S &= -Q_D - Q_G &
\end{flalign}

\begin{figure}[ht!]
\centering
\includegraphics[width=\textwidth]{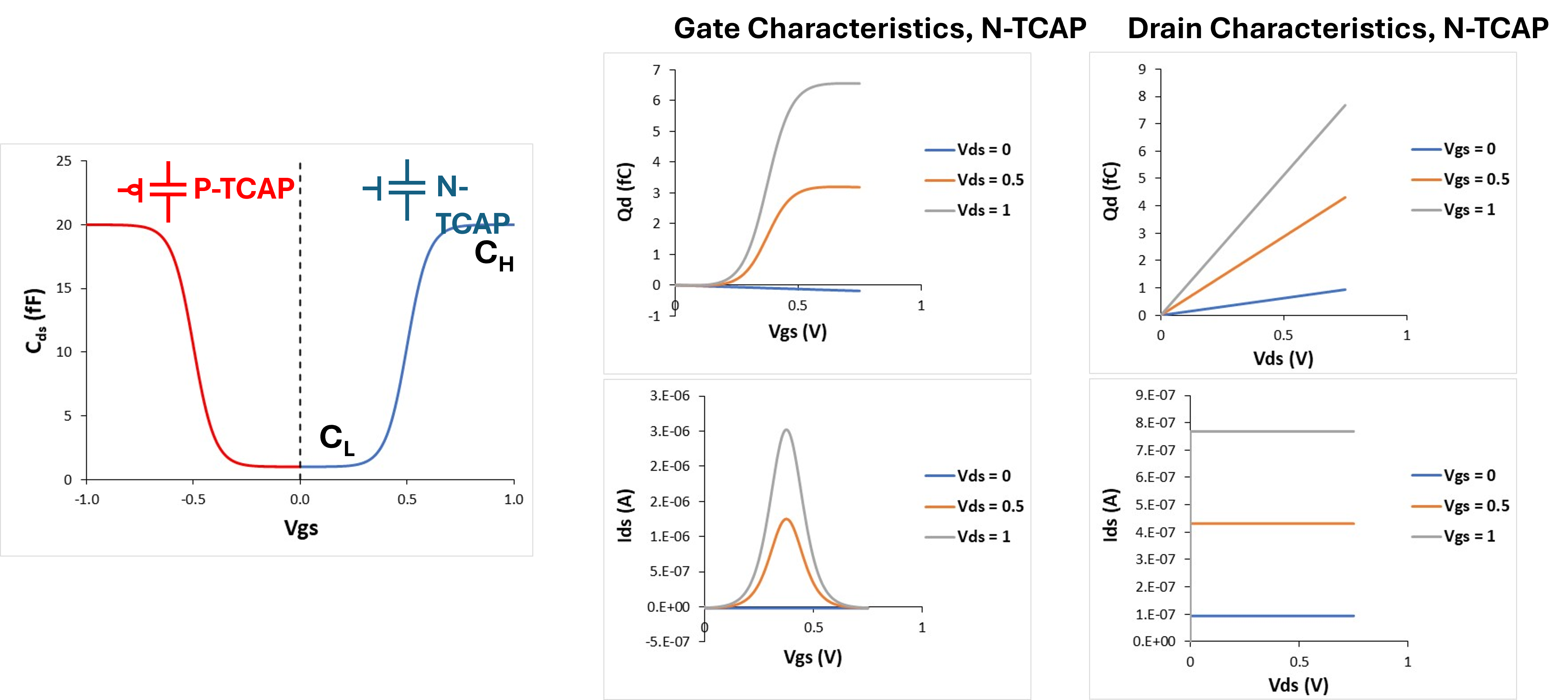}
\caption{ Ideal TCAP model described by the modulation of $C_{DS}$ from $C_L$ to $C_H$ with the application of $V_{GS}$, including drain charge and drain current characteristics.
}
\label{fig:ideal_tcap_CHCL_transfer}
\end{figure}

\begin{figure}[ht!]
\centering
\includegraphics[width=\textwidth]{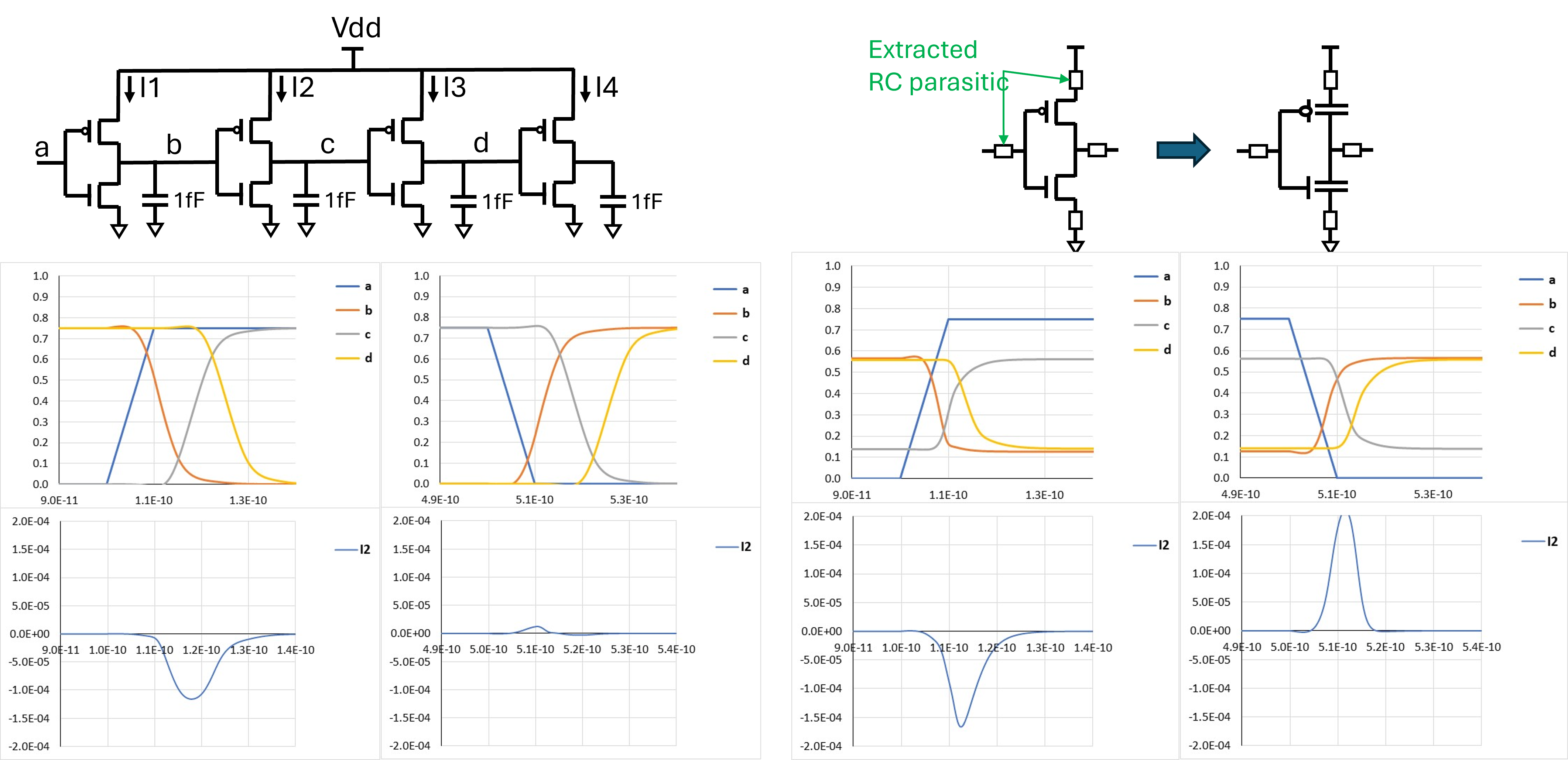}
\caption{ Inverter chain simulations of CMOS and TCAP, with extracted parasitics and FO=3 load. TCAP is described by an ideal model with the modulation of $C_{DS}$.}
\label{fig:ideal_tcap_CHCL_sims}
\end{figure}

Here, $t = 1$ for N-TCAP and $t = -1$ for P-TCAP to enable the concept in the negative and positive voltage regimes for $V_{GS}$ operations. Parameters values chosen for the simulations are $C_H=10fF$, $C_L=1fF$ and $C_{GG}=0.5fF$. We simulate a chain of inverters with an additional load of 1fF on each inverter (equivalent to driving an overall FO-3) in N5 environment. 
Transfer functions of TCAP given by this ideal model are given in
Fig.~\ref{fig:ideal_tcap_CHCL_transfer}. 
Simulation results of an inverter chain are described in Fig.~\ref{fig:ideal_tcap_CHCL_sims}. The simulation results show $~940aJ$ of energy consumption in CMOS for 0.75V supply voltage, combined for a pull up and a pull down cycle. For TCAP, the energy consumption in parasitic is limited to approximately $0.1 aJ$. The simulation shows that the energy spent in parasitics with a TCAP device is minimal at the advanced process nodes. 

\section{Material mechanism for coercive voltage reduction with stress in Barium Titanium Oxide (BTO) }
\label{device_dft}

\begin{figure}[ht!]
  \centering
  \begin{subfigure}[t]{0.5\linewidth}
    \includegraphics[width=\linewidth]{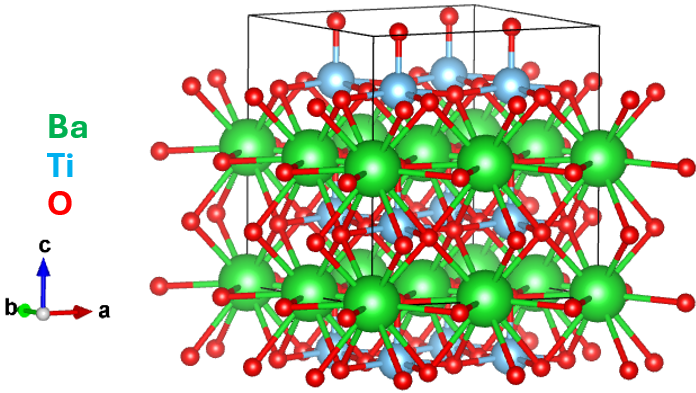}
    \caption{}\label{fig:deviceFig_dft_a}
  \end{subfigure}\hfill
  \begin{subfigure}[t]{0.5\linewidth}
    \includegraphics[width=\linewidth]{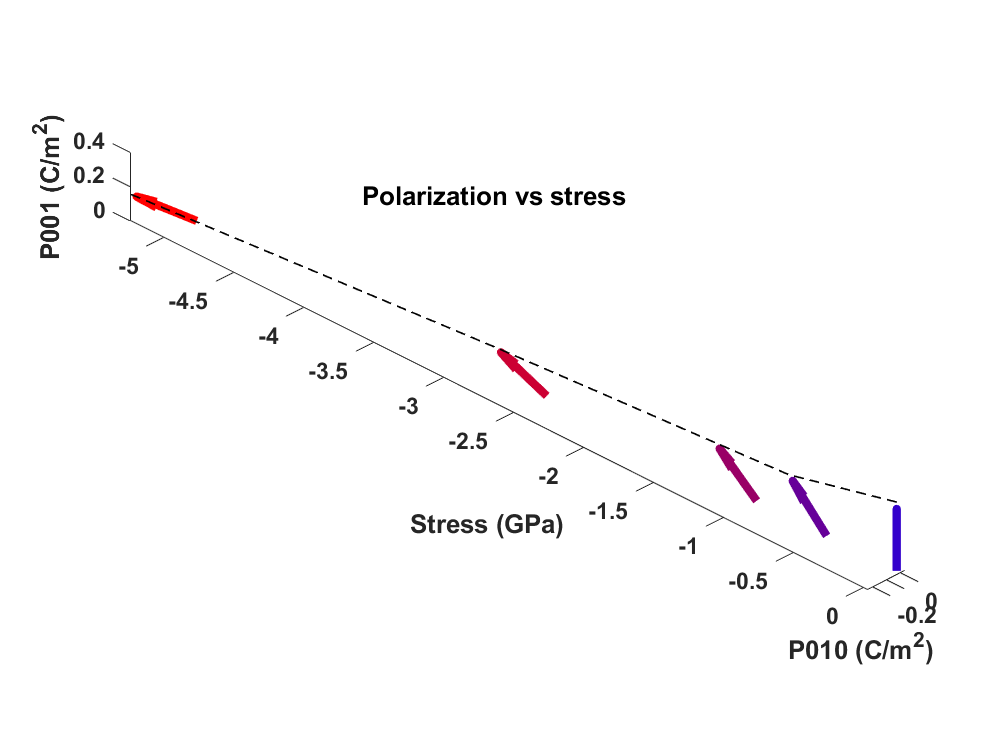}
    \caption{}\label{fig:deviceFig_dft_b}
  \end{subfigure}
  \\
  \begin{subfigure}[t]{0.5\linewidth}
    \includegraphics[width=\linewidth]{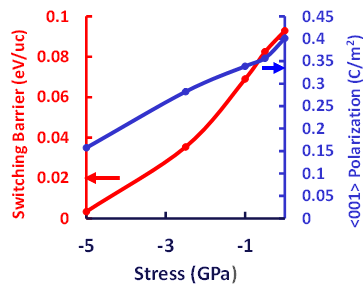}
    \caption{}\label{fig:deviceFig_dft_c}
  \end{subfigure}\hfill
   \begin{subfigure}[t]{0.5\linewidth}
    \includegraphics[width=\linewidth]{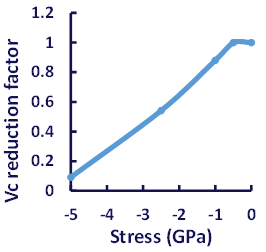}
    \caption{}\label{fig:deviceFig_dft_d}
  \end{subfigure}
  \caption{\textbf{Material mechanism of effect of stress on switching barrier and polarization in bulk Barium Titanium Oxide (BTO). \textbf{a} A 2x2x2 model of BTO unit cell; \textbf{b} Spontaneous polarization ($P_s$) rotation vs applied $\langle 001 \rangle$ compressive stress; \textbf{c} Corresponding reduction of switching barrier (right axis) and 001 component of polarization (left axis); \textbf{d} expected coercive voltage reduction from switching barrier and $P_s$ change.}}
\label{fig:deviceFig_dft}
\end{figure}

 We  use the first-principles Density-Functional Theory within the PBE Generalized Gradient approximation
(GGA), simulated with Quantum Espresso \cite{giannozzi2009quantum}, to study the effect of applied stress on electric polarization and polarization switching barrier in Barium Titanium Oxide (BTO). We employ ultra-soft pseudopotentials using the Rappe-Rabe-Kaxiras-Joannopoulos scheme \cite{rappe1990optimized}. We include as valence electrons the 3s2, 3p6, 3d2, 4s2  for Ti, 5s2, 5p6, and 6s2 for Ba within a semi-core treatment, and 2s2, 2p4 for O. The energy cutoff for the plane-wave expansion of single-particle Kohn–Sham states is 64 Ry, and for charge density
it is 782 Ry. We use a tetragonal 1x1x1 $BaTiO_3$ 5-atom unit cell with P4mm 
symmetry and 6x6x6 k-grid for 001 polymorphs with polarization pointing along $\langle 001 \rangle$ and $\langle 00\overline{1} \rangle$ in our calculations as detailed further. (1) We first perform the full lattice and atomic coordinates structural optimization at zero stress for two polarization polymorphs. (2) We impose strain on a converged lattice and atomic positions  for a given stress state by solving for strain using this mechanical equilibrium condition: 
\begin{equation}
\sigma_{i} = C_{ij} \epsilon_{j} ,
\label{eq:stress_strain_eqn}
\end{equation}
where $\sigma_{i}$, $\epsilon_{i}$ are stress and strain components in the Voight notation and C is the stiffness tensor with nonzero elements listed in Table  \ref{tab:material parameters}. Strained lattice and initial atomic coordinates of strained unit cell ($X^\prime$) are obtained from unstrained  lattice and coordinates ($X$) using this tensor transformation 
\begin{equation}
X^\prime = (1+\epsilon) X.
\label{eq:strain_transform_eqn}
\end{equation}
(3) We do relaxation of atomic coordinates under a fixed strained lattice. (4) To calculate the switching barrier between opposite polarization states we perform Automated Nudged Elastic Band (NEB)  to locate transition trajectory \cite{giannozzi2009quantum}. (5) To calculate electric polarization we first do self-consistent field simulation on relaxed lattices, and then do phonon calculations which utilize Density-Functional Perturbation Theory to get effective Born charges \cite{resta1992theory}. Then the electric polarization is computed as 
\begin{equation}
\Delta P_{i} = \frac{e}{\Omega} \sum\limits_{k,j} Z_{k,ij}^{*} \Delta R_{k,j},
\label{eq:born_charges_eqn}
\end{equation}
where $i,j$ are x, y, z components of vectors, $Z_{k,ij}^{*}$ is effective Born  charge component for an atom $k$, $\Omega$ is a unit cell volume. $\Delta R_{k,j}$ is a spatial component of displacement for an atom $k$ from its position in a reference  $\mathrm{Pm}\overline{3}\mathrm{m}$ centrosymmetric nonpolar unit cell of $BaTiO_3$.
Electric polarization rotates away from from $\langle 001 \rangle$ direction plotted in Fig. \ref{fig:deviceFig_dft_b}, its 001 component ($P_s$) and polarization switching barrier (Vb) reduce as a function of applied vertical 001 compressive stress in Fig. \ref{fig:deviceFig_dft_c}. This reduction of $P_s$ and Vb with stress forms a material basis for the effect of stress on coercive voltage ($V_c$) in ferroelectric BTO capacitors which we utilize in proposed devices. Coercive voltage $V_c$ scales as 
$\propto V_b / P_s$. 
Density-Functional Theory predicts the reduction of coercive voltage as a function of vertical compressive stress as shown in Fig. \ref{fig:deviceFig_dft_d}.

\section{Models and Equations}
\label{sec:ModelsEquations}
The free energy density \textit{G} in a  Barium Titanium Oxide (BTO) ferroelectric that is one of candidates for a device channel material is described by these state variables $v_i$: the three components $P_{i}$ of polarization, nine components of elastic stress tensor $\sigma_{ij}$, and a coupling between them \cite{wang2010temperature}. In the presence of electric field $\vec{E}$, the free energy is further lowered by the coupling between an electric field and polarization:
\begin{equation}
G = G_{P}-G_{\sigma}-G_{P,\sigma}-G_{P,E}-G_{dw,P},
\label{eq:LKfree energy abstract}
\end{equation}
where $G_{P}$ is the Landau potential expanded into 8th power of polarization components based on crystal symmetry, $G_{\sigma}$ is the elastic energy, $G_{P,\sigma}$ is the energy of electro-mechanical coupling, and $G_{P,E}$ is the energy lowering in the presence of electric field.
\begin{equation}
\begin{split}
G_{P} = \alpha_1 (P_1^2+P_2^2+P_3^2) + 
\alpha_{11} (P_1^4+P_2^4+P_3^4)+
\alpha_{12} (P_1^2P_2^2+P_1^2P_3^2+P_2^2P_3^2) \\
+\alpha_{123} (P_1^2P_2^2P_3^2) 
+\alpha_{111} (P_1^6+P_2^6+P_3^6) 
+\alpha_{112} (P_1^2(P_2^4+P_3^4)+  P_2^2(P_1^4+P_3^4) \\ +P_3^2(P_1^4+P_2^4))
+\alpha_{1111} (P_1^8+P_2^8+P_3^8)+
\alpha_{1122} (P_1^4P_2^4+P_1^4P_3^4+P_2^4P_3^4)+ \\
\alpha_{1112} (P_1^6(P_2^2+P_3^2)+  P_2^6(P_1^2+P_3^2)  +P_3^6(P_1^2+P_2^2))+ \\
\alpha_{1123} (P_1^4P_2^2P_3^2+P_1^2P_2^4P_3^2+P_1^2P_2^2P_3^4) 
\label{eq:LKfree energy gp}
\end{split}
\end{equation}
\begin{equation}
\begin{split}
G_{\sigma} = S_{11} (\sigma_1^2+\sigma_2^2+\sigma_3^2)/2
+ S_{12} (\sigma_1\sigma_2+\sigma_1\sigma_3+\sigma_2\sigma_3)+
S_{44} (\sigma_4^2+\sigma_5^2+\sigma_6^2)/2, 
\label{eq:LKfree energy gs}
\end{split}
\end{equation}
where $\sigma_{i}$ are stress components in the Voight notation.
\begin{equation}
\begin{split}
G_{P,\sigma} = Q_{11} (\sigma_1P_1^2+\sigma_2P_2^2+ \sigma_3P_3^2)+
Q_{12} (\sigma_1(P_2^2+P_3^2)+ \\ \sigma_2(P_1^2+P_3^2)+\sigma_3(P_2^2+P_1^2)) 
+Q_{44} (\sigma_4P_2P_3+\sigma_5P_1P_3+\sigma_6P_2P_1),
\label{eq:LKfree energy gps}
\end{split}
\end{equation}
where the tensor Q describes the electrostrictive coupling between stress and polarization. 
\begin{equation}
\begin{split}
G_{P,E} = P_1E_1+ P_2E_2 + P_3E_3,
\label{eq:LKfree energy gpe}
\end{split}
\end{equation}

In the presence of non-uniform polarization, free energy $G_{dw,P}$ of domain wall formation \cite{hu1998three}  is added:
\begin{equation}
\begin{split}
G_{dw,P} = \frac{1}{2}G_{11} (P_{1,1}^2+P_{2,2}^2+P_{3,3}^2)+
G_{12} (P_{1,1}P_{2,2}+P_{2,2}P_{3,3}+P_{3,3}P_{1,1})+ \\ 
+\frac{1}{2} G_{44}((P_{1,2}+P_{2,1})^2+(P_{2,3}+P_{3,2})^2+
(P_{3,1}+P_{1,3})^2) +\\
+\frac{1}{2} G_{44}^{\prime}((P_{1,2}-P_{2,1})^2+(P_{2,3}-P_{3,2})^2+
(P_{3,1}-P_{1,3})^2),  
\label{eq:LKfree energy dwe}
\end{split}
\end{equation}
where $P_{i,j}=\frac{\partial P_i}{\partial r_{j}}$ with $r_{j}$ being a Cartesian coordinate at a point. 
The gradient of total free energy density with respect to state variables, leads to their dynamics described by Landau-Khalatnikov equation \cite{landau2013course}:
\begin{equation}
\begin{split}
\rho_{v_i}\frac{\partial ({v}_i(t, \vec{r}))}{\partial t} = -\frac{\partial G}{\partial {v}_i},
\label{eq:LKfree energy dynamics}
\end{split}
\end{equation}
where $\rho_{v_i}$ is the damping parameter which controls the switching speed of the state variable change. 
For a uniform ferroelectric with spontaneous polarization $P_s$ under applied electric field, the polarization switching time is $\tau_{sw}=\rho_{P} P_s/E_c$ with $E_c$ being a ferroelectric coercive field. 
In the simulations presented in this paper we had solved time-dependent Landau-Khalatnikov equations for the three components of polarization $P_x$, $P_y$, and $P_z$. 

The ferroelectric BTO material has a complex energy landscape as shown in Fig. \ref{fig:deviceFig1_energybarriers_a} with six minima along principal crystal directions indicating six spontaneous polarization states of the material. To model a paraelectric channel which has zero spontaneous polarization at the zero electric field, we had developed a simplified energy landscape model by keeping only $\alpha_1$ and $\alpha_{11}$ coefficients nonzero in Eq. \ref{eq:LKfree energy gp} and calibrated this model to the stress response of the material in the nanoindenter simulations which we discuss in the main text and in Sec. \ref{device_nidetails}.

For stress ($\sigma$) response we used the equilibrium Solid Mechanics module in COMSOL to solve for the three spatial components of mechanical displacement $\mathbf{u}$. We include this Solid Mechanics equation for completeness here: 
\begin{equation}
\begin{split}
\nabla \cdot \left(F \sigma  \right)^{T} = 0, \text{where} \, F=\left[\mathbf{I} + \nabla \mathbf{u}\right].
\label{eq:SMcomsolequation}
\end{split}
\end{equation}
A channel material is modeled with anisotropic compliance  $S$ and electrostrictive $Q$ tensors, where electrostrictive strain $\varepsilon_{i}=Q_{ij}P^{2}_{j}$ (in Voight notation) is a source of inelastic strain which couples the solid mechanics and the polarization dynamics physics. A stressor material is assumed piezoelectric modeled with equations below using the stress-charge form:
\begin{equation}
\begin{split}
\sigma_i = c_{ij} \varepsilon_{j} - e_{ki} E_k, \\ 
D_k = e_{kj} \varepsilon_j + \epsilon_{kl} E_l,
\label{eq:SMpiezo}
\end{split}
\end{equation}
where $C$ is a stiffness tensor (an inverse of a compliance tensor), $e_{ki}$ are piezoelectric stress coefficients, $\epsilon_{kl}$ is a dielectric permittivity tensor, $E_k$ and $D_{k}$ are electric field and displacement components, respectively. A reference system of coordinates is chosen for a piezoelectric such that its polar axis is oriented up parallel to the heterostructure growth direction. All electrode layers are described as isotropic elastic materials. An additional strain corresponding to an in-plane lattice mismatch between a given layer material and an assumed  substrate is included when indicated as an initial strain tensor \cite{comsol2025}. For example, for the full strain due to lattice mismatch of the stack pseudomorphically  grown on  DyScO3, dysprosium scandate substrate, compressive strains of 1.35 \% in BTO and 2.38 \% in PZT are included as initial strains for the respective layers in the device. We use substrate strains as tuning parameters to set a baseline width of the channel hysteresis characteristic to achieve an optimal device performance. For memory devices two-dimensional simulations we assumed 20 \% of the full substrate strain which physically corresponds to a partial relaxation of the full strain. In the three-dimensional simulations and in logic devices two-dimensional simulations, we assumed that substrate strain is relaxed fully.

\begin{figure}[ht!]
  \centering
\begin{subfigure}[t]{0.5\linewidth}
    \includegraphics[width=\linewidth]{figs/deviceFigures/deviceFig1_energybarriers_e.png}    
    \caption{}\label{fig:deviceFig_simflow_a}
  \end{subfigure}\hfill
  \begin{subfigure}[t]{0.5\linewidth}
    \includegraphics[width=\linewidth]{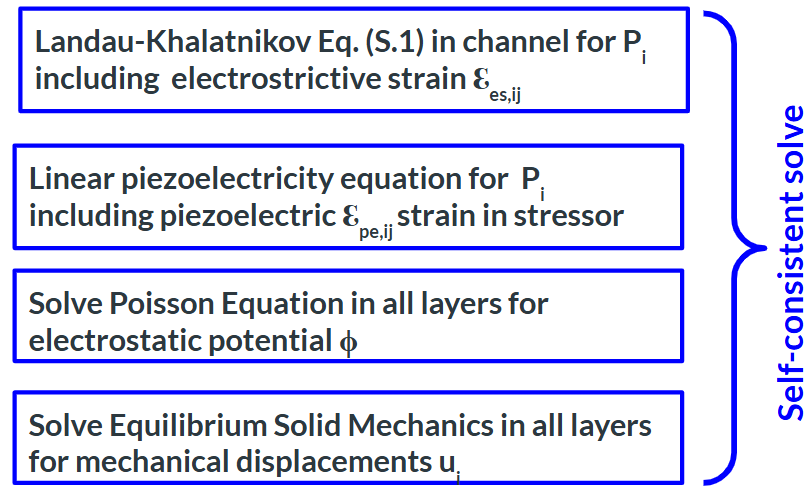}
    \caption{}\label{fig:deviceFig_simflow_b}
  \end{subfigure}
 
  \caption{\textbf{Multi-physics Simulation flow.}
           }
\label{fig:deviceFig_simulationflow}
\end{figure}

The simulation flow for a device simulated with multi-physics model in COMSOL is summarized in Fig. \ref{fig:deviceFig_simulationflow}.
We solve self-consistently time-dependent Landau-Khalatnikov (LK) equation Eq. \ref{eq:LKfree energy dynamics} for polarization in the ferroelectric or paraelectric channel and time-dependent Poisson equation for all layers. The equilibrium mechanics is solved for all layers as discussed above where a piezoelectric effect in the stressor or an electrostrictive effect in the channel couples electrostatics through included polarization charge with mechanics and polarization dynamics. The mechanical and electrical displacement is taken continuous across material interfaces. In two-dimensional simulations, which are shown in Fig. \ref{fig:deviceFig_simulationflow}, periodic boundary conditions for mechanical displacement and electrostatic potentials are set on side boundaries, modeling the limit of large area devices. In the two-dimensional simulations we use a plane strain approximation in COMSOL which assumes isotropic  strain in out-of-plane direction. For an optimal stress transfer the interfaces between channel or stressor layers with drain and gate electrode materials are mechanically pinned, by fixing the displacement to zero at those boundaries in the two-dimensional simulations. This can be thought of being achieved by pinning large area devices by the substrate on the bottom, and by wafer bonding process on the top of the device stack as discussed in the main text. All other surfaces that are not periodic or pinned, have free surface boundary conditions in the two-dimensional simulations with zero normal stress at those boundaries. The simulated three dimensional nanopillar structures are only pinned at the substrate as described in Section \ref{device_3ddev} with all other surfaces kept as free surfaces.

Selected materials and geometry parameters are listed in 
Table~\ref{tab:material parameters}. 
All other parameters in these equations are taken either from Ref. \cite{wang2010temperature} or from the standard Comsol material library \cite{comsol2025}. Geometry parameters which change for specific devices are stated in appropriate simulations descriptions.

\begin{table}[!ht] 
\centering 
\caption{Material parameters by device layers.}
\label{tab:material parameters}
\begin{tabularx}{\textwidth}{
    |l|X}
    \hline
\textbf{Layers} & 
\textbf{Gate or Word Line (WL), Drain or Bit Line (BL) }  \\
 \end{tabularx}   
\begin{tabularx}{\textwidth}{
    |l|X|c|} 
\hline
\textbf{Parameter} & 
\textbf{Value } & 
\textbf{Description} \\ 
\hline
\textbf{material} &
 Aluminum &
Properties taken  from the COMSOL Material Library \cite{comsol2025}\\
\hline
\textbf{thickness} &
 10 nm &
vertical height of a layer \\
\hline
\textbf{width} &
 20 nm &
horizontal device width used in 2D simulations\\
\hline
\textbf{Y} &
 68.9 GPa &
Young's Modulus \\
\hline
\end{tabularx}
\begin{tabularx}{\textwidth}{
    |l|X}
    \hline
\textbf{Layers} & 
\textbf{Source or Plate Line (PL) }  \\
 \end{tabularx}   
\begin{tabularx}{\textwidth}{
    |l|X|c|} 
\hline
\textbf{Parameter} & 
\textbf{Value } & 
\textbf{Description} \\ 
\hline
\textbf{material} &
 Aluminum &
Properties taken  from the COMSOL Material Library \cite{comsol2025}\\
\hline
\textbf{thickness} &
 2 nm &
vertical height of a layer in 2D simulations\\
\hline
\textbf{width} &
 20 nm &
horizontal device width used in 2D simulations\\
\hline
\textbf{Y} &
 138.8 GPa &
Young's Modulus of Aluminum Metal Matrix Composite is used \cite{ekpu2024characterising}\\
\hline
\end{tabularx}
\begin{tabularx}{\textwidth}{
    |l|X}
    \hline
\textbf{Layers} & 
\textbf{Channel: Ferroelectric (FE) or Paraelectric (PAE) }  \\
 \end{tabularx}   
\begin{tabularx}{\textwidth}{
    |l|X|c|} 
\hline
\textbf{Parameter} & 
\textbf{Value } & 
\textbf{Description} \\ 
\hline
\textbf{material} &
 BTO &
Properties specified below \cite{wang2010temperature}\\
\hline
\textbf{thickness} &
 10 nm &
vertical height of a layer in 2D simulations\\
\hline
\textbf{width} &
 20 nm &
horizontal device width used in 2D simulations\\
\hline
\textbf{S11} &	$9.07 \times 10^{-12}$ 1/Pa & Compliance tensor element \\
\hline
\textbf{S12} &	$-3.19 \times 10^{-12}$ 1/Pa & Compliance tensor element \\
\hline
\textbf{S44} &	$8.20 \times 10^{-12}$ 1/Pa & Compliance tensor element\\
\hline
\textbf{Q11} &	0.11 m$^{4}$/C$^{2}$ & Electrostrictive tensor element \\
\hline
\textbf{Q12} &	-0.045 m$^{4}$/C$^{2}$ & Electrostrictive tensor element \\
\hline
\textbf{Q44} &	0.029 m$^{4}$/C$^{2}$ & Electrostrictive tensor element\\
\hline
\textbf{G11} &	$5.0 \times 10^{-10}$ m$^{4}$/[N C$^{2}]$ & Domain wall energy tensor parameter \\
\hline
\textbf{G12} &	0 m$^{4}$/[N C$^{2}]$ & Domain wall energy tensor parameter \\
\hline
\textbf{G44} &	$1.0 \times 10^{-11}$ m$^{4}$/[N C$^{2}$]& Domain wall energy tensor parameter\\
\hline
$\mathbf{{\rho}_{P, FE}}$ &	0.001 m s/F
 & Landau damping for polarization dynamics in FE \\
 \hline
$\mathbf{{\rho}_{P, PAE}}$ &	0.0001 m s/F
 & Landau damping for polarization dynamics in PAE \\
  \hline
$\mathbf{{\alpha}_{1, PAE}}$ &	$2.31 \times 10^{7}$ m/F
 & Landau coefficient ${\alpha}_{1}$  in PAE \\
 \hline
$\mathbf{{\alpha}_{11, PAE}}$ &	25879
 m$^{6}$ N/ C$^{4}$
 & Landau coefficient ${\alpha}_{11}$  in PAE \\ 
\hline
$\mathbf{{\varepsilon}_{BTO}}$ &	136 & Static dielectric constant in units of  $\varepsilon_{0}$\\
\hline
$\mathbf{{\varepsilon}_{\parallel}}$ &	-0.0135 & In-plane strain of BTO   on DSO substrate \\
\hline
\end{tabularx}
\begin{tabularx}{\textwidth}{
    |l|X}
    \hline
\textbf{Layers} & 
\textbf{Stressor: Piezoelectric (PE)  }  \\
 \end{tabularx}   
\begin{tabularx}{\textwidth}{
    |l|X|c|} 
\hline
\textbf{Parameter} & 
\textbf{Value } & 
\textbf{Description} \\ 
\hline
\textbf{material} &
 PZT &
Properties taken  from the COMSOL Material Library \cite{comsol2025}\\
\hline
\textbf{thickness} &
 10 nm &
vertical height of a layer in 2D simulations\\
\hline
\textbf{width} &
 20 nm &
horizontal device width used in 2D simulations\\
\hline
\textbf{eES31=eES32} &	-5.2028  C/m$^{2}$ & Piezoelectric stress coupling tensor element \\
\hline
\textbf{eES33} &	15.0804 C/m$^{2}$ & Piezoelectric stress coupling tensor element \\
\hline
\textbf{eES24=eES15} &	12.7179 C/m$^{2}$ & Piezoelectric stress coupling tensor element \\
\hline
$\mathbf{{\varepsilon}_{33}}$ &	199 & Dielectric constant $\mathbf{{\varepsilon}_{33}}$ in units of  $\varepsilon_{0}$\\
\hline
$\mathbf{{\varepsilon}_{\parallel}}$ &	-0.0238
 & In-plane strain of PZT   on DSO substrate \\
\hline
\textbf{dpz} &	1 - 6 & Scaling coefficient multiplying eES tensor \\
\hline
\end{tabularx}
\end{table}

\begin{figure}[ht!]
  \centering
  \begin{subfigure}[t]{0.5\linewidth}
    \includegraphics[width=\linewidth]{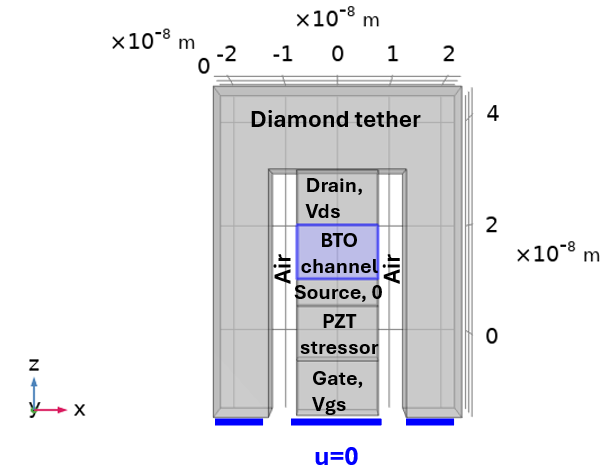}
    \caption{}\label{fig:deviceFig_3ddev_a}
  \end{subfigure}\hfill
  \begin{subfigure}[t]{0.5\linewidth}
    \includegraphics[width=\linewidth]{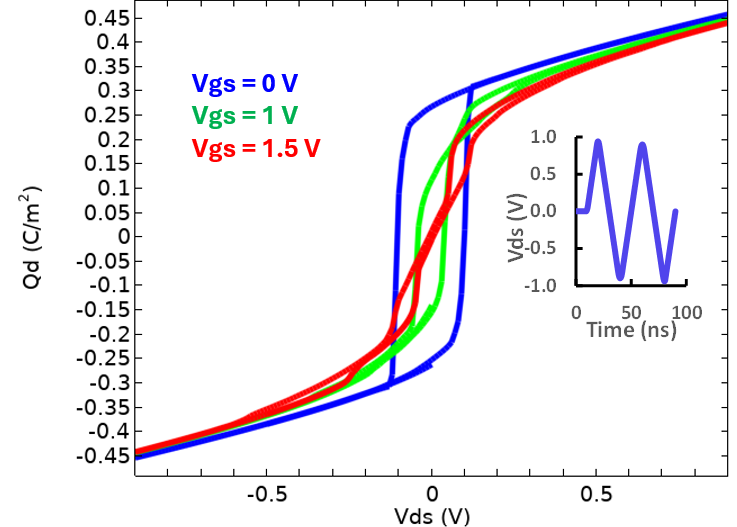}
    \caption{}\label{fig:deviceFig_3ddev_b}
  \end{subfigure}
  \\
  \begin{subfigure}[t]{0.5\linewidth}
    \includegraphics[width=\linewidth]{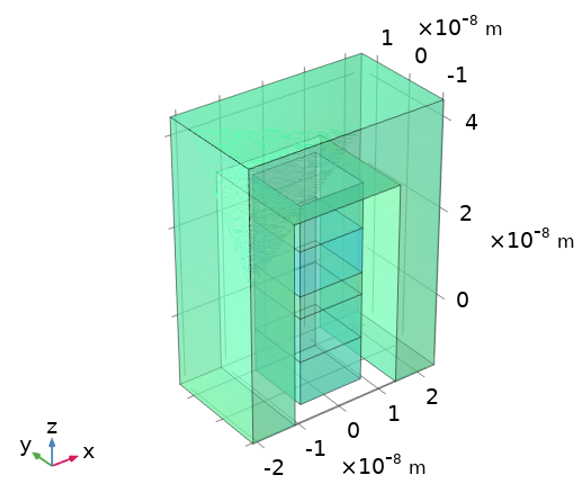}
    \caption{}\label{fig:deviceFig_3ddev_c}
  \end{subfigure}\hfill
   \begin{subfigure}[t]{0.5\linewidth}
    \includegraphics[width=\linewidth]{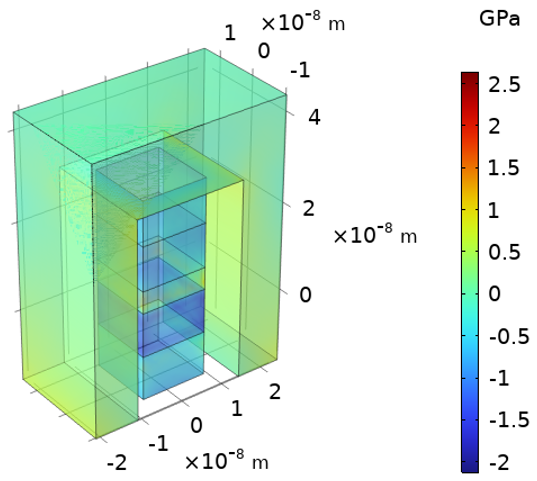}
    \caption{}\label{fig:deviceFig_3ddev_d}
  \end{subfigure}
  \caption{\textbf{Three-dimensional design of transcapacitor with tether and airgaps. \textbf{a} An xz crossection of the nanopillar; \textbf{b} Drain charge $Q_D$ hysteresis vs. 
  $V_{DS}$
  signal (in the inset) at 
  $V_{GS} = 0 V, 1 V, 1.5 V$;  Vertical stress (Szz) distribution in the device at
  $V_{GS}=0V$ 
  in \textbf{c} and at 
  $V_{GS}=1.5V$ 
  in \textbf{d}. } }
\label{fig:deviceFig_3ddev}
\end{figure}

\begin{figure}[ht!]
  \centering
    \includegraphics[width=7cm]{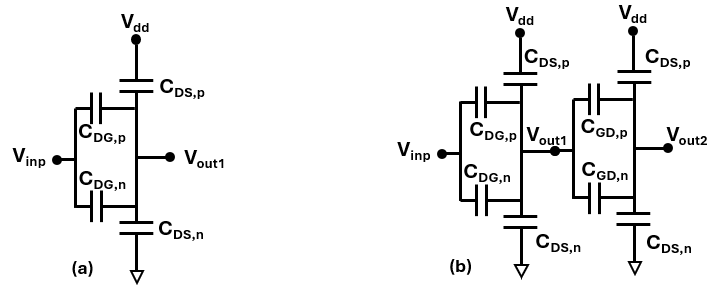}
  \caption{\textbf{ A circuit model of an unloaded transcapacitor inverter in (a) and of an inverter driving another inverter in (b).}}
  \label{fig:deviceFig_circuitinv}
\end{figure}

\section{Piezo-transcapacitor device operation in nanopillar/tether geometry and a circuit model of an unloaded inverter}
\label{device_3ddev}

Individual transcapacitor devices for memory or logic applications can be realized using nanopillar with side airgaps design simulated in Fig. ~\ref{fig:deviceFig_3ddev}. The 15 $\times$ 15 nm$^2$  area vertical nanopillar consists of 10 nm  metallic Aluminum bottom gate and top drain electrodes (with Young's Modulus of 69 GPa), 10 nm PZT stressor poled vertically up along z-direction, 5 nm middle source electrode metallic hardened composite Aluminum material (with Young's
Modulus of 138 GPa), 10  nm BTO ferroelectric channel, 5 nm airgaps on each side along the width (x-direction in the figure), and a pi-shaped tether with 15 nm on top of drain electrode and 10 nm on airgap sides (see xz-crossection of the simulated device in Fig. \ref{fig:deviceFig_3ddev_a}). Diamond material with Young's Modulus of  1050 GPa is used for the tether. The otherwise free standing structure is fixed to the substrate at tether sides and a bottom gate electrode with the boundary condition of mechanical displacement u = 0. The substrate is simulated to be lattice matched to channel BTO. Gate voltage Vgs is applied to the gate electrode, source is kept at 0V, source-drain voltage Vds is applied to the drain electrode. For Vgs = 0 V, the nanopillar is stress-free as simulated in Fig. \ref{fig:deviceFig_3ddev_c}. Under an application of the gate voltage of Vgs = 1.5 V, the created stress in the PZT stressor is transferred to the BTO channel, reaching  values of -1 GPa for the vertical stress.   The drain charge channel response to the Vds signal in the inset in Fig. \ref{fig:deviceFig_3ddev_b} evolves from a full ferroelectric hysteresis 
with $V_c=0.1 V$ at 
$V_{GS}=0 V$
to the fully collapsed hysteresis with $V_c=0V$
at $V_{GS}=1.5V$
in Fig. \ref{fig:deviceFig_3ddev_b}. We have obtained similar device characteristics for a range of device areas from 15 $\times$ 15 nm$^2$ to 30 $\times$ 30 nm$^2$ (not shown). This demonstrated in simulation device functionality for a finite size device forms a basis for logic and memory applications of transcapacitors. 

A circuit model of an unloaded inverter is shown in Fig. \ref{fig:deviceFig_circuitinv} (a).
A circuit model of it consists of p-device and n-device  source-drain capacitances $C_{DS,p}$, and  $C_{DS,n}$ of channel capacitors connected in series. The effect of stressors on their electro-mechanically coupled channel charges is expressed through the effective capacitances $C_{DG,p}$, and  $C_{DG,n}$ of two capacitors connected in parallel. In a circuit of an inverter driving another inverter an output drain charge from the first inverter becomes the gate input charge of the second inverter in Fig. \ref{fig:deviceFig_circuitinv} (b).

\begin{figure}[ht!]
  \centering
  \begin{subfigure}[t]{0.5\linewidth}
    \includegraphics[width=\linewidth]{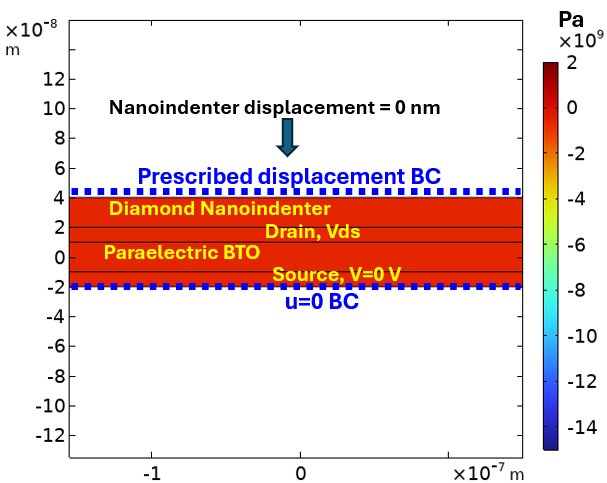}
    \caption{}
    \label{fig:deviceFig_NI_a}
  \end{subfigure}\hfill
  \begin{subfigure}[t]{0.5\linewidth}
    \includegraphics[width=\linewidth]{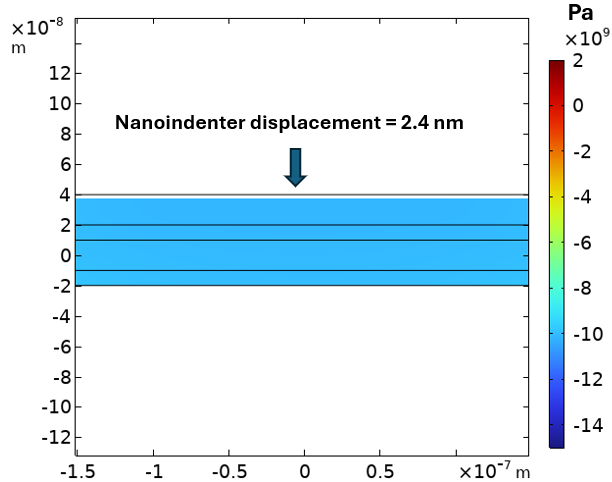}
    \caption{}
    \label{fig:deviceFig_NI_b}
  \end{subfigure}
  \\
  \begin{subfigure}[t]{0.5\linewidth}
    \includegraphics[width=\linewidth]{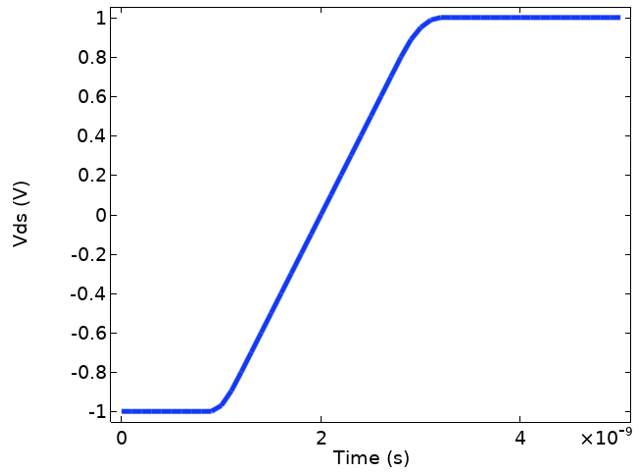}
    \caption{}
    \label{fig:deviceFig_NI_c}
  \end{subfigure}\hfill
   \begin{subfigure}[t]{0.5\linewidth}
    \includegraphics[width=\linewidth]{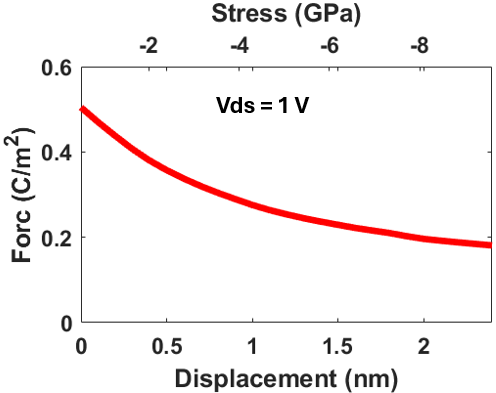}
    \caption{}
    \label{fig:deviceFig_NI_d}
  \end{subfigure}
  \caption{\textbf{Nanoindenter experiment simulations for paraelectric BTO. \textbf{a-b}: A vertical stress across the stack for 0 nm and 2.4 nm vertical displacement down of the nanoindenter; \textbf{c}: a Vds voltage pulse signal for each FORC simulation for a given magnitude of offset voltage and a nanoindenter displacement; \textbf{d}: collected FORC charge at a fixed voltage pulse magnitude vs nanoindenter displacement (top x-axis) and vertical stress (bottom x-axis).}}
\label{fig:deviceFig_NI}
\end{figure}

\section{Nanoindenter simulations}
\label{device_nidetails}

We reproduced the experimental charge response behavior by performing simulations of the paraelectric stack under vertical compressive stress applied by the nanoindenter (overlayed lines in Fig.~\ref{fig:demo_stress_effect}B) in Comsol. We considered a heterostructure stack  consisting of the four layers of 10 nm Aluminum source electrode, 20 nm BTO paraelectric,  10 nm Aluminum drain electrode and 20 nm Diamond nanoindenter as shown in Fig. \ref{fig:deviceFig_NI_a}.  The mechanical properties of simulated electrodes are expected to be approximating properties of experimental electrode layers. The whole stack is pinned (mechanical displacement u=0 boundary condition) to the bottom DSO substrate. The substrate strain is not included in these materials here, because we assume that under growth condition of our experiments, the paraelectric BTO is mostly relaxed. The area of simulated two-dimensional devices is 500 nm wide (x-direction) and 1 micron in depth (z-direction). We included the 5 nm airgaps on each side of the width. Air gaps do not impact results for this wide structure, but improve the convergence of the solution. For BTO polarization response, we solve Landau-Khalatnikov equation developed for a paraelectric as described in previous sections. Source-drain voltage is applied by grounding the source electrode, and applying Vds voltage to the drain electrode. The effect of the nanoindenter force is modeled using the prescribed vertical displacement boundary condition in equilibrium solid mechanics set on the top of the structure: $u_{y} = u_{0y}$. We vary vertical down displacement $u_{0y}$ from 0 to 2.5 nm, obtaining a range of compressive vertical  stresses across the stack from 0 to -10 GPa. For example, displacing nanoindenter by 2.4 nm results in -10 GPa as simulated in \ref{fig:deviceFig_NI_b}. We simulated Forc characteristics by applying source-drain voltage pulses in Fig. \ref{fig:deviceFig_NI_c} and collecting the integrated charge for various nanoindenter displacements and voltage pulse magnitudes. FORC charge for a 1V magnitude of voltage pulse is plotted vs displacement on the bottom x-axis, shown with the corresponding vertical stress on the top of x-axis in Fig. \ref{fig:deviceFig_NI_d}. 

\section{Memory details}
\label{sec:memory_details}



Table ~\ref{tab:memory_params_suppl} shows the parameters used for device simulations with 
COMSOL in memory simulations. Table  ~\ref{tab:2x2ArrayStates} shows the initial and final states and read-out charge for memory operations in the simulated 2x2 array.  


\begin{table}[ht]
\captionsetup{width=\linewidth}   
\caption{Parameters used for COMSOL device simulations for memory.}
\label{tab:memory_params_suppl}
\centering
\begin{tabular}{lll}
\toprule
\multicolumn{3}{c}{\textbf{Memory Parameters}}\\
\midrule
Rho                     & \multicolumn{2}{l}{\quad 0.001} \\ 
$P_s$                      & \multicolumn{2}{l}{\quad 0.25 C/m2} \\

Dielectric constant PZT & \multicolumn{2}{l}{\quad 198.6} \\ 
Dielectric constant BTO & \multicolumn{2}{l}{\quad 136} \\
Strainscale             & \multicolumn{2}{l}{\quad 0.2} \\ 
DPZ                     & \multicolumn{2}{l}{\quad 1} \\ 
Thickness BTO               & \multicolumn{2}{l}{\quad 10 nm} \\ 
Thickness PZT               & \multicolumn{2}{l}{\quad 10 nm} \\  
Ramp rate               & \multicolumn{2}{l}{\quad 0.1 ns} \\ 
WL Voltage                      & \multicolumn{2}{l}{\quad 1.25 Volt} \\ 
BL Voltage                     & \multicolumn{2}{l}{\quad 0.14 Volt} \\ 
Vc\_original            & \multicolumn{2}{l}{\quad 0.22 Volt} \\ 
With gate $V_c$            & \multicolumn{2}{l}{\quad $< 5$\,mV} \\ 
\bottomrule
\end{tabular}
\end{table}

\begin{table}[ht]
\caption{Table with the polarization states corresponding to a 2x2 memory read non-destructive operation}
\label{tab:2x2ArrayStates}
\centering
\begin{tabular}{lcccc}
\toprule
Polarization state & (11) & (12) & (21) & (22) \\
\midrule
Initial          & 0          & 1          & 1          & 0          \\
After Write      & 1          & 0          & 0          & 1          \\
Delta Q at Read  & $-\Delta Q$ & $+\Delta Q$ & $+\Delta Q$ & $-\Delta Q$ \\
\bottomrule
\end{tabular}
\end{table}
\section{Analytical Treatment of Piezoelectric Transcapacitors}
\label{sec:PiezoTpolModel}

The general expression of the piezoelectric coupling in a material is as follows and includes 
$D$ =electric displacement, 
$E$ =electric field, 
$S$=strain, 
$T$=stress, 
$C$=elastic stiffness, 
which is the inverse of the elastic compliance at constant electric field, $s^E$ , 
$d$=piezoelectric, and 
$\epsilon^T$=dielectric coefficient at constant stress. 
Vectors are designated by \{\} and tensors by []:
\begin{eqnarray}
\{ S \} = [s^E ] \{ T \} + [ d^t ] \{ E \} \\
\{ D \} = [d] \{ T \} + [\epsilon^T ] \{ E \}
\end{eqnarray}
where the subscript 't' designates the transpose of the matrix.
The equations for single components of vectors and tensors can be related by approximate equations relating to the gate and channel capacitors as per Fig.~\ref{fig:1}c.
\begin{eqnarray}
S_g=T_g/C_g+d_g E_g \\
D_g=d_gT_g+\epsilon _g E_g \\
S_c=T_c/C_c+d_c E_c \\
D_c=d_cT_c+\epsilon _cE_c
\end{eqnarray}
The variable voltage is applied to the gate capacitor. It results in the creation of stress in it. According to the equations below, we assume that the stress is transferred without a loss to the channel capacitor. Constant voltage is applied to the channel capacitor, and this $E_c$ can be set to zero without loss of generality. Also, the strain can be clamped in the gate capacitors and thus $S_g$ can be set to zero:
\begin{eqnarray}
\Delta E_c = 0 \\
\Delta T_c=\Delta T_g \\
h_g \Delta S_g = - h_c \Delta S_c \\
\Delta V_g = - h_g \Delta E_g \\
\Delta V_c = h_c \Delta E_c \\
\Delta Q_g = A \Delta D_g \\
\Delta Q_d = - A \Delta D_c
\end{eqnarray}
As a result, some charge will flow out of the channel capacitor even at a constant voltage applied to it. 
The above solution results in the following figures of merit for the piezo transcapacitor.

Transcapacitance
\begin{equation}
TC = 
\frac{\Delta Q_d}{\Delta V_g}
= - \frac
{A d_c  d_g}
{\left(h_g/C_g+h_c / C_c \right)}
\end{equation}

Voltage gain
\begin{equation}
VG = 
\frac{\Delta V_c}{\Delta V_g}
= \frac
{d_c  d_g}
{ 2 \epsilon _c \epsilon_0
\left(1/C_c+h_g / (h_c C_g) \right)}  
\end{equation}

Charge gain
\begin{equation}
CG = 
\frac{\Delta D_c}{\Delta D_g}
= \frac
{d_c  d_g}
{ d_g^2 - \epsilon _g \epsilon_0
\left(1/C_g+h_c / (h_g C_c) \right)}
\end{equation}

Current modulation
\begin{equation}
CM =  \frac{2V_c}{log_{10}
\left(2 P_s / (t_{sw} J) \right) }
\end{equation}

The material parameters 
pertinent to the gate material such as PZT,
and the channel material such as BTO,
used in the calculations as well as the resulting values of the figures of merit are collected in
Table~\ref{tab:piezoFOM}.

\begin{table}[ht!] 
\centering
\caption{
Parameters and figures of merit for a piezo transcapacitor
}
\label{tab:piezoFOM}
\begin{tabularx}{\textwidth}{
| >{\hsize=1\hsize\raggedright\arraybackslash}X
| >{\hsize=0.5\hsize\raggedcenter\arraybackslash}X
| >{\hsize=0.5\hsize\raggedcenter\arraybackslash}X
| >{\hsize=0.5\hsize\raggedcenter\arraybackslash}X |
}
\hline
Quantity & Symbol & Value & Unit  \\
\hline
Polarization & $P_s$  & 0.25 & $C/m^2$   \\
\hline
Piezoelectric coefficient, channel & $d_c$ & 200 & $pC/N$  \\
\hline
Piezoelectric coefficient, gate & $d_g$ & 300 & $pC/N$ \\
\hline
Elastic stiffness, channel & $C_c$ & 170 & GPa  \\
\hline
Elastic stiffness, gate & $C_g$ & 150 & GPa  \\
\hline
Dielectric constant, channel & $\epsilon_c$ & 150 &   \\
\hline
Dielectric constant, gate & $\epsilon_g$ & 700 &   \\
\hline
Switching time & $t_{sw}$ & 10 & ps  \\
\hline
Leakage current density & $J$ & 50 & $A/m^2$  \\
\hline
Coercive voltage & $V_c$ & 40 & mV  \\
\hline
Thickness, channel & $h_c$ & 10 & nm  \\
\hline
Thickness, gate & $h_g$ & 10 & nm  \\
\hline
Transcapacitance per area & $TC/A$ & 0.48 & $F/m^2$  \\
\hline
Voltage gain & $VG$ & 1.8 &   \\
\hline
Charge gain & $CG$ & 4.9 &   \\
\hline
Current modulation & $CM$ & 8.9 & $mV/decade$  \\
\hline
\end{tabularx}
\end{table}

\section{More Details on Various Transcapacitors}
\label{sec:DetailsVariTpols}

\begin{figure}[ht!]
\centering
\includegraphics[width=\textwidth]{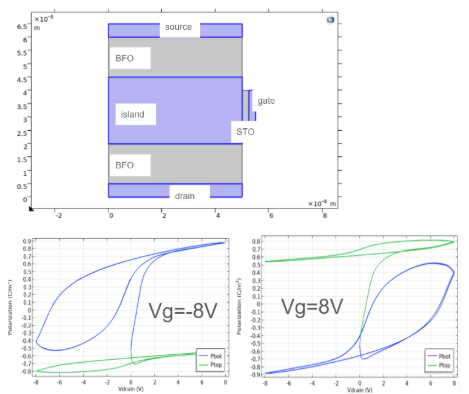}
\caption{
Operating principle of the floating island transcapacitor. (Left) Scheme of stacked capacitors and electrodes. (Right) Polarization in top and bottom capacitors vs. gate voltage.}
\label{fig:tpol_float_island}
\end{figure}

Let us describe the types of transcapacitors in more detail.
A) A floating island transcapacitor Fig.~\ref{fig:tpol_float_island}
consists of two polarized layers and a floating electrode between them. The potential in the floating island is partially controlled by the gate. Varying the gate voltage leads to a shift and a change in the shape of the charge-voltage curve of the two capacitors. Its advantage is the conceptual simplicity that allowed us to demonstrate transcapacitance in it (see the following section). Its disadvantages are that the charge gain is likely not to exceed 1 and the voltage inversion does not seem feasible.

\begin{figure}[ht!]
\centering
\includegraphics[width=5cm]{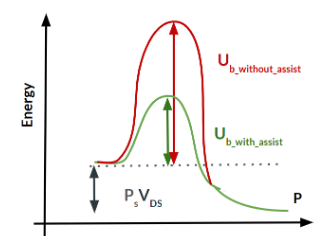}
\caption{
Concept of changing the coercive voltage by the gate via lowering the energy barrier between the switched states.}
\label{fig:barrier_lowering}
\end{figure}

\begin{figure}[ht!]
\centering
\includegraphics[width=\textwidth]{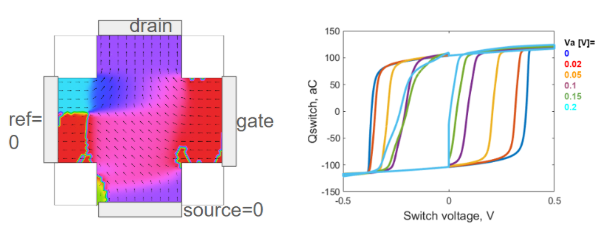}
\caption{
Operating principle of the field assist transcapacitor. (Left) the scheme of electrodes and polarization distribution in the ferroelectric layer. (Right) Charge at the drain electrode vs. the drain-to-source voltage at various values of the gate (‘assist’) voltage.}
\label{fig:tpol_field_assist}
\end{figure}

B) A field assist transcapacitor operates on the basis of the influence of an additional field, created by the gate, on the coercive value of the switching field, applied between the source and the drain. The assist effect relies on a non-trivial landscape of energy of the ferroelectric across the direction of the polarization. The assist field lowers the switching energy barrier 
(Fig.~\ref{fig:barrier_lowering}) and thus was predicted in simulations to shift the charge-voltage curve and narrow its hysteresis width 
(Fig.~\ref{fig:tpol_field_assist}). 
We were able to measure the transcapacitance in our samples \ref{sec:experim_demo}. The charge gain ~3 was predicted in simulations. However, voltage inversion might not be feasible since this type of transcapacitors still relies on the electrostatic effect. A promising direction for field-assist transcapacitors is to explore structures with multiple different ferroelectric layers.

\begin{figure}[ht!]
\centering
\includegraphics[width=\textwidth]{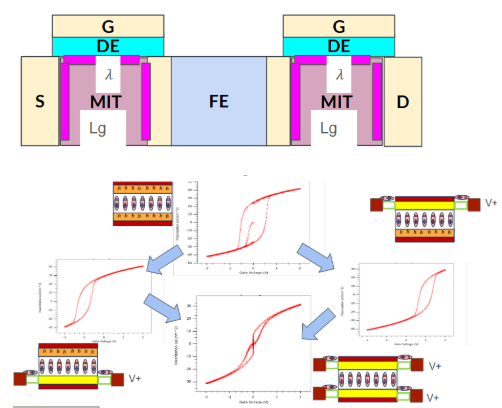}
\caption{
Operating principle of the field-effect (metal-insulator transition) transcapacitor. (Top) Scheme of the ferroelectric and surrounding layers controlled by gates. (Bottom) Change of the drain-to-source polarization vs. voltage in the states of the surrounding layers are switched between conducting and undulating states.}
\label{fig:tpol_mit}
\end{figure}

C) A field-effect (metal-insulator transition) transcapacitor relies on the gate voltage to change the conductance of a semiconducting layer adjacent to a polarizer layer (Fig.~\ref{fig:tpol_mit}). An application of the gate voltage changes the carrier density in the channel layer at the surface or throughout the volume of the semiconductor (depending on its geometry). If a metal insulator transition is induced, the impact will be stronger, but it is not required. The change of a layer from conducting to insulating affects the field distribution in the overall stack of materials between the source and the drain. This results in a shift and/or narrowing of the hysteresis (Fig.~\ref{fig:tpol_mit}). Analytical estimates predicted the charge gain $>10$ and the voltage inversion. The concern about this scheme is that the transcapacitance effect has not yet been experimentally demonstrated. The key step in developing this transcapacitor is the integration of the field-effect gates and the ferroelectric layer in a monolithic process.

\begin{figure}[ht!]
\centering
\includegraphics[width=5cm]{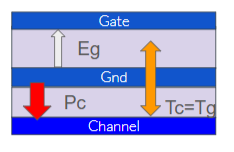}
\caption{
Operation of a piezo-gated transcapacitor.  Scheme of a piezoelectric transcapacitor including the gate ‘g’ and channel ‘c’ capacitors.}
\label{fig:tpol_piezo}
\end{figure}

D) A piezoelectric transcapacitor relies on the transfer of stress between two capacitors (Fig.~\ref{fig:tpol_piezo}). The application of voltage to one piezoelectric capacitor results in a change of strain in it. This creates stress which is exerted on an adjacent piezoelectric capacitor. The stress can change the polarization in the second capacitor and thus produce a change of the electrode charge even without a change of voltage applied to the second capacitor.
Our simulations predicted gain $>3$ and the voltage inversion in this type of a transcapacitor. A concern about this type is the efficiency of the stress transfer and the effect of the strain clamping by the substrate. A key step in its development would be the process incorporating multiple nanoscale-thickness layers with sufficient piezoelectric coefficients. 

\begin{figure}[ht!]
\centering
\includegraphics[width=\textwidth]{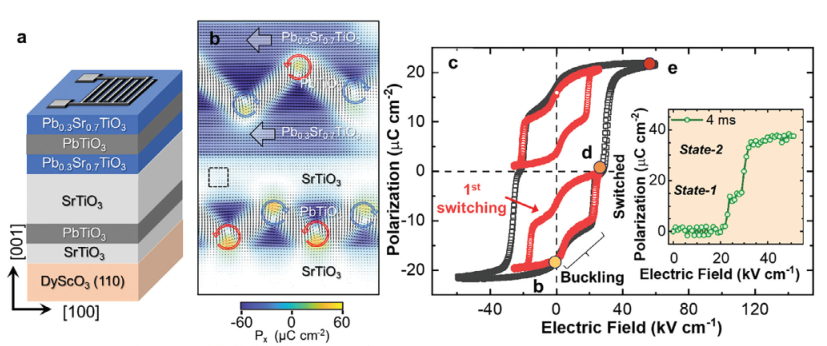}
\caption{
Operating principle of the phase transition transcapacitor. (Left) Pattern of polarization in the ferroelectric layer surrounded by insulating layers. 
(Right) Variation of the hysteresis of in-plane polarization vs. in-plane electric field. The vertical displacement of the vortices is switched with an in-plane electric field.
Reproduced from \cite{kavle2024highly} [\copyright , get permission]}
\label{fig:tpol_phase_transition}
\end{figure}

E) Phase transition transcapacitor, operating by switching the phases of polarization, e.g. from the the state with to a state without vortices \cite{kavle2024highly} (Fig.~\ref{fig:tpol_phase_transition}).
Vortices are formed in a thin FE layer. A gate field destroys them and varies polarizability.
Potentially large transcapacitance.
Vortex state difficult to observe.
Buckled ‘vortex tubes’ in the plane of the layer are shown, giving rise to a second order of ferroelectricity that emerges from the domain structure \cite{behera2025anisotropic}.

F) In this paper, we have mainly focused on solid-state devices. In addition, the MEMS-based transcapacitor 
has been proposed \cite{galisultanov2017capacitive,pillonnet2017adiabatic}. 
It is based on the mechanical motion of one of the electrodes of the capacitor (placed in a cantilever) relative to the other electrode (placed on a substrate). When the cantilever touches the substrate, the capacitance is greater, and when it is moved away, the capacitance is lower. The motion of the cantilever is controlled by the electrostatic force applied from a different gate electrode. It is understandable that this transcapacitor can have both charge gain and voltage inversion if the electrode geometries are engineered correctly. The downside of this type is the difficulty to achieve the necessary reliability, as the experience with other devices with moving parts shows. For the integrated circuit implementation, scaling the size to the nanometer range is highly desired.

\end{document}